\documentclass[twoside,12pt]{article}

\usepackage{lineno}
\modulolinenumbers[1]
\usepackage{hyperref}
\usepackage{xcolor}
\usepackage{bbding}
\usepackage{pifont}
\usepackage{amssymb}
\usepackage{amsmath}
\usepackage{wasysym}
\usepackage{multirow}
\usepackage{graphicx}
\usepackage{slashed}
\usepackage{xspace}
\usepackage{siunitx}

\def\mum{\mu{\rm m}}  
\def\cm{{\rm cm}}  
\def\mm{{\rm mm}}  
\def\gev{{\rm GeV}}   
\def\tev{{\rm TeV}}   
\def\invfb{{\rm fb}^{-1}}
\def\invpb{{\rm pb}^{-1}}

\def\pt{\ensuremath{p_{\mathrm{T}}} \xspace} 
\def\pT{\ensuremath{p_{\mathrm{T}}} \xspace}

\graphicspath{{figures/}}

\begin{document}


\title{
Collider Searches for Long-Lived Particles\\Beyond the Standard Model}

\author{Lawrence Lee$^{1}$, Christian Ohm$^{2,3}$, Abner Soffer$^{4}$, Tien-Tien Yu$^{5,6}$\
\\
\\
$^1$Harvard University, Cambridge, MA, USA\\
$^2$KTH Royal Institute of Technology, Stockholm, Sweden\\
$^3$Oskar Klein Centre, Stockholm, Sweden\\
$^4$School of Physics and Astronomy, Tel Aviv University, Tel Aviv, Israel\\
$^5$European Organisation for Nuclear Research (CERN), Geneva, Switzerland\\
$^6$Institute of Theoretical Science, University of Oregon, Eugene, Oregon, USA\\
}


\maketitle

\begin{abstract}
Experimental tests of the Standard Model of particle physics (SM) find excellent agreement with its predictions. Since the original formation of the SM, experiments have provided little guidance regarding the explanations of phenomena outside the SM, such as the baryon asymmetry and dark matter. Nor have we understood the aesthetic and theoretical problems of the SM, despite years of searching for physics beyond the Standard Model (BSM) at particle colliders. Some BSM particles can be produced at colliders yet evade being discovered, if the reconstruction and analysis procedures not matched to characteristics of the particle. An example is particles with large lifetimes. As interest in searches for such long-lived particles (LLPs) grows rapidly, a review of the topic is presented in this article. The broad range of theoretical motivations for LLPs and the experimental strategies and methods employed to search for them are described. Results from decades of LLP searches are reviewed, as are opportunities for the next generation of searches at both existing and future experiments. 
\end{abstract}


\newpage
\tableofcontents
\newpage


\section{Introduction}
\label{sec:intro}

The Standard Model of particle physics (SM) is a mathematically elegant theory
that describes fundamental physics and provides high-precision predictions consistent with decades of experimental studies. 
Nonetheless, it has several important shortcomings that are of primary interest for current research in the field. Of particular relevance to the research reported here is the fact that the SM offers no explanation for the gauge hierarchy problem, the existence of dark matter, the baryon asymmetry of the universe, and the origin of neutrino masses~\cite{Tanabashi:2018oca}. 

To address these inadequacies, many theories and models beyond the Standard Model (BSM) have been proposed. These generically predict new particles, in addition to those of the SM, that are in many cases observable at particle colliders. However, despite decades of searches, direct evidence for BSM particles has not been seen. This situation has resulted in the development of new ideas and methods, both theoretical and experimental, that push the search for BSM physics beyond previously studied regimes. One such frontier involves new particles with long lifetimes. This review summarizes developments and results from searches with collider experiments for long-lived particles (LLPs) that can be detected through
\begin{itemize}
\vspace{-0.2cm}
    \item their direct interactions with the detector, or\vspace{-0.2cm}

    \item their decay, occurring at a discernible distance from their production point. \vspace{-0.2cm}

\end{itemize}

\begin{figure}[b!]
\centering
\includegraphics[width=4.5in]{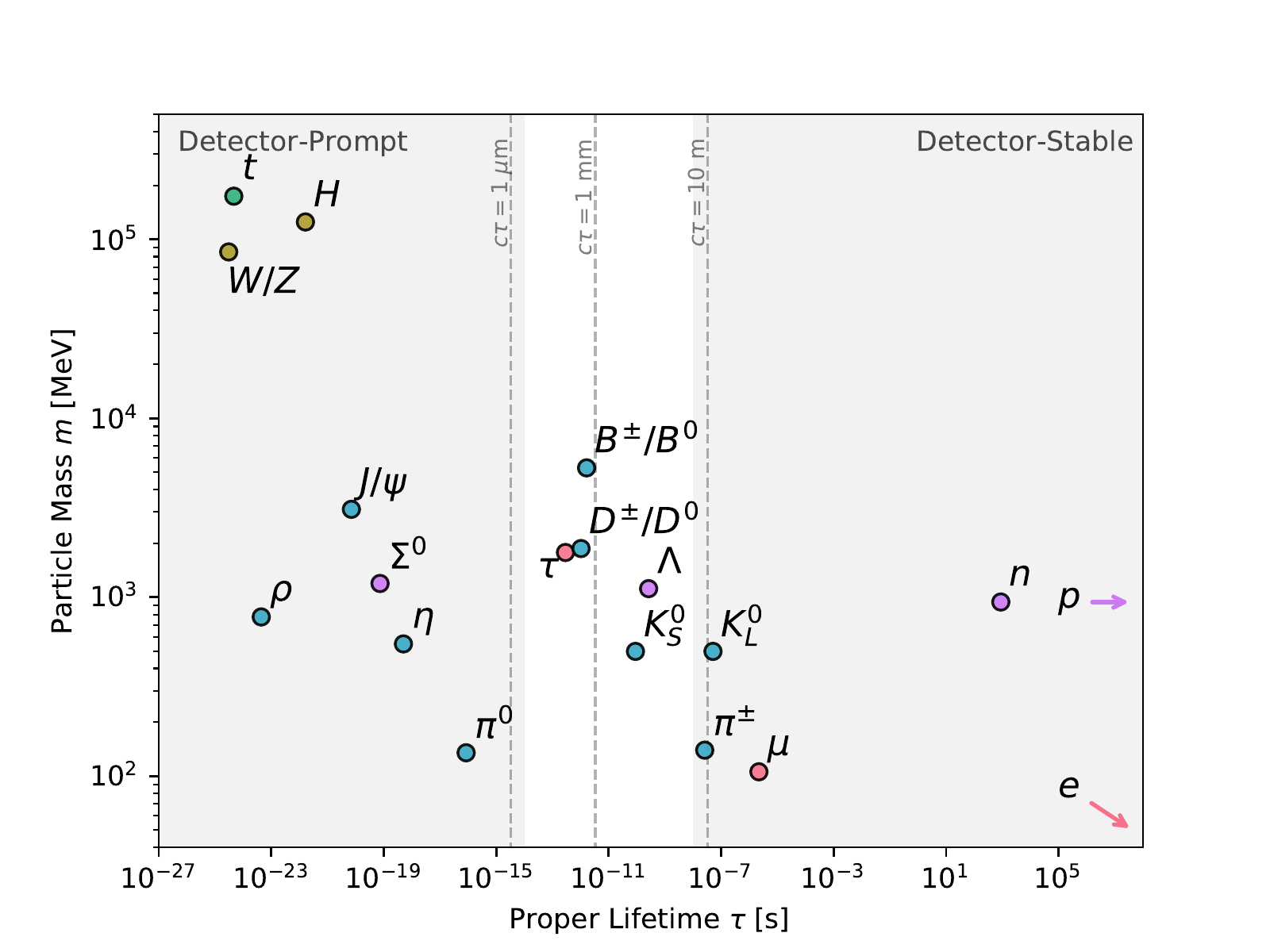}
\caption{The SM contains a large number of metastable particles. A selection of the SM particle spectrum is shown as a function of mass and proper lifetime. Shaded regions roughly represent the detector-prompt and detector-stable regions of lifetime space, for a particle moving at close to the speed of light.}
\label{fig:smllpsummary}
\end{figure}

Collider searches for BSM phenomena motivated by the problems of the SM have largely assumed that decays of new particles occur quickly enough that they appear prompt. This expectation has impacted the design of the detectors, as well as the reconstruction and identification techniques and algorithms. However, there are several mechanisms by which particles may be \emph{metastable} or even stable, with decay lengths that are significantly larger than the spatial resolution of a detector at a modern collider, or larger than even the scale of the entire detector. The impact of such mechanisms can be seen in the wide range of lifetimes of the particles of the SM, some of which are highlighted in Fig.~\ref{fig:smllpsummary}. Decay-suppression mechanisms are also at play in a variety of BSM scenarios. Thus, it is possible that BSM particles directly accessible to experimental study are long-lived, and that exploiting such signatures would discover them in collider data. 

The realization that LLPs are a crucial part of the BSM collider search program has led to development of theoretical models that give rise to LLPs, reconstruction techniques exploiting their signatures, and experimental searches aiming to discover LLPs at previous accelerator facilities~\cite{Fairbairn:2006gg}.
More recently, many searches for such LLPs have been conducted by experiments at the \emph{Large Hadron Collider} (LHC)~\cite{1748-0221-3-08-S08001} at CERN. Since 2010, the LHC  has been collecting data from proton-proton collisions at four primary experiments: ATLAS~\cite{1748-0221-3-08-S08003}, CMS~\cite{1748-0221-3-08-S08004}, LHCb~\cite{1748-0221-3-08-S08005}, and ALICE~\cite{1748-0221-3-08-S08002}. 
So far, ATLAS and CMS have each collected a data sample of about $150~\invfb$ at a center-of-mass energy of $\sqrt{s}=13~\tev$. Only part of this sample has already been used for LLP searches, reflecting the time required to complete such an analysis. LLP searches have also been performed at $\sqrt{s}=7$ and~$8~\tev$, with the full data samples of about $5$ and $20~\invfb$, respectively. 
LHCb, which is more sensitive to low-mass particles, has searched for LLPs with $3~\invfb$ of $\sqrt{s}=7$ and $8~\tev$ data, and has yet to use the $5.7~\invfb$ of data collected at $\sqrt{s}=13~\tev$ for this purpose. 

Experiments at other colliders have also searched for LLPs.
Until 2011, the CDF~\cite{cdf1,cdf2} and D0~\cite{dzero} experiments at the Tevatron at Fermi National Accelerator Laboratory~\cite{tevatron} collected a total of about $10~\invfb$ at $\sqrt{s}=1.96~\tev$~\cite{Bandurin:2014bhr}, with smaller samples at lower center-of-mass energies. Samples of up to $3.6~\invfb$ have been used for LLP searches.
The \emph{Large Electron-Positron Collider} (LEP)~\cite{lep,lep2} at CERN operated from 1989 to 2000 with four main experiments, ALEPH~\cite{aleph}, DELPHI~\cite{delphi}, OPAL~\cite{opal}, and L3~\cite{l3}. A sample of $208~\invpb$ was delivered at the $Z$ resonance mass, and a total of $785~\invpb$ were recorded at other energies, up to $\sqrt{s}=209~\gev$~\cite{Assmann:2002th}. 
The $B$-factory experiments BABAR~\cite{Aubert:2001tu} and Belle~\cite{Abashian:2000cg} collected data from $e^+e^-$ collisions at the $\Upsilon(4S)$ resonance mass of $10.59~\gev$, at other $\Upsilon$ resonances, and in the continuum regions off the resonances. Operating between 1999 and 2010, the two experiments collected data samples totaling about $1600~\invfb$. The largest sample used for LLP searches was $711~\invfb$.

In many LLP search analyses performed to date, the SM backgrounds have been extremely small, sometimes much less than one event. In such cases, the search sensitivity grows roughly linearly with the integrated luminosity of the data sample. This is in contrast to background-dominated BSM searches, where sensitivity is proportional to the square root of the integrated luminosity. Therefore, LLP searches are especially attractive for high-luminosity colliders. In particular, this includes the future runs of the LHC~\cite{hl-lhc}, but also those of Belle~II~\cite{Abe:2010gxa} and proposed high-energy $e^+e^-$ facilities such as FCC-ee~\cite{Gomez-Ceballos:2013zzn}.

As the focus of this review is BSM LLP searches at particle colliders, we aim to cover the broad range of theoretical models, their experimental signatures at such facilities, and published searches pursuing them. Thus, other than an occasional mention when relevant, we do not discuss experiments at non-collider facilities or results from astrophysical observations\footnote{For a review of implications of collider-accessible LLPs on cosmology and astroparticle physics, see Ref.~\cite{Fairbairn:2006gg}}.
Furthermore, following the definition of LLP signatures stated above, we do not include signatures without detectable features of the LLP or its decay. 

Basic distance-scale definitions used throughout the review are indicated in Fig.~\ref{fig:smllpsummary}. A particle decay is considered \emph{prompt} if the distance between the particle's production and decay points is smaller than or comparable to the spatial resolution of the detector. By contrast, a distance significantly larger than the spatial resolution characterizes a \emph{displaced} decay. Depending on the relevant detector subsystem, the typical resolution scale is between tens of micrometers to tens of millimeters. The second distance scale of relevance is the typical size of the detector or relevant subsystem, ranging from about $10~\cm$ to 10~m. A particle is \emph{detector stable} if its decay typically occurs at larger distances.

In Sec.~\ref{sec:theory} we review the theoretical motivation and a variety of BSM scenarios that give rise to LLPs. The experimental methods used for identifying LLPs, which frequently give rise to non-standard signatures, are described in Sec~\ref{sec:signatures}. The existing experimental results are summarized in Sec.~\ref{sec:searches}. In Sec.~\ref{sec:constraints} we summarize a selection of experimental constraints on theoretical scenarios. A discussion of the future outlook given planned and proposed experiments appears in  Sec.~\ref{sec:future}. We end with concluding remarks in Sec.~\ref{sec:conclusions}, and a glossary of acronyms in Sec.~\ref{sec:glossary}.

\section{Theoretical Motivation for Long-Lived Particles}
\label{sec:theory}

The proper lifetime of a particle, $\tau$, is given by 
\begin{equation}
\tau^{-1}=\Gamma=\frac{1}{2m_X}\int d\Pi_f|{\mathcal M}(m_X\to \{p_f\})|^2
\end{equation}
where $m_X$ is the mass of the particle, ${\mathcal M}$ is the matrix element for the particle's decay into the decay products $\{p_f\}$, and $d\Pi_f$ is the Lorentz-invariant phase space for the decay. We use $\hbar=c=1$.
For a particle to be long-lived, there must be a small matrix element and/or limited phase space for the decay. 
There are several mechanisms that typically lead to a small matrix  element. One is an approximate symmetry which would, if exact, forbid the operator that mediates the decay. Small breaking of the symmetry results in a small coupling constant for this operator. Another mechanism arises from an effective higher dimension operator. In this case, the coupling constant is suppressed by powers of the scale $\Lambda >>m_X$ at which the decay is mediated. This, in fact, is the mechanism for long lifetimes in the case of weakly decaying particles in the SM. To summarize, for a model to predict LLPs, it must satisfy at least one of the following: 
\begin{itemize}
\item (nearly) mass-degenerate spectra
\item small couplings
\item highly virtual intermediate states.
\end{itemize}
These conditions, and the LLPs that result from them, are generic features of many BSM models developed to address the big open questions of particle physics mentioned in Sec.~\ref{sec:intro}.

In what follows, we categorize the discussion of LLP mechanisms into models of supersymmetry (SUSY), models of Neutral Naturalness, mechanisms of producing dark matter (DM), and portal interactions between a hidden sector and the SM. We also briefly discuss magnetic monopoles.
This section is meant to provide theoretical context for the experimental searches described later in the report, and not as an exhaustive summary of theoretical models. A more detailed description of theoretical models can be found in Ref.~\cite{Curtin:2018mvb}.
Therefore, we give the most attention to those models in which there are existing searches, in particular models of SUSY. 
Note that the different mechanism categories are not mutually exclusive. For example, models of SUSY can also give rise to DM, produced via the mechanisms described in Sec.~\ref{sec:DM}. We summarize the dominant features that gives rise to long lifetimes for the different scenarios in Table~\ref{tab:models}. 

\begin{table}
\centering
\begin{tabular}{c l||c|c|c}
& & Small coupling & Small phase space & Scale suppression\\ \hline
\multirow{4}{*}{\rotatebox[origin=c]{90}{\large{SUSY}}}&GMSB & & & \Checkmark\\ 
&AMSB & &\Checkmark & \\ 
&Split-SUSY & & & \Checkmark\\ 
&RPV &\Checkmark & & \\ \hline
\multirow{3}{*}{\rotatebox[origin=c]{90}{\large{NN}}}&Twin Higgs & \Checkmark & & \\ 
&Quirky Little Higgs & \Checkmark && \\ 
&Folded SUSY & &\Checkmark & \\ \hline
\multirow{3}{*}{\rotatebox[origin=c]{90}{\large{DM}}}&Freeze-in &\Checkmark & & \\ 
&Asymmetric & & & \Checkmark\\ 
&Co-annihilation & & \Checkmark&  \\ 
\hline
\multirow{4}{*}{\rotatebox[origin=c]{90}{\large{Portals}}}
&Singlet Scalars & \Checkmark & & \\
&ALPs &  & & \Checkmark\\
& Dark Photons & \Checkmark & &\\
&Heavy Neutrinos&  & &\Checkmark
\end{tabular}
\caption{Dominant feature that gives rise to long-lived particles in the theoretical models and mechanisms discussed in the text.}
\label{tab:models}
\end{table}

\subsection{Supersymmetry}
\label{sec:susytheory}
The mass of the SM Higgs boson is required to be around the electroweak scale ($M_{EW}\sim 100$ GeV) due to arguments of perturbative unitarity~\cite{Tanabashi:2018oca}. Since the Higgs boson is a scalar particle, it is sensitive to quantum corrections that are proportional to the cutoff energy scale below which the SM is a good effective field theory. In particular, the Higgs boson mass squared diverges quadratically with the cutoff scale. 
For a cutoff scale far above the weak scale, maintaining the Higgs mass at its physical value requires fine tuning of the corresponding SM parameter. This is known as the \emph{Hierarchy Problem}.  Of the solutions to the Hierarchy Problem, supersymmetry is the most well-known and well-studied~\cite{Martin:1997ns}. The dominant contribution of the quadratic divergence of the Higgs mass come from the top quark loop. SUSY protects the weak-scale value of the Higgs mass by introducing a colored scalar partner to the top, $\tilde t$, which cancels out the quadratic divergence. Many models of SUSY give rise to naturally long-lived particles, and have thus served as standard benchmarks in many of the LHC LLP searches. The simplest variation of SUSY is the Minimal Supersymmetric Standard Model (MSSM). If SUSY were an exact symmetry, we would have a spectrum of superpartners that would be mass degenerate with the SM particles. Since we have not observed these particles, we know that SUSY must be a broken symmetry. Within the MSSM, one has a variety of options for breaking SUSY, which in turn determine the phenomenology.
\subsubsection{Gauge-Mediated SUSY Breaking}
The simplest SUSY model that gives rise to LLP signatures is \emph{Gauge-Mediated SUSY Breaking} (GMSB)~\cite{Giudice:1998bp}.
In GMSB, SUSY is broken via the gauge interactions of the chiral messenger superfields, $\Phi$, which interact with the goldstino superfield $X$ through the superpotential
\begin{equation}
W=\lambda_{ij}\bar\Phi_i X \Phi_j.
\end{equation}
SUSY is broken when $X$ acquires a \emph{vacuum expectation value} (vev) along the scalar and auxiliary components,
\begin{equation}
\langle X \rangle = M+\theta^2 F,
\end{equation}
where $M$ is the messenger mass scale and $\sqrt{F}$ is proportional to the mass splitting inside the supermultiplet. 

One feature of GMSB is that the gravitino, $\widetilde G$, is typically the lightest supersymmetry partner (LSP), and that the attributes that give rise to LLP signatures depend only on $F$. In particular, the next to lightest superpartner (NLSP) decays to the gravitino and a SM particle via higher dimensional operators that are suppressed by $1/F$. The mass of the gravitino is given by
\begin{equation}
m_{\widetilde G}=\frac{F}{k\sqrt{3} M_{Pl}},
\label{eq:mgravitino}
\end{equation}
where $M_{Pl}=(8\pi G_N)^{-1/2}\simeq 2.4\times 10^{18}$ GeV is the reduced Planck mass and $G_N$ is the gravitational constant. The constant $k\equiv F/F_0 < 1$, where $F_0$ is the fundamental scale of SUSY breaking, depends on how SUSY breaking is communicated to the messengers. The suppression by $M_{Pl}$ results in the gravitino being very light. 

\begin{figure}
    \centering
    \includegraphics[width=0.35\textwidth]{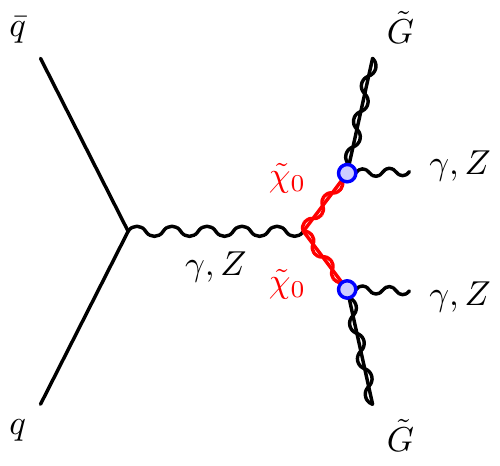}
    \includegraphics[width=0.35\textwidth]{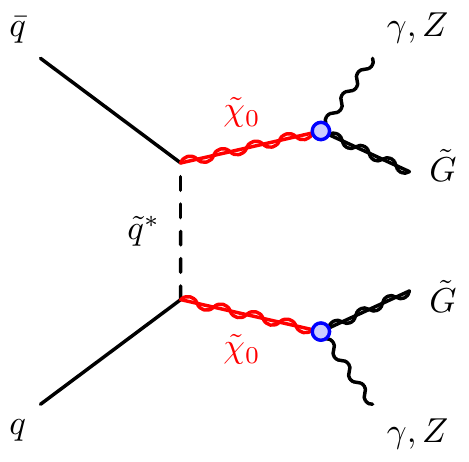}
    \caption{Example of long-lived neutralino NLSP ($\tilde\chi_0$)
    production at a hadron collider through either an $s$-channel $Z$ (left) or $t$-channel squark exchange (right). The neutralino decays predominantly into a gravitino $\tilde G$ and a $\gamma$ or $Z$ for a Bino-like or Higgsino-like neutralino, respectively. The blue circle denotes the vertex that makes $\tilde\chi_0$ long-lived.
    }
    \label{fig:GMSB}
\end{figure}

If the neutralino $\widetilde\chi_1^0$ is the NLSP, its inverse decay width is given by 
\begin{equation}
 \Gamma^{-1}(\widetilde\chi_1^0\to\widetilde G +{\rm SM})=\frac{16\pi F^2}{k^2\kappa_i m_\chi^5}\sim  \frac{1}{\kappa_i}\left(\frac{\sqrt{F/k}}{10^6~{\rm GeV}}\right)^4\left(\frac{300~{\rm GeV}}{m_\chi}\right)^5\times10^{-2} ~{\rm ns} ,
\label{eq:tau-gmsb}
\end{equation}
where $m_\chi$ is the mass of the neutralino and $\kappa_i$ is a parameter that depends on the neutralino mixing matrix. For example, if $\tilde\chi_1^0$ is a pure Bino, the superpartner of the SM $U(1)$ gauge boson, then the decay is dominantly into a photon with $\kappa_i=\kappa_\gamma\equiv |N_{11}\cos\theta_W+N_{12}\sin\theta_W|^2$ where $\theta_W$ is the weak-mixing angle and $N_{1i}$ are the components of $\tilde\chi_0$ in standard notation~\cite{HABER198575}.
We see that $\sqrt{F/k}\sim 10^6$~GeV gives rise to a long-lived neutralino that decays to a displaced photons or $Z$ via the diagrams shown in Fig.~\ref{fig:GMSB}.  
In general, the long-lived neutralino NLSP can be a mixture of the Bino, Wino, and Higgsino gauge eigenstates, leading to a wider variety of final states than described in this section~\cite{Ruderman:2011vv}. 
Although the Higgsino and Wino are not the NLSP in the most minimal version of GMSB, they can occur in General Gauge Mediation~\cite{Meade:2008wd,Buican:2008ws,Cheung:2007es} and with potentially interesting long-lived signatures~\cite{Meade:2010ji,ATL-PHYS-PUB-2017-019}.

\subsubsection{Anomaly-Mediated SUSY Breaking}
\label{sec:amsb}
One can also break SUSY through a combination of anomaly and gravity effects. This is known as \emph{Anomaly-mediated SUSY breaking} (AMSB) and gives rise to a different pattern of masses and signatures from those of GMSB. In general, the superconformal anomaly will give rise to soft mass parameters that break SUSY~\cite{Randall:1998uk,Giudice:1998xp}.
In fact, this effect is present in any model with SUSY breaking, but is subdominant if there are other mechanisms for SUSY breaking, such as GMSB.

In pure anomaly mediation, the gaugino masses are given by 
\begin{equation}
M_i=\frac{\beta(g_i^2)}{2g_i^2}m_{\tilde G},
\end{equation}
where $g_i$ is the gauge coupling constant for gauge groups $i=1,2,3$, corresponding to $U(1), SU(2)$, and $SU(3)$, respectively,
$\beta(g_i^2)$ is the corresponding renormalization group beta-function~\cite{Gherghetta:1999sw}, and $m_{\tilde G}$ is the gravitino mass.

\begin{figure}
    \centering
    \includegraphics[width=0.3\textwidth]{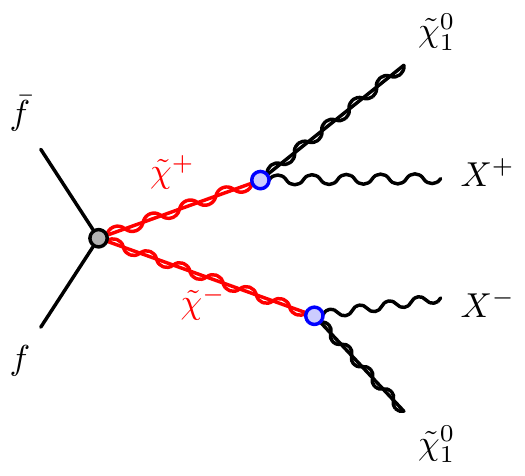}
    \includegraphics[width=0.3\textwidth]{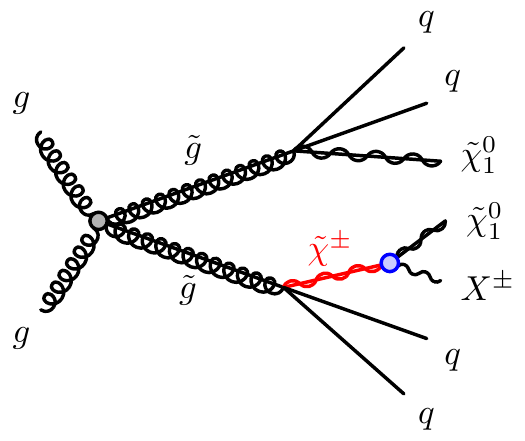}
    \includegraphics[width=0.3\textwidth]{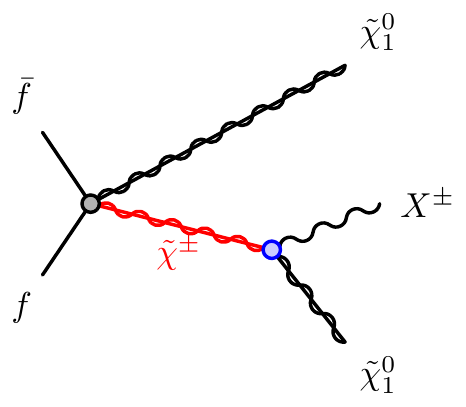}
    \caption{Various production modes for the long-lived chargino ($\chi^\pm$) at a hadron collider in models of AMSB. The chargino then decays to a neutralino $\tilde\chi_0$ and a soft SM particle $X^\pm$.
    }
    \label{fig:AMSB}
\end{figure}

AMSB predicts mass ratios of $M_1:M_2:M_3\simeq 3:1:7$, so that the Wino is the LSP. 
One consequence of this mass hierarchy, and a defining feature of AMSB, is that the lightest chargino is nearly mass degenerate with the lightest neutralino due to an approximate custodial symmetry~\cite{SIKIVIE1980189}. The mass difference is given by~\cite{Giudice:1998xp}
\begin{equation}
m_{\widetilde\chi_\pm}-m_{\widetilde \chi_0}=\frac{M_W^4}{\mu^3}\sin 2\beta+\frac{M_W^4\tan^2\theta_W}{(M_1-M_2)\mu^2}\sin^2 2\beta,
\end{equation}
where $M_W$ is the mass of the $W$ boson, $\mu$ is the supersymmetric Higgs mass, and $\tan\beta$ is the ratio of up and down-type Higgs vevs. 
$\widetilde\chi_\pm$ decays with inverse decay width~\cite{Asai:2008sk}
\begin{equation}
 \Gamma^{-1}(\widetilde \chi^\pm\to\widetilde \chi_0+X^\pm)\sim \left(\frac{800~{\rm MeV}}{m_{\widetilde \chi_\pm}-m_{\widetilde \chi_0}}\right)^3\times10^{-3}~{\rm ns},
\end{equation}
where $X$ is a SM particle, \emph{e.g.} $\widetilde W^\pm\to\widetilde W_0+\pi^\pm$. 
 In contrast to the GMSB scenario above, where the NLSP was a neutralino, the NLSP here is the chargino. Being long-lived and charged, it directly interacts with the detector, leaving a unique track signature. Several production modes for the chargino are shown in Fig.~\ref{fig:AMSB}.

\subsubsection{Split-SUSY}
\label{sec:split-susy}
Models of \emph{split-SUSY}~\cite{ArkaniHamed:2004fb,Giudice:2004tc} give rise to long-lived gluinos, which can have interesting signatures at the LHC~\cite{Hewett:2004nw,Kilian:2004uj}. In these models, SUSY is no longer the solution to the hierarchy problem. Instead, SUSY breaking occurs at a scale of $m_S\gg 1000~\tev$, 
and all the scalars are ultra-heavy, except for one, which serves as the Higgs boson. By contrast, the fermions, particularly the gluino,
can have weak-scale masses due to chiral symmetries. This setup solves some of the issues in other SUSY models, including the absence of experimental evidence of superpartners, 
avoids proton decay, 
solves the SUSY flavor and CP problems, as well as the cosmological gravitino and moduli problems, but at the expense of a fine-tuning~\cite{Martin:1997ns}. 

\begin{figure}
    \centering
    \includegraphics[width=0.35\textwidth]{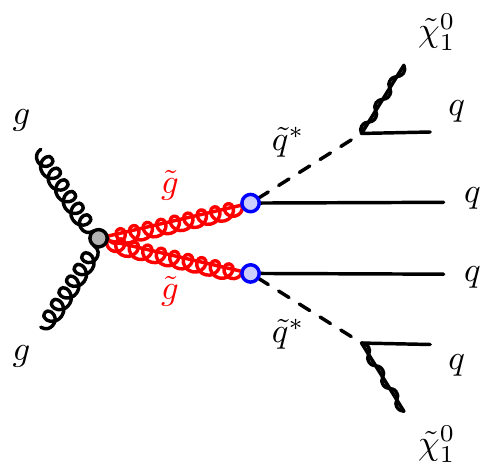}
    \caption{Production of long-lived gluinos ($\tilde g$) in a hadron collider, which subsequently decay through an off-shell squark in models of split-SUSY.
    }
    \label{fig:splitSUSY}
\end{figure}

The long lifetime of the gluino arises due to the fact that it can only decay through a virtual squark, as shown in Fig.~\ref{fig:splitSUSY}.
Since the squarks are ultra-heavy by construction, this decay is highly suppressed.
The effective operators 
that give rise to gluino decay are higher-dimensional and suppressed by the squark mass scale, $m_S$:
\begin{equation}
{\cal O}^{(6)}\sim \frac{g_s^2}{m_S^2}\bar q\tilde g\bar{\tilde \chi} q,~~~{\cal O}^{(5)}\sim \frac{g_s^2}{16\pi^2m_S}\tilde g\sigma_{\mu\nu}\tilde \chi G^{\mu\nu},
\end{equation}
where $g_s$ is the strong-coupling constant. These operators lead to the gluino decays $\tilde g\to\chi^0 g, \tilde g\to\chi^0 q\bar q$, and $\tilde g\to\chi^\pm q\bar q^\prime$. An example of the second process is given in Fig.~\ref{fig:splitSUSY}.
Parametrically, the inverse of the gluino decay width is~\cite{Gambino:2005eh}
\begin{equation}
 \Gamma^{-1}\sim\frac{4}{N}\left(\frac{m_S}{10^3~{\rm TeV}}\right)^4\left(\frac{1~{\rm TeV}}{m_{\tilde g}}\right)^5\times10^{-4}~\rm{ns},
\end{equation}
where $N$ is an ${\cal{O}}(1)$ normalization factor that depends on the exact parameters of the theory, as well as on the particular decay channel.

There are several mass scales to note. For $m_S>10^3~\gev$, the gluino is long-lived enough to hadronize into a color-singlet state known as an $R$-hadron before decaying~\cite{Farrar:1978xj}. The bosonic gluino $R$-baryon is composed of a gluino and $qqq$ states, while the fermionic gluino $R$-meson and $R$-glueball are formed through a gluino binding to $q\bar q$ and gluon states, respectively. The $R$-hadron flavor structure is analogous to that of ordinary baryons, mesons, and glueballs~\cite{Tanabashi:2018oca}.  For $m_S>10^6~\gev$, the gluino travels macroscopic distances before decaying, and for $m_S>10^7~\gev$ it typically decays outside the detector or is stopped in the detector material. At $m_S>10^9~\gev$, the $R$-hadrons may begin to affect nucleosynthesis in the early universe, and at $m_S>10^{13}~\gev$ it is effectively stable, since it has a lifetime longer than the age of the universe. 

The mass spectrum of allowed $R$-hadron states has been studied in a variety of ways. Simple models based on constituent-quark and gluon masses give predictions for mass splittings between various states~\cite{PhysRevD.12.147,Kraan:2004tz,Farrar:2010ps,BUCCELLA1985311}. Limited calculations from lattice QCD also exist for certain simplified states~\cite{Marsh:2013xsa}. The phenomenology of $R$-hadron detection can depend greatly on the mass spectrum, especially for the identity of the lightest of these states. Heavier states will tend to cascade to the lightest state in interactions with material, and the charge of this lightest state impacts the character of allowed signatures. Neutral $R$-hadrons lose energy through hadronic interactions, while charged ones also lose energy via ionization. Due to hadronic scattering, $R$-hadrons can change electric charge as they pass through detector material, and can also become doubly charged, giving rise to unique signatures~\cite{Kraan:2004tz,deBoer:2007ii,Mackeprang:2006gx,Mackeprang:2009ad}. $R$-hadrons that decay inside the detector can be detected via displaced or delayed decays, as well as ``disappearing'' tracks. These signatures are discussed in Sec.~\ref{sec:signatures}.

\subsubsection{SUSY Models with \texorpdfstring{$R$}{R}-Parity Violation}
In all the models discussed in the previous sections, there is an implicit global $\mathbb{Z}_2$ symmetry known as $R$-parity, with quantum number $R_p=(-1)^{3(B-L)+2S}$, where $B$, $L$, and $S$ are baryon number, lepton number, and spin, respectively. All SM particles have $R_p=+1$ and their superpartners have $R_p=-1$. This forbids dangerous tree-level renormalizable operators that violate baryon and lepton number, which can lead to proton decay and flavor violation. However, the $\mathbb{Z}_2$ symmetry is not theoretically required for supersymmetry. Therefore, one can remove it and allow for more general \emph{$R$-parity-violating} (RPV) interactions, as long as the experimental constraints are satisfied.
Models of RPV SUSY have been studied extensively~\cite{Barbier:2004ez,Kon:1994xe,Dreiner:1997uz,Barry:2013nva}.

The most general renormalizable Lagrangian with
RPV operators is, 
in term of the left-handed chiral superfields, 
\begin{equation}
W_{\rm RPV}=\mu_i L_i H_u + \frac{1}{2}\lambda_{ijk} L_i L_j\bar e_k +\lambda'_{ijk}L_iQ_j\bar d_k +\frac{1}{2}\lambda{''}_{ijk}\bar u_i\bar d_j\bar d_k,
\end{equation}
where $\mu_i, \lambda_{ijk}, \lambda'_{ijk}, \lambda''_{ijk}$ are the coefficients for
the RPV interactions. For example, non-zero values for both $\lambda'$ and $\lambda''$ 
lead to proton decay. In models of dynamical RPV (dRPV)~\cite{Csaki:2013jza}, $R$-parity is conserved at some high-scale, and its breaking is communicated to the visible sector at a mediating scale $M$. As a result, additional non-holomorphic
operators are generated in the K\"ahler potential part of the superpotential.
These take the form
\begin{equation}
    W_{nh{\rm RPV}}=\kappa_i\bar e_i H_d H_u^\dagger+\kappa_i^\prime L_i^\dagger H_d+\eta_{ijk}\bar u_i\bar e_j\bar d_k^\dagger+\eta^\prime_{ijk}Q_i\bar u_j L^\dagger_k+\frac{1}{2}\eta^{\prime\prime}_{ijk} Q_i Q_j\bar d_k^\dagger,
\end{equation}
and couple to the SUSY-breaking field $X=M+\theta^2 F_X$.

Since RPV operators 
are highly constrained from flavor measurements and non-observation of proton decay~\cite{Chemtob:2004xr}, the RPV coefficients must be small. In the case of the non-holomorphic operators, the coefficients are small since they are suppressed by $\epsilon_X\equiv F_X/M^2$, which can be as small as ${\mathcal O}(10^{-16})$, depending on the SUSY-breaking mediation scheme. 
As a result, particles with long-lifetimes are a generic feature of RPV theories. 

\begin{figure}
    \centering
    \includegraphics[width=0.35\textwidth]{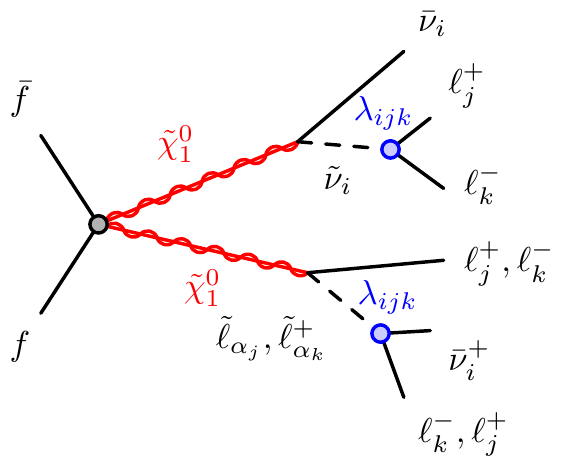}~~
     \includegraphics[width=0.35\textwidth]{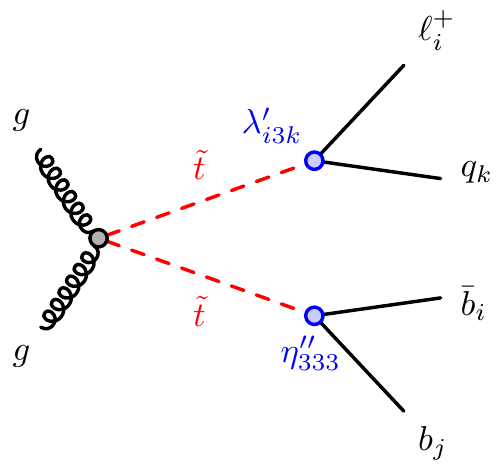}
    \caption{(Left) Decays of the long-lived neutralino $\tilde{\chi}^0_1$ into 3 leptons via the RPV coupling $\lambda_{ijk}$ (circles). (Right) Decay of a long-lived $\tilde t$ via either the RPV $\lambda^\prime$ or $\eta^{\prime\prime}$ couplings (circles).}
    \label{fig:RPVlambda}
\end{figure}

One experimental signature of RPV can be displaced decays of the LSP~\cite{Aad:2012zx,Khachatryan:2014mea}. 
For example, a neutralino can decay into a lepton and two quarks via an off-shell slepton (shown in Fig.~\ref{fig:RPVlambda}), with a mean inverse decay width of~\cite{Dreiner:1991pe}
\begin{equation}
     \Gamma^{-1}(\widetilde\chi_0\to \ell_i q_j q_k)\sim \left(\frac{m_{\tilde {\ell}i}}{750{~\rm GeV}}\right)^4\left(\frac{100{~\rm GeV}}{m_{\widetilde \chi_0}}\right)^5\left(\frac{10\times 10^{-5}}{\lambda_{ijk}^\prime}\right)^2\times0.1{~\rm ns},
\end{equation}
where the indices $i, j, k$ denote the lepton and quark generations. 
The lightest stop can decay directly into a lepton and quark with a inverse decay width of
\begin{equation}
     \Gamma^{-1}(\widetilde t_1\to \ell^+_i q_k)\sim \left(\frac{500{~\rm GeV}}{m_{\widetilde t_1}}\right)\left(\frac{10^{-7}}{\lambda'_{ijk}}\right)^2\left(\frac{0.12}{\cos^2\theta_t}\right)\times 10^{-3}{~\rm ns},
\end{equation}
where $\theta_t$ is the mixing angle between the left- and right-handed stops.  
Recently, there have been several studies on the LHC signature of LLPs with hadronic RPV decays~\cite{Csaki:2015uza,Liu:2015bma}. For example, the $\eta_{333}^{\prime\prime}$ coefficient 
induces a $\tilde t\to\bar b\bar b$ decay with inverse decay width
\begin{equation}
     \Gamma^{-1}(\tilde t\to\bar b\bar b)\sim \left(\frac{300{~\rm GeV}}{m_{\tilde t}}\right)\left(\frac{M}{10^9{~\rm GeV}}\right)^2\left|\frac{1}{\eta_{333}^{\prime\prime}}\right|^2\times 1.5{~\rm ns}.
\end{equation}
We also note that if the LSP is a long-lived stop, it forms an $R$-hadron.

\subsection{Neutral Naturalness}
\label{sec:neutralnaturalness}
An alternative class of models to solve the Hierarchy Problem involves models of \emph{Neutral Naturalness}. These models rely on discrete symmetries that result in colorless top partners that protect the weak-scale, in contrast to the colored top partners in traditional SUSY models. 

Neutral naturalness includes the Twin Higgs~\cite{Chacko:2005pe,Chacko:2005un}, Folded SUSY~\cite{Burdman:2006tz}, and Quirky Little Higgs~\cite{Cai:2008au} models. These models lead naturally to \emph{Hidden Valley} scenarios~\cite{Strassler:2006im,Strassler:2006ri,Han:2007ae}, in which there is a confining hidden sector that is neutral under the SM and only interacts with the SM through so-called portal-type interactions,
which we discuss in more details in Sec.~\ref{sec:portals}. Such models can lead to various signatures at colliders, and have been studied in the context of emerging jets~\cite{Schwaller:2015gea}, \emph{Soft Unclustered Energy Patterns} (SUEPs, also known as \emph{soft bombs})~\cite{Knapen:2016hky}, 
and semi-visible jets~\cite{Cohen:2015toa,Cohen:2017pzm} to name a few.

\emph{Twin Higgs} is a class of pseudo-Nambu-Goldstone Boson (pNGB) models with two exact copies of the SM 
related by a discrete $\mathbb{Z}_2$ symmetry, and a scalar potential 
\begin{equation}
V = \frac{\lambda}{2} \left( |H|^2 + |H^T|^2 \right)^2  ~~,
\label{eq:portal}
\end{equation}
where $H$ and $H^T$ are the Higgs doublets of the SM and the Twin sector, respectively.
The key feature is that this scalar potential respects a $\rm{SU}(4)$ symmetry. In the vacuum, this symmetry is spontaneously broken,
giving rise to a vev $\langle H^T \rangle = f/\sqrt{2}$, and the SM Higgs emerges as a light pNGB. 
Since the quadratic corrections to both doublets are equal and respect the $\rm{SU}(4)$ symmetry, they do not contribute to the mass of the SM Higgs. As a result, the weak scale is protected from quadratic corrections. The most minimal version of the Twin Higgs is one in which the Twin sector contains only the third generation fermions, and is known as the \emph{Fraternal Twin Higgs}~\cite{Craig:2015pha}. In this scenario, the Twin gluons are the lightest objects charged under the Twin color. These Twin gluons can hadronize into long-lived glueball states, which then decay back to SM particles through the Higgs portal~\cite{Juknevich:2009gg,Juknevich:2009ji}. Models of \emph{Folded SUSY} are similar in spirit to those of the Twin Higgs, but instead of having a twin copy of the SM gauge groups, they only have a twin $SU(3)$. 

 In the \emph{Quirky Little Higgs} model, the top partner is an uncolored ``top quirk" charged under a hidden $SU(3)$ gauge group. Quirks and anti-quirks are stable, heavy
 particles that are connected by a flux tube of the dark gluons of the hidden $SU(3)$~\cite{Kang:2008ea}. In QCD, the quark mass is much smaller than the confining scale, $m_q\ll \Lambda_{\rm QCD}$, and so the gluon flux tube easily breaks into multiple bound states.
 By contrast, quirky models have the opposite hierarchy between the quirk mass and the confining scale, $m_Q\gg \Lambda$. As a result, the dark gluon flux tube does not break easily and instead causes the quirks to have macroscopic oscillations before they eventually annihilate. This leads to exotic signatures~\cite{Knapen:2017kly}, particularly when the quirk is electrically charged.  

The coupling between the SM Higgs $h$ and the top partners in models of neutral naturalness induces a loop-level coupling of the Higgs to the hidden gluons. The resulting effective coupling is of the form
\begin{equation}
{\cal L}\supset\theta^2\frac{\widehat \alpha_3}{12\pi}\frac{h}{v}\widehat G^a_{\mu\nu}\widehat G^{\mu\nu}_a,
\label{eq:NNoperator}
\end{equation}
where  $\widehat\alpha_3$ is the Twin $SU(3)$ coupling, $\widehat G_a^{\mu\nu}$ is the Twin gluon field strength, 
$\theta$ is a model-dependent mixing angle, and $v$ is the vev of $H$. For the Twin or Quirky Little Higgs models, $\theta^2\simeq v^2/f^2$, where $f$ is the scale of spontaneous global symmetry breaking. For Folded SUSY, $\theta^2\simeq m_t^2/2 m_{\tilde t}^2$, where $\tilde t$ is the scalar top partner. 

The hidden $SU(3)$ sector contains a spectrum of glueball states. The lightest of these is typically the scalar $G_{0^{++}}$ (where $0^{++}$ indicates its $J^{PC}$ quantum numbers), which can decay back into SM states through the Higgs-portal interaction of Eq.~\ref{eq:NNoperator}. This can result in exotic Higgs decays 
at the LHC~\cite{Curtin:2015fna} or displaced decays of the glueball with a mean inverse decay width of~\cite{Craig:2015pha} 
\begin{equation}
 \Gamma^{-1}(G_{0^{++}}\to h^*\to XX)\sim \left(\frac{10{~\rm GeV}}{m_0}\right)^7\left(\frac{f}{5{~\rm TeV}}\right)^4\times 10^3{~\rm ns},
\end{equation}
where $m_0$ is the mass of $G_{0^{++}}$,
$X$ is a SM state, and $h^*$ is an off-shell higgs.

\begin{figure}
    \centering
    \includegraphics[width=0.35\textwidth]{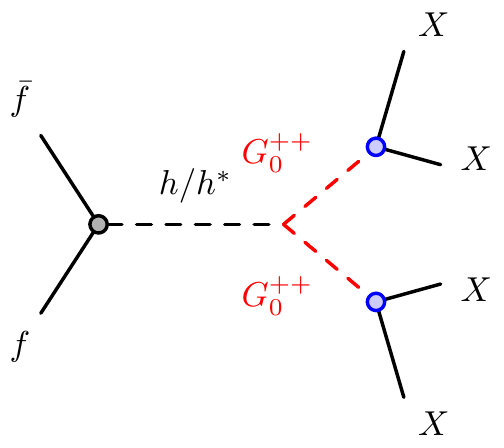}  ~~~~~~~  \includegraphics[width=0.35\textwidth]{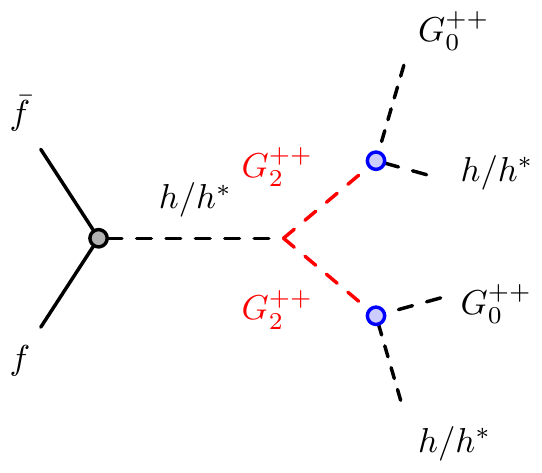}
    \caption{Production of the long-lived glueballs $G_{0^{++}}$ (left) and $G_{2^{++}}$ (right) at a hadron collider through Higgs decay, and their subsequent decay into SM particles $X$ or $G_0^{++}$ and SM particles via Higgs decay.
    }
    \label{fig:TH}
\end{figure}

The next lightest glueball, $G_{2^{++}}$, has a mass of $m_2 \sim 1.4\, m_0$, and is metastable. It predominantly decays through radiative Higgs production, $G_{2^{++}} \to G_{0^{++}} h$. 
Depending on the parameters, 
the $G_{2^{++}}$ can be long-lived and give rise to displaced vertices at the LHC. 
Diagrams for production and decay of these glueballs are shown in Fig.~\ref{fig:TH}.

In addition, there is a spectrum of twin-quarkonia states. In particular, the phenomenology is enriched if there is also a sufficiently light twin bottom quark. This may lead to an overall enhancement in the twin hadron production rate, giving rise to some combination of twin glueballs and twin bottomonium~\cite{Juknevich:2009ji,Craig:2015pha}.

\subsection{Dark Matter}
\label{sec:DM}
In this section, we move from a discussion on models to one on mechanisms related to dark matter that give rise to LLPs. We present different ways to populate the observed cosmological DM relic abundance, which also give rise to LLPs at colliders. Specifically, we discuss Freeze-in DM, Asymmetric DM, and Co-annihilating DM, and give explicit examples of how these mechanisms are manifested in models of SUSY. 

\subsubsection{Freeze-in DM}
Models of \emph{Freeze-in} DM~\cite{Hall:2009bx} generically give rise to LLPs in colliders, and have been studied in much detail (see {\emph{e.g.}}~\cite{Hall:2010jx,Cheung:2010gk,Cheung:2011nn,Co:2015pka,Garny:2018ali,Heeba:2018wtf}). 
The freeze-in mechanism is effectively the inverse of the well-known thermal ``freeze-out'' mechanism~\cite{Kolb:1990vq}, 
and works by populating the DM abundance through $\chi_2\to\chi_1 + X$ decays, where $\chi_2$ is in thermal equilibrium in the early universe, $\chi_1$ is the DM particle, and $X$ represents one or more SM particles. For example, in one specific realization of ``freeze-in" in the context of SuperWIMP theories~\cite{Feng:2003uy,Feng:2004zu}, $\chi_2$ is a charged slepton that decays into a lepton and the gravitino, $\tilde\ell^\pm\to\ell^\pm\tilde G$, as discussed in~\cite{Cheung:2010gk}. 

The key feature of freeze-in models is that the interaction between $\chi_2$ and $\chi_1$ is given by a very feeble coupling $g_{12}$, such that $\chi_1$ is thermally decoupled from the plasma. The feebleness of $g_{12}$ results in a long lifetime for $\chi_2$, which can be seen via displaced signatures at colliders. 

The relic abundance of $\chi_1$ is related to the $\chi_2$ decay width $\Gamma_{\chi_2}$ through 
\begin{equation}
 \Omega_{\chi_1} h^2  = \frac{10^{27}}{g_\star^{3/2}} \frac{ m_1 \Gamma_{\chi_2}}{m_2^2}\,,
\end{equation}
where $\Omega_{\chi_1} h^2$ is the cosmological density of $\chi_1$, 
and $g_\star$ is the number of
relativistic degrees of freedom at a temperatures $T \approx m_2$ around the $\chi_2$ mass. In the SM, $g_\star(100{~\rm GeV})\simeq 100$ while 
$g_\star(100{~\rm MeV})\simeq 10$~\cite{Tanabashi:2018oca}.  Taking $\chi_1$ to constitute all of the DM today,  \emph{i.e.} $\Omega_{\chi_1} h^2=0.11$~\cite{Tanabashi:2018oca}, one obtains a prediction for the inverse decay width of $\chi_2$, 
\begin{equation}
 \Gamma^{-1}(\chi_2\to\chi_1+X)\sim\left(\frac{m_1}{100~{\rm{GeV}}}\right)\left(\frac{200~{\rm{GeV}}}{m_2}\right)^2\left(\frac{100}{g_*(m_2)}\right)^{3/2}\times10^6{~\rm{ns}}.
\end{equation}
Thus, the $\chi_2$ is practically stable on detector scales, and can be detected directly if it is electrically charged. 
This direct correlation between the cosmological abundance of dark matter and the lifetime of the NLSP allows for precision collider tests of the freeze-in origin of dark matter. The production of $\chi_2$ at colliders depends on the specific implementation of the freeze-in mechanism.

\subsubsection{Asymmetric DM}
Models of \emph{Asymmetric DM} (ADM)~\cite{Kaplan:2009ag,Zurek:2013wia,Petraki:2013wwa} connect the observed DM abundance to the baryon abundance, and thus explain the relatively similar abundances in the dark and visible sectors. The asymmetry is transferred between the visible and dark sectors through higher dimensional operators of the form

\begin{equation}
    {\cal O}_{ADM}=\frac{{\cal O}_{B-L}{\cal O}_{X}}{\Lambda^{n+m-4}},
\end{equation}
where ${\cal O}_{B-L}$ is a SM operator that contains baryon number minus lepton number but no gauge quantum numbers, ${\cal O}_{X}$ is an operator that contains DM number, and $n,m$ are the dimensions of ${\cal O}_{B-L}$ and ${\cal O}_{X}$, respectively. ADM can be realized in a variety of different ways. For example, in SUSY, the simplest operators giving rise to ADM are given by 
\begin{equation}
    W_{ADM}=XLH\, , \; \frac{XU_i^c D_j^c D_k^c}{\Lambda_{ijk}}\, , \; \frac{XQ_i^c L_j D_k^c}{\Lambda_{ijk}}\, ,\; \mbox{and}~\frac{X L_i L_j E_k^c}{\Lambda_{ijk}}\, , 
\end{equation}
 where $X$ is the supermultiplet containing the DM candidate, $U^c, D^c,E^c$ are the right-handed anti-quarks and charged anti-leptons, $Q,L$ are the left-handed quark and lepton doublets, and $H$ is the Higgs doublet. $i,j,k$ are the flavor indices. 

\begin{figure}
    \centering
    \includegraphics[width=0.35\textwidth]{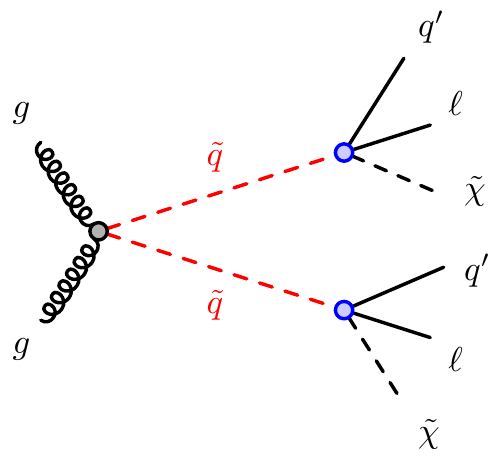}~~~~~~~~
     \includegraphics[width=0.35\textwidth]{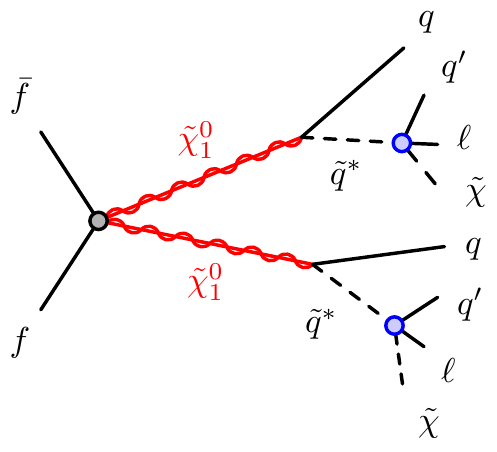}
    \caption{(Left) Decay of the long-lived LSP squark $\tilde{q}$, and 
    (Right) Decay of a long-lived LSP neutralino $\tilde{\chi}_1^0$
    in asymmetric DM scenarios.
}
    \label{fig:ADM}
\end{figure}

These interactions allow the LSP to decay into the $X$-sector 
plus SM particles. Depending on the size of $\Lambda$, this decay can be long-lived. 
As an example, the fermionic operator ${\cal O}_{B-L}=qld^c$ leads to a 3-body decay of the squark LSP, with inverse decay width
\begin{equation}
     \Gamma^{-1} (\tilde q\to q' \ell \tilde x)\sim \left(\frac{(F^{(3-\rm{body})})^{-1}}{10^{-5}~\rm{mm}}\right)\left(\frac{100~\rm{GeV}}{m_{\tilde q}}\right)^3\left(\frac{\Lambda_{ijk}}{100~\rm{TeV}}\right)^2 \times10^{-3}{~\rm ns},
\end{equation}
where we have ignored the final state particle masses. If the neutralino is the LSP, then it 4-body decay proceeds through an off-shell squark, and has inverse decay width 
\begin{eqnarray}
      \Gamma^{-1} (\widetilde\chi_0\to q' \ell \tilde x)\sim  \left(\frac{F^{(4-\rm{body})^{-1}}}{100~\rm{mm}}
     \right)\left(\frac{100~\rm{GeV}}{m_{\widetilde\chi_0}}\right)^7\left(
     \frac{m_{\tilde q} }{500~\rm{GeV}}\right)^4\left(\frac{\Lambda_{ijk}}{100~\rm{TeV}}\right)^2&&\\
     \times x^5[(10x^3-120 x^2-120x)+60(1-x)(2-x)\log(1-x)]^{-1}\times10^{-3}{~\rm ns},&&\nonumber
\end{eqnarray}
where $x=(m_{\tilde\chi_0}/m_{\tilde q})^2$, and $F^{(3-\rm{body})}$, $F^{(4-\rm{body})}$ are the 3-body and 4-body coefficients found in~\cite{Kim:2013ivd}. 
Diagrams for production and decay of these LSPs are shown in Fig.~\ref{fig:ADM}.

\subsubsection{Co-annihilating DM}
For 
models with more than one particle species in the dark sector, the dark matter relic abundance can be set by annihilation between two different species. This is known as \emph{co-annihilation}. The effective annihilation cross-section between a DM particle $\chi_1$ and its co-annihilation partner $\chi_2$, 
taking them to be in thermal and chemical equilibrium, is given by~\cite{Griest:1990kh}
\begin{eqnarray}
\sigma_{\rm eff}=\frac{g_{1}^2}{g_{\rm eff}^2}\left[\sigma_{11}+2\sigma_{12}\frac{g_2}{g_{1}}(1+\Delta)^{3/2}{\rm exp}(-x\Delta)\right.\nonumber\\
\left. +\sigma_{22}\frac{g_2^2}{g_{1}^2}(1+\Delta)^{3}{\rm exp}(-2x\Delta)\right],
\end{eqnarray}
where $\Delta=(m_2-m_1)/m_{1}$, $x=m_{1}/T$, $g_{1,2}$ is the number of degrees of freedom for $\chi_{1,2}$, and $g_{\rm eff}=\sum_{i=1}^N g_i(1+\Delta_i)^{3/2}{\rm exp}(-x\Delta_i)$ is the number of effective degrees of freedom of the dark sector with $\Delta_2=\Delta$ and $\Delta_1=0$.

This co-annihilation process, which determines the current dark matter relic abundance, also plays a crucial role in the phenomenology of dark matter production at colliders~\cite{Izaguirre:2015zva,Baker:2015qna,Buschmann:2016hkc}. If the mass-splitting $\Delta$ is small compared to the masses of the decay products, $\chi_2$ can be long-lived. 

\begin{figure}
    \centering
    \includegraphics[width=0.35\textwidth]{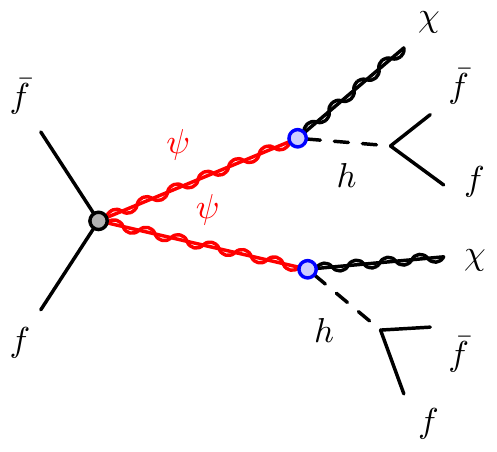}
    \caption{Production of the long-lived $\psi$ which decays into the DM $\chi$ in a model of co-annihilation.
    }
    \label{fig:coann}
\end{figure}

As an explicit example, let us consider the low-energy Lagrangian
\begin{equation}
{\cal L}\supset \bar\chi(i\partial-m_\chi)\chi+\bar\psi(i\partial-m_\psi)\psi+(yh\bar\chi\psi+h.c.),
\end{equation}
where $\chi$ is the DM particle, $\psi$ is its co-annihilation partner, and $m_\psi>m_\chi$. This scenario can occur in models of supersymmetry in which both the LSP and NLSP are predominantly mixtures of Bino-Higgsinos (see, \emph{e.g.}~\cite{ArkaniHamed:2006mb,Cheung:2012qy}). In this scenario, the $\psi$ can be long-lived with inverse decay width
\begin{equation}
    \Gamma^{-1}(\psi\to\chi \bar f f)\sim \frac{74}{N_c y^2}\left(\frac{10^{-3}}{y_f}\right)^2\left(\frac{10^{-2}}{\Delta}\right)^5\left(\frac{100{~\rm GeV}}{m_\chi}\right)^5\times 10{~\rm ns},
\end{equation}
where $f$ is a SM fermion, $y_f$ is its Yukawa coupling to the Higgs, and $N_c=3$ for quark final states and 1 for leptons. This decay is depicted in Fig.~\ref{fig:coann}.

\subsection{Effective Portals to a Hidden Sector}
\label{sec:portals}
The concept of a hidden sector appears in some of the models reviewed above. LLPs also arise when details of the hidden sector are not known, and one posits only the existence of a feeble interaction between the SM and the hidden sector, mediated by a single new field.
The interactions of this type between new physics and the SM are greatly restricted by the gauge and Lorentz symmetries of the SM, and can be categorized into various ``portals" defined by the mediating particle. The dominant interaction terms that give rise to these portals are 

\begin{equation}
\cal L\supset
\begin{cases}
     (\mu S+\lambda S^2)H^\dagger H & {\rm scalar} \\[13pt]
    \displaystyle{\frac{a}{f}}\,\widetilde F_{\mu\nu}F^{\mu\nu} & {\rm pseudoscalar}\\[13pt]
    -\displaystyle{\frac{\epsilon}{2\cos\theta_W}}\,F^{\prime}_{\mu\nu}F^{\mu\nu} & {\rm vector}\\[13pt]
    y_n LHN & {\rm neutrino}   
\end{cases}~~.
\end{equation}
In this section, we introduce each of these portals, as well as the various channels in which the LLP manifests itself.  

\subsubsection{Scalar Portal}
\label{sec:scalar-portal}
The simplest extension to the SM is to add a real singlet scalar, $S$, which interacts with the SM Higgs $H$ doublet 
through the renormalizable Lagrangian
\begin{equation}
{\cal L}\supset -\frac{\epsilon}{2} S^2 |H|^2+\frac{\mu_S}{2} S^2-\frac{\lambda_S}{4!}S^4+\mu_H^2 |H|^2-\lambda_H|H|^4,
\end{equation}
where we have imposed a discrete $\mathbb{Z}_2$ symmetry $S\to -S$ that removes the linear and cubic in $S$ terms~\footnote{See \emph{e.g.}~\cite{Curtin:2013fra,Evans:2017lvd} for a discussion of this model in the context of exotic Higgs decays}. 
If both $S$ and $H$ have nonzero vevs, $S=s+v_s$ and $H=(h+v_h)/\sqrt{2}$, then the two physical scalar particles,
$h$ and $s$, can mix with a mixing angle $\sin\theta=\epsilon v_h v_s/(m_h^2-m_s^2)+{\cal O}(\epsilon^3)$. As a result, $s$ couples to SM fermions $f$ through the term 
\begin{equation}
    {\cal L}\supset \sin\theta\frac{m_f}{v_h}sf\bar f.
\end{equation}

For sufficiently small $\sin\theta$, the singlet scalar $s$ can be long-lived,
as long as its rapid decay to hidden-sector states is forbidden, \emph{e.g.} due to kinematics.
Its lifetime is then given by
\begin{equation}
     \Gamma^{-1}(s\to f\bar f)\sim \left(\frac{0.2}{\sin\theta}\right)^2\left(\frac{100{~\rm MeV}}{m_s}\right)\left(\frac{0.511{~\rm MeV}}{m_f}\right)^2\left(1-\frac{4m_f^2}{m_s^2}\right)^{-3/2}\times 3.8~{\rm ns}.
\end{equation}

If the scalar mass is less than half the Higgs mass, it can be pair-produced at the LHC through exotic Higgs decays with partial width
\begin{equation}
    \Gamma(h\to ss)=\frac{\lambda_S\sin^2\theta m_h^3}{48\pi m_s^2}\left(1+2\frac{m_s^2}{m_h^2}\right)^2\sqrt{1-4\frac{m_s^2}{m_h^2}}.
\end{equation}
If $s$ is light enough, it can also be produced through rare meson decays, in particular $\Upsilon\to s\gamma$ and the penguin decays such as $B\to sK$~\cite{GRINSTEIN1988363}. 
\begin{figure}
    \centering
    \includegraphics[width=0.35\textwidth]{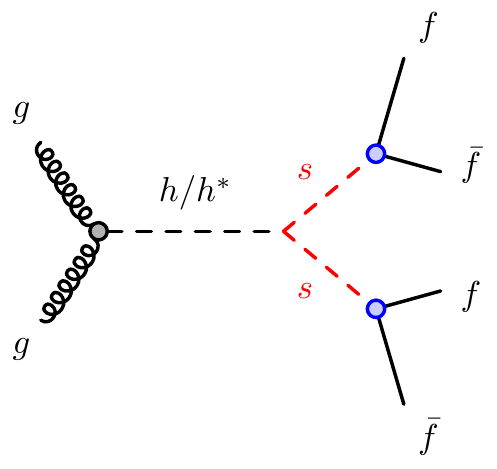}
    \caption{Di-scalar production from an exotic Higgs decay.}
    \label{fig:scalar-schannel}
\end{figure}
\subsubsection{Pseudoscalar Portal}

One can also consider a pseudoscalar particle, $a$, which arises as a pNGB in theories with a spontaneously broken global symmetry. One famous example of such a particle is the \emph{axion}, which was introduced to solve the strong CP problem of QCD and arises from the breaking of the $U(1)$ Peccei-Quinn symmetry~\cite{Peccei:1977ur,Wilczek:1977pj}. The QCD axion has a fixed relationship between its mass $m_a$ and decay constant $f$. More general models allow for these two parameters to be independent, in which case the $a$ is commonly known as an \emph{axion-like particle} (ALP). A key feature of ALPs is that they have derivative couplings to fermions, and therefore their masses are protected from radiative corrections through a shift-symmetry. As a result, one can have naturally light ALPs. 

The ALP Lagrangian, including interaction with the SM fields up to dimension~5, is given by
\begin{eqnarray}
    {\cal L}&\supset& \frac{(\partial_\mu a)^2}{2}- \frac{a}{f}\left[c_G\frac{g_S^2}{16\pi^2}\widetilde{G}_{\mu\nu}^AG^{A\mu\nu}+c_W\frac{g^2}{16\pi^2}\widetilde{W}_{\mu\nu}^AW^{A\mu\nu}+c_Y\frac{{g_Y}^2}{16\pi^2}\widetilde{B}_{\mu\nu}B^{\mu\nu}\right]\nonumber\\
    &+&\frac{\partial_\mu a}{f}\sum_i\frac{c_i}{2}\bar\psi_i\gamma_\mu\gamma_5\psi,
\label{eq:alpL}
\end{eqnarray}
where the notation $\widetilde X_{\mu\nu}=\frac{1}{2}\epsilon^{\mu\nu\alpha\beta}X_{\alpha\beta}$ represents a dual field strength tensor.
The $c_j$ are model-dependent coupling constants and $f$ is a scale for the UV completion. Generically, the fermion couplings of the ALPs are subdominant to the gauge couplings.

The $a$ can be produced in $Z\to a\gamma$, through inclusive $\gamma^*\to\gamma a$ at $e^+e^-$ colliders,
and via flavor-changing neutral current meson decays, such as $B \to aK$ and $K \to a\pi$~\cite{Izaguirre:2016dfi}. At beam-dump experiments, the ALP can be produced via Primakoff production, $\gamma\gamma\to a$~\cite{Dobrich:2015jyk}, 
or via emission from a fermion (see~\cite{Essig:2013lka} for a review). One can also search for ALPs in diphoton initial states in ultra-peripheral heavy-ion collisions~\cite{Knapen:2016moh}. The ALP subsequently decays to either fermions or photons, depending on the exact model parameters, with inverse decay widths

\begin{eqnarray}
     \Gamma^{-1}(a\to \gamma\gamma)&\sim&\left(\frac{f}{15{~\rm TeV}}\right)^2\left(\frac{300{~\rm MeV}}{m_a}\right)^3\times 10^{-3}{~\rm ns},\\
     \Gamma^{-1}(a\to f_i\bar f_i)&\sim&4\left(\frac{300{~\rm MeV}}{m_a}\right)\left(\frac{f}{15{~\rm TeV}}\right)^2\left(\frac{100{~\rm MeV}}{m_{f_i}}\right)^2\times 10^{-4}{~\rm ns},
\end{eqnarray}
where we have taken $c_j=1$. 

At dimension-6 and higher, there are additional couplings between the ALP and the Higgs,
\begin{equation}
    {\cal L}\supset \frac{c_{ah}}{f^2}(\partial^\mu a)(\partial_\mu a)H^\dagger H+\frac{c_Zh}{f^3}(\partial^\mu a)\left(H^\dagger iD_\mu H +{\rm h.c.}\right)H^\dagger H+\ldots,
\end{equation}
which lead the exotic Higgs decays $h\to aa$ and $h\to Za$~\cite{Bauer:2017ris}. These couplings can also contribute to the meson decays mentioned above.

\subsubsection{Vector Portal}
Adding a dark-sector Abelian gauge group, $U(1)_D$, to the SM leads to a vector portal interaction between the ``dark photon" $A^\prime$ and the SM hypercharge gauge boson $A$ through kinetic mixing,
\begin{equation}
    {\cal L}\supset-\frac{\epsilon}{2\cos\theta_W}F^\prime_{\mu\nu} F^{\mu\nu}
\end{equation}
where $F^{\mu\nu}$ and $F^{\prime\mu\nu}$ are  the field strength tensors for the SM hypercharge $U(1)_Y$ and for the $U(1)_D$ gauge groups, respectively. 
The coefficient $\epsilon$ parameterizes the strength of the mixing between the two gauge fields, and in principle can have  arbitrary value. However, values in the range $\epsilon^2\sim10^{-8}-10^{-4}$ are favored if the interaction is generated by heavy particles charged under both $U(1)_D$ and $U(1)_Y$~\cite{Holdom:1985ag}.
The dark photon can obtain mass through either a dark Higgs~\cite{Batell:2009yf,Curtin:2014cca} or the Stueckelberg mechanism~\cite{Stueckelberg:1900zz}. In the former scenario, either the dark photon or the dark Higgs can be long-lived. 

A result of the kinetic mixing between $A^\prime$ and the SM photon is that the $A^\prime$ can be produced, when kinematically allowed, in any scenario in which a photon is produced. Therefore, one can  search for the $A^\prime$ at $B$-factories~\cite{Batell:2009yf,Essig:2013vha}, electron beam dump experiments~\cite{Essig:2010xa}, and at both lepton~\cite{Essig:2009nc} and hadron~\cite{Curtin:2014cca} colliders. For $m_A^\prime<m_\pi$, $A^\prime$ is produced in the decay $\pi^0\to A^\prime\gamma$, and can be searched for at proton fixed-target experiments, where pions are copiously produced~\cite{Batell:2009di}. 

Once produced, the $A^\prime$ will decay into any charged SM particle pair through its kinetic mixing with the SM photon. The width for this decay is
\begin{equation}
    \Gamma(A^\prime\to \bar f f)=\frac{1}{3}\epsilon^2\alpha\left(1+\frac{2m_f^2}{{m_A^\prime}^2}\right)\sqrt{{m_A^\prime}^2-4m_f^2}.
\end{equation}
In the case $m_A^\prime\gg m_f$, the lifetime of the dark photon decaying into fermions is given by
\begin{equation}
 \Gamma^{-1}(A^\prime\to \bar f f)\sim 2.6 \left(\frac{10^{-5}}{\epsilon}\right)^2\left(\frac{100{~\rm MeV}}{m_A^\prime}\right)^2\times 10^{-2}{~\rm ns}.
\end{equation}

If $m_{A^\prime}<2m_e$, then its only possible decay channel is $A^\prime\to 3\gamma$ with inverse decay width~\cite{Pospelov:2008jk,McDermott:2017qcg}
\begin{equation}
    \Gamma^{-1}(A^\prime\to 3\gamma)\sim \left(\frac{0.003}{\epsilon}\right)^2\left(\frac{m_e}{m_{A'}}\right)^9~\rm{s}.
\end{equation}
In this case the dark photon flies hundreds of meters before decaying, even for $\epsilon\sim 1$, and is seen only as missing energy in colliders.

\subsubsection{Heavy-Neutrino Portal}
The SM predicts that neutrinos are massless, but the observation of neutrino oscillations provides evidence that neutrinos do have small, non-zero masses~\cite{Tanabashi:2018oca}. A simple
way to generate neutrino masses is to add a set of three right-handed neutrinos, $n_i$, 
with Majorana masses $M_{i}$ and no SM-gauge quantum numbers. They can couple to the SM via neutrino-Higgs Yukawa interactions in the Lagrangian,
\begin{equation}
    {\cal L}_{\rm Type-I}\supset y_{\alpha i} L^\alpha n_i^c H-\frac{M_{ij}}{2}n_i^cn_j^c+{\rm h.c.},
    \label{eq:seesaw-lag}
\end{equation}
where $i=1,2,3$, 
$L^\alpha \equiv \genfrac(){0pt}{2}{\nu^\alpha}{\ell^\alpha}$
is the left-handed lepton doublet with generation index $\alpha=e,\mu,\tau$, and $y_{\alpha i}$ are the Yukawa couplings. $M_{ij}=M_{ji}$ are the elements of a $3\times 3$ right-handed neutrino Majorana mass matrix. After electroweak symmetry breaking, the Yukawa terms generate a 
Dirac mass matrix
for the neutrinos, $m_D^{\alpha i}=y^{\alpha i}v$, where $v$ is the SM Higgs vev. This results in 6 potentially massive Majorana fermions, which are linear combinations of $\nu_\alpha$ and $n_i$. In this basis, the 
"$6\times 6$"
neutrino Majorana mass matrix is
\begin{eqnarray}
m_\nu^{\alpha i}=\begin{pmatrix}
0&m_D^{\alpha i}\\
m_D^{i\alpha} & M^i
\end{pmatrix}.
\label{eq:neutrino-mass-matrix}
\end{eqnarray}
Diagonalizing this matrix and taking the limit where the $m_D\ll M$, we end up with 3 heavy neutrinos, $N$, which are predominantly the right-handed $n_i$ states and have masses of order $M$, and 3 light neutrinos, $\nu$,
which are predominantly the $\nu_\alpha$ 
states and have masses of order $m_{\nu}\sim m_D^2/M$. These light states are the ones which are observed experimentally.
 This is the simplest example of the seesaw mechanism~\cite{Minkowski:1977sc,Yanagida:1979as,Mohapatra:1979ia,GellMann:1980vs} and is known as the Type-I seesaw\footnote{See~\cite{King:2003jb,dGneutrino} for a review of additional neutrino mass mechanisms.}. A result of the mixing between the mass and flavor eigenstates is that the heavy neutrino states $N$ acquire a small coupling under the weak interactions. 
The small mixing angle $\theta^2\simeq m_\nu/M$ characterizes the strength of the interaction of $N$ with the SM. 

\begin{figure}
    \centering
    \includegraphics{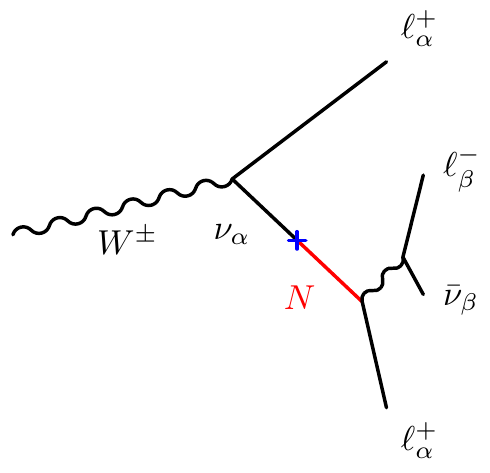} ~~~~~~~
    \includegraphics{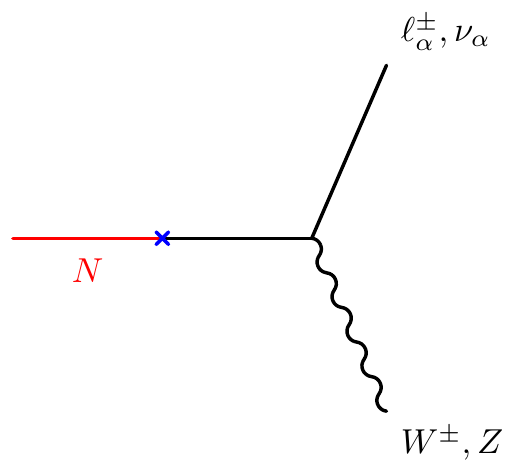}
    \caption{Example processes for the production and decay of a light, right-handed neutrino (left) and decay of a heavy $N$ into and on-shell gauge boson and a lepton (right).
    }
    \label{fig:RHN}
\end{figure}

Since $N$ couples to the SM through weak interactions, it can be
produced 
in rare decays of 
ground-state mesons that are heavier than the $N$. Heavier $N$ states can be produced in the vector-boson decays $W^\pm\to\ell^\pm N$ and  $Z\to \nu N$. Likewise, all decays of $N$ are mediated by either neutral- or charged-current interactions~\cite{Johnson:1997cj,Gorbunov:2007ak,Helo:2010cw,Bondarenko:2018ptm}. For sufficiently small $\theta$, the $N$ flight distance is macroscopic. For example, if $M_N\ll m_W$, then its inverse decay width into leptons is given by
\begin{equation}
     \Gamma^{-1}(N\to\ell_\alpha^-\ell_\beta^+\nu_\beta) \sim  \left(\frac{12{~\rm GeV}}{M_N}\right)^5\left(\frac{10^{-4}}{|\theta|^2}\right)\times10^{-3}{~\rm ns}.
\end{equation}
As with all weak processes, the semileptonic decays 
$N\to\ell q_\alpha \bar q_\beta$
and 
$N\to\nu q_\alpha \bar q_\alpha$
occur as well. If $N$ is heavy enough, its decay final states include an on-shell boson, particularly $N\to \ell^\pm W^\mp $,
with smaller branching fractions for $N\to Z\nu$ and $N\to h\nu$~\cite{Basso:2008iv}. Fig.~\ref{fig:RHN} shows diagrams for production and decay of the $N$.

\subsection{Magnetic Monopoles}
A strong theoretical motivation for the existence of monopoles was proposed by Dirac as a way to explain charge quantization in quantum electrodynamics (QED)~\cite{Dirac:1931kp,Dirac:1948um}\footnote{More comprehensive reviews on magnetic monopole solutions can be found in~\cite{Milton:2006cp,Balestra:2011ks,Rajantie:2012xh,Acharya:2014nyr}. A summary of the recent status of searches can be found in~\cite{Patrizii:2015uea}}. Dirac demonstrated that adding a magnetic monopole, now commonly referred to as the Dirac monopole, to the theory and quantizing angular momentum leads to the following relationship between electric charge $q_e$ and magnetic charge $q_m$,
\begin{equation}
    q_m q_e=\frac{n}{2},
\label{eq:mag-charge}
\end{equation}
where $n$ is an integer. This
results in a magnetic charge $q_m = n Q_D$ where $Q_D\equiv Q_e/{2\alpha}$ is the Dirac charge, $Q_e$ is the electron's electric charge, and $\alpha\simeq1/137$ is the fine-structure constant. We can then define an analogous magnetic fine-structure constant $\alpha_m\equiv Q_D^2/(4\pi)\simeq 34.25$. An experimental consequence of the large magnetic coupling is that the monopole is a highly ionizing particle (HIP), which experiences large electromagnetic energy losses as it traverses through matter. A theoretical consequence is that  calculations of monopole processes are pushed into the non-perturbative regime. 

Monopoles arise naturally in Grand Unified Theories (GUTs) as topological defects of space-time whenever a gauge group is spontaneously broken into an exact $U(1)$ subgroup~\cite{tHooft:1974kcl,Polyakov:1974ek}. An example of this is~\cite{Dokos:1979vu}
\begin{equation}
    SU(5)\to SU(3)\otimes SU(2)\otimes U(1),
\end{equation}
which results in a monopole with mass
\begin{equation}
    M_{\rm mon}\sim\frac{\Lambda_{\rm GUT}}{\alpha}.
\end{equation}
For a GUT unification scale of $10^{16}$ GeV, this yields $M_{\rm mon}\sim 10^{17}-10^{18}$ GeV. One can produce intermediate-mass monopoles with $M_{\rm mon}\sim 10^7-10^{14}$ GeV through additional symmetry-breaking schemes~\cite{Huguet:1999bu,Wick:2000yc}. However, these are still far above the reach of current collider probes. Lower-mass monopoles, known as Cho-Maison monopoles, can be produced through the electroweak symmetry-breaking and can be interpreted as a hybrid between the Dirac monopole and the 't~Hooft-Polyakov GUT monopole~\cite{Cho:1996qd,Bae:2002bm}. Assuming that the Cho-Maison monopole is a topological soliton, one can estimate its mass to be in the 1-10~TeV range~\cite{Cho:2012bq,Cho:2013vba}. The non-perturbative nature of the monopole makes a more accurate estimate of the mass difficult. A priori, such monopoles could be pair-produced electromagnetically at colliders. However, their large couplings might cause them to annihilate immediately or form bound monopole-antimonopole states known as monopolonium~\cite{HILL1983469,Dubrovich:2002gp,Epele:2007ic}. These states can be produced through photon-fusion at the LHC~\cite{Epele:2008un}.

Another class of defect solutions is known as electroweak strings. These were proposed in the context of the SM by Nambu, who suggested that they have a monopole and antimonopole at either end~\cite{Nambu:1977ag}. The mass of the monopole and the tension of the string are roughly in the TeV range\footnote{See~\cite{Achucarro:1999it} for further discussion of electroweak strings.}.
The estimated mass of the Nambu monopole is given by~\cite{Nambu:1977ag}
\begin{equation}
    M_N\sim\frac{4\pi}{3e}\sin^{5/2}\theta_W\sqrt{\frac{m_h}{M_W}}\mu\simeq 689~\rm{GeV},
\end{equation}
where $M_W$ is the $W$ boson mass, $m_h$ is the Higgs boson mass, $\sin\theta_W$ is the weak mixing angle, $\mu=M_W/g$, and $g$ is the $SU(2)$ gauge coupling. 
The dumbbell configuration can rotate and emit electromagnetic radiation, and can possibly have a lifetime long enough to be observed at the LHC~\cite{James:1992wb,James:1992zp}.

\section{Detector Signatures in Collider Experiments}
\label{sec:signatures}

This section discusses the methods by which collider-based particle detectors are used to search for LLPs.  We begin in in Sec.~\ref{sec:dir-indir} with the definition of the two types of LLP detection, direct and indirect. Typical detector subsystems and their use for SM particle detection are described in Sec.~\ref{sec:detectors}. In Sec.~\ref{sec:llp-det} we review the various detector signatures produced by LLPs of different types. In Sec.~\ref{sec:acceptance-comparison} We compare the sensitivities of the different detector subsystems for detection of a LLP that decays within them using a simple acceptance-based approach. In Sec.~\ref{sec:recoconsiderations} we discuss complications arising from differences between reconstruction of prompt and displaced particles.

\subsection{Direct and Indirect LLP Detection}
\label{sec:dir-indir}

Collider searches for LLPs can be roughly categorized into two classes based on the LLP detection method: \emph{direct} detection and \emph{indirect} detection. The direct category uses experimental signatures arising from direct  interaction of the LLP with the detector. By contrast, indirect searches reconstruct the decay of the LLP to SM particles. This classification is closely related to the the one used to categorize experimental searches for dark matter based on the same principle.

\subsection{Typical Detector Subsystems and Particle Detection}
\label{sec:detectors}

\begin{figure}[hb]
\centering
\includegraphics[width=0.8\textwidth]{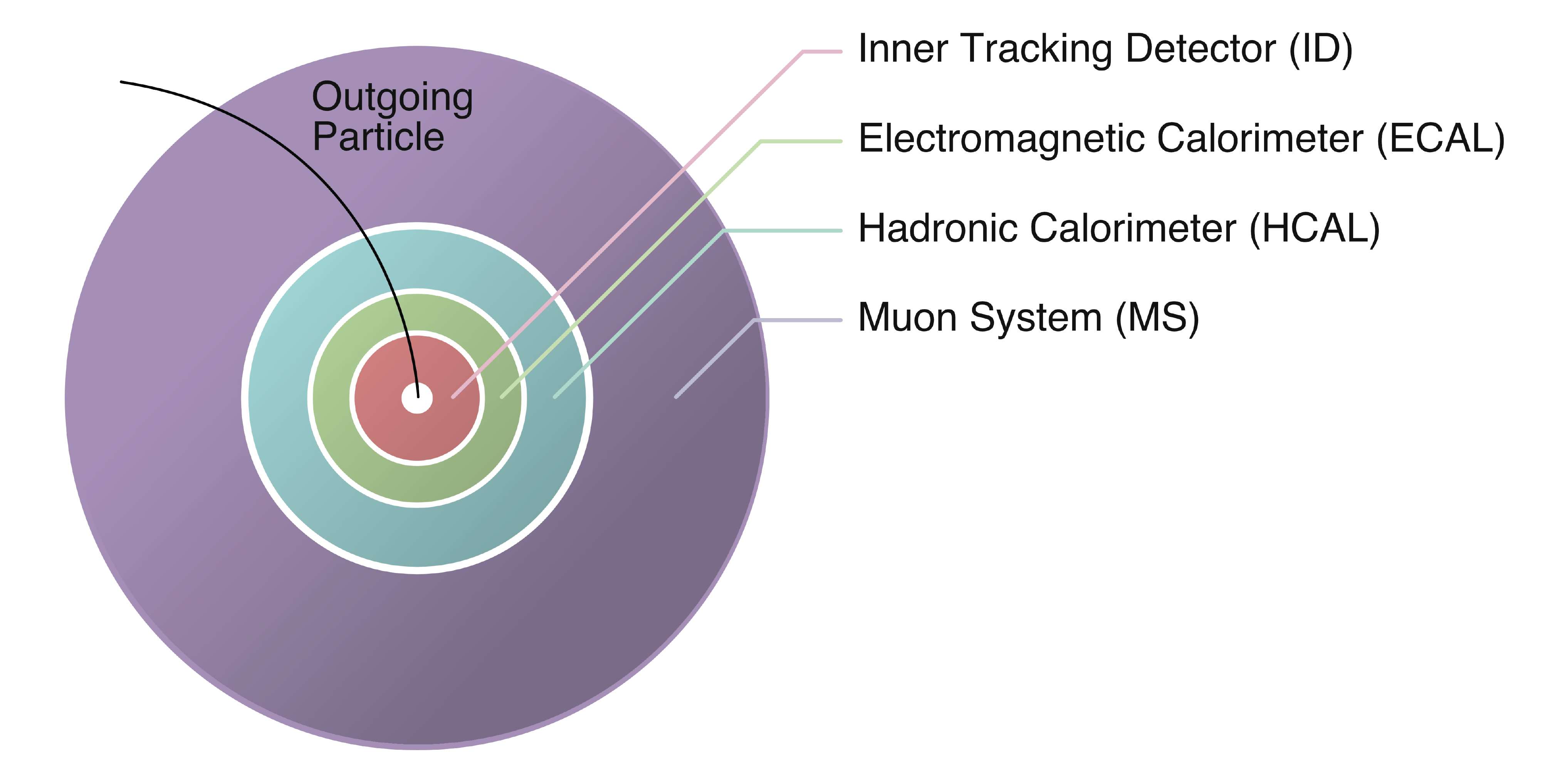}
\caption{A cross-sectional view of a schematic collider experiment is shown in the plane transverse to the beam direction. From the center outwards, this figure shows a generic inner tracking detector (ID), an electromagnetic calorimeter (ECAL), a hadronic calorimeter (HCAL), and muon system (MS).}
\label{fig:genericdetector}
\end{figure}

A typical collider experiment comprises several main detector subsystems that are used jointly to detect and measure the properties of particles produced in the collision. A schematic representation of such a generic detector is shown in Fig.~\ref{fig:genericdetector}. We note that this figure and all other schematic detector representations in this review are intended only for illustration. In particular, they do not accurately represent the relative spatial dimensions of detector subsystems or the magnetic field configurations in any specific experiment. Therefore, illustrations of charged-particle trajectories and their passage through detector subsystems are not to be understood literally.  

The innermost subsystem, called the inner detector (ID), is designed to detect electrically charged particles that are long-lived enough to traverse the ID. The most common such particles from the SM are two charged leptons (the electron $e$ and the muon $\mu$) and three hadrons (the pion $\pi$, kaon $K$, and proton $p$). Regions of ionization produced by such a particle in solid-state or gaseous detector sensors are detected as spatial \emph{hits} that are fit into a trajectory, referred to as a \emph{track}. The direction and curvature of the track in a magnetic field yield the particle's momentum vector and electric charge. In some detectors, the ID is enclosed in a Cherenkov-light detector used to measure the velocity of the tracked particles. Combined with the momentum measurement in the ID, this yields the particle mass with sufficient resolution to differentiate between pions, kaons, and protons in a relevant momentum range.

After passing through the tracker, particles produced in the collisions typically enter an electromagnetic calorimeter (ECAL), designed to measure the energies of photons,  electrons and positrons. The energy measurement exploits the properties of electromagnetic shower production via photon radiation and $e^+e^-$ pair production, resulting from the interaction of energetic particles with the ECAL material. 

Hadrons deposit energy via hadronic interactions with the detector material. Since this process involves large fluctuations and a variety of energy-deposition mechanisms, precise hadron-energy measurement is achievable only at high-energy colliders, where fluctuations are effectively averaged out. In particular, high-energy quarks and gluons hadronize into a collimated spray of hadrons known as a \emph{jet}. Containing the jet requires use of a deep hadronic calorimeter (HCAL) beyond the ECAL. While a jet can be identified solely in the calorimeters, its energy is nowadays measured from a combination of the momenta of tracks in the ID and the signals integrated in the ECAL and HCAL.

Muons do not undergo hadronic interactions, and are heavy enough that they lose energy due to ionization at a low rate. Therefore, they lose only a few GeV while traversing a typical LHC-detector calorimeter. Using this property to identify them, a muon system (MS) is built outside the calorimeter. In high-energy collider detectors, the MS is usually immersed in a magnetic field in order to measure the momenta of muons. Tracks reconstructed in the MS are often combined with tracks in the ID to obtain a high-quality momentum measurement.

When studying final states that include long-lived, weakly interacting particles, such as neutrinos in the SM, an important reconstructed quantity is missing momentum. Using three-momentum conservation and the approximate hermeticity of the detector, it is possible to measure the momentum imbalance in the event and to infer the combined momentum of the invisible set of particles. Since the interacting partons in proton collisions generally carry different fractions of the momenta of the incoming hadrons and many of the particles produced fall outside of the acceptance of the sensitive detector, the summed momenta of measured final-state particles along the beam axis $z$ are not expected to cancel. Therefore, experiments at the LHC and Tevatron measure the \emph{missing transverse momentum}, denoted $E_{\mathrm{T}}^{\mathrm{miss}}$ or MET, where momentum balance is assumed only in the $x$-$y$ plane transverse to the beam direction~\cite{1748-0221-3-08-S08003,1748-0221-3-08-S08004}\footnote{Missing momentum is the primary signature for a neutral LLP that traverses the detector without decaying or interacting, which is outside the scope of this review.}.

Collider detectors are mostly designed and constructed for optimal detection of SM particles produced in the collision. However, LLPs or their decay products would also interact with and deposit energy in the detector, with  characteristics that are impacted by the long lifetimes and often high masses of the LLPs. Generally, LLP detection is less efficient and measurement of LLP properties is less precise than those of SM particles, with performance degrading as particle displacement increases. Nonetheless, collider detectors have proven to be powerful instruments for LLP searches, once experimenters take these differences into account. We return to this subtlety in Sec.~\ref{sec:recoconsiderations} after describing in Sec.~\ref{sec:llp-det} the ways in which LLPs can be studied with collider detectors.

\subsection{LLP Detector Signatures}
\label{sec:llp-det}

With the detectors at hand, there are several categories of signatures that can be used for discovering LLPs and measuring their properties. Typical signatures used for direct detection are reviewed in Secs.~\ref{sec:dedx} through~\ref{sec:anomalous-tracking}, while  Secs.~\ref{sec:displaced-tracks} through~\ref{sec:displaced-cal} describe the building blocks of indirect-detection searches. The use of combination of signatures is discussed in Sec.~\ref{sec:ddcllps}.

\subsubsection{Anomalous Ionization}
\label{sec:dedx}
A detector-stable, charged LLP (CLLP) is directly detectable via the track that it forms in the ID. If the CLLP is much heavier than the proton, its speed $\beta$ will be markedly lower than that of any track-forming SM particle of the same momentum. One way to detect this is via specific ionization. The average ionization energy loss per unit distance traveled by a charged particle in material of a particular density has a $\beta$ dependence given by the Bethe-Bloch formula~\cite{Tanabashi:2018oca},
\begin{equation}
\left\langle\frac{dE}{dx}\right\rangle \sim -\frac{z^2}{\beta^2} \cdot \left[\ln \left(\frac{\beta^2}{(1-\beta^2)}\right) - \beta^2 + C \right] ,
\label{eq:bethebloch}
\end{equation}
where $C$ is a near-constant that depends on the properties of the material traversed and $z$ is the electric charge of the traversing particle.
Thus, a CLLP that is slow-moving or has charge greater than 1 can be identified via anomalously large $\left\langle\frac{dE}{dx}\right\rangle$. 

Silicon-based and gaseous tracking detectors are routinely used to measure the charge deposition associated with a hit. Gas-based detectors used in the MS also have this capability. Calorimeters may also be used for identification of anomalous ionization relative to that of muons, although this has not yet been utilized in any collider CLLP search.

Magnetic monopoles are another example of LLPs that give rise to high specific energy loss through ionization. 
As discussed in Sec.~\ref{sec:anomalous-tracking}, they follow anomalous trajectories in the magnetic field of the ID and are thus difficult to track. Nonetheless, their high $\left\langle dE/dx\right\rangle$ signature can be identified in ID tracking detectors as well as in calorimeters segmented to measure the shower development. For a more detailed review of the detector signatures expected for magnetic monopoles in the LHC experiments, and projected sensitivities for searches there, see Ref~\cite{DeRoeck:2011aa}.

\subsubsection{Delayed Detector Signals}
\label{sec:delayed-signals}

A heavy LLP traveling at low speed relative to a SM particle of the same momentum takes more time to cover the distance from its production vertex to a distant detector subsystem, particularly the calorimeter or MS. This ``late'' arrival constitutes a unique LLP signature. Measurement of the time of flight provides a measurement of the speed of the LLP candidate and, in conjunction with its momentum measurement, gives the LLP mass. 
Since the bunch spacing at newer colliders is only of order a few meters, the detectors' subsystems are designed with high timing resolution in order to associate the detector signals to the correct bunch crossings. At the LHC, where the bunch crossings are typically separated by 25~ns, many detector subsystems have timing resolutions of $\mathcal{O}(1)$~ns. This resolution, combined with the sheer physical dimensions of the detectors, enables identification of slow moving particles with high precision. This is particularly the case for the muon spectrometer systems, which have excellent timing resolution and are located at the outermost radii. In addition, the calorimeters are finely segmented and are sensitive enough to pick up modest energy deposits along the path of a charged LLP with $\mathcal{O}(1)$~ns precision. 
As the LHC luminosity increases, precise timing measurement of these relatively weak calorimeter signals in the presence of growing background from SM particles becomes increasingly challenging. 

If a LLP is very slow, or if it is completely stopped in the detector and decays long afterwards, it generally gives rise to a detector signal that occurs after the triggering and readout time windows associated with the collision that produced the LLP. As a result, the LLP signal will not be associated to the correct trigger, and may be lost or misidentified. For such cases, a dedicated triggering technique has been developed for searches at LHC, utilizing the possibility that the detector signal occurs during gaps in the proton bunch train. Bunch crossings at LHC occur every $25$~ns, but some bunches are intentionally not filled and thus contain no protons. Since no collisions take place during such \emph{empty bunch crossings}, detector signals occurring at these times by the late arrival or decay of a LLP are not masked by the presence of high collision background. Therefore, searches looking for such a LLP trigger on  these \emph{out-of-time} signals, and generally have very low background levels.

\begin{figure}[tb]
\centering
\includegraphics[width=5in]{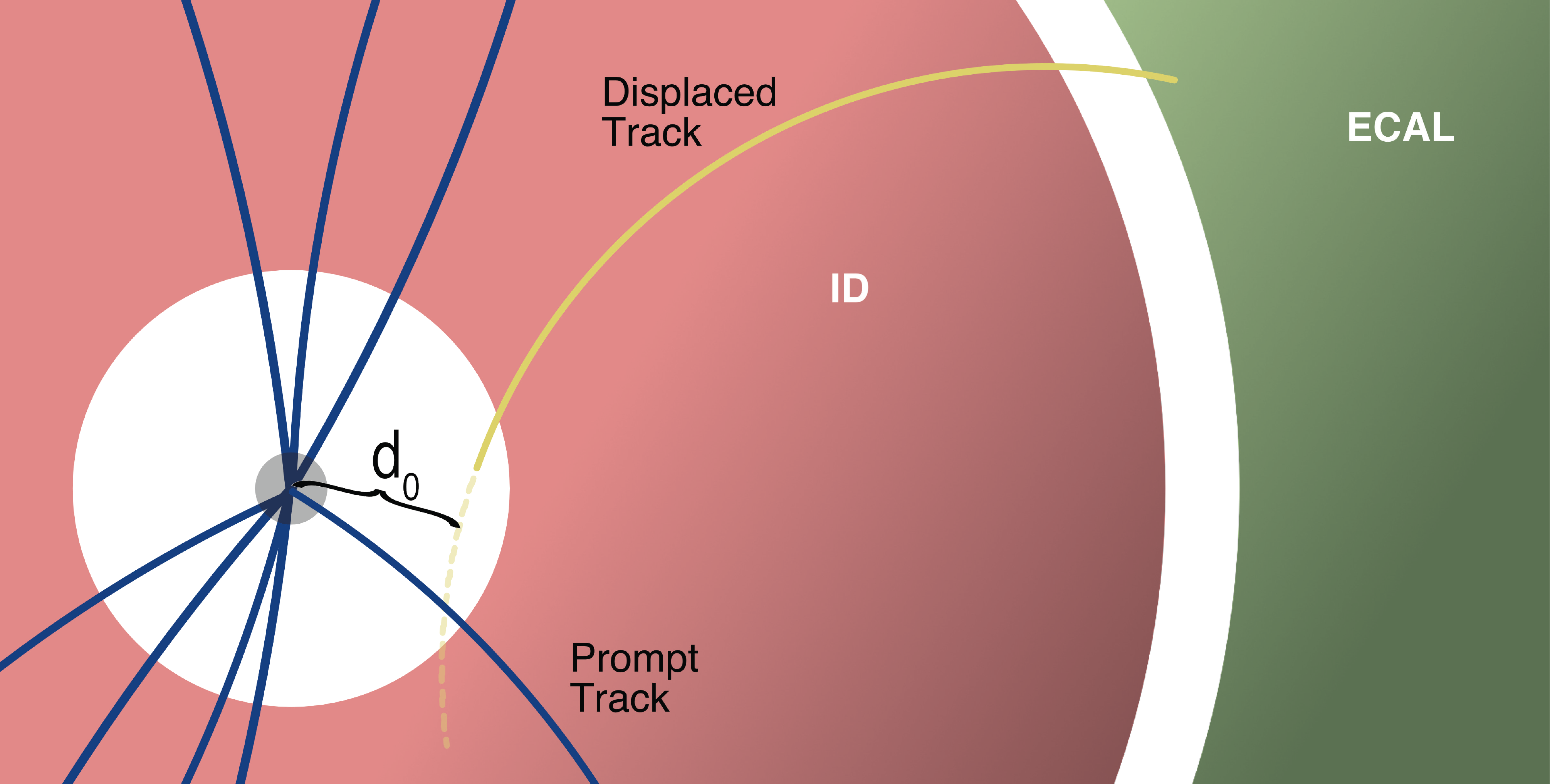}
\caption{A standard track is shown traversing the entirety of the ID. A high-momentum, disappearing track is shown at the bottom of the diagram with a missing hit in the outer region of the ID indicated with an X. In the example shown here, the charged particle is decaying to a low-momentum charged particle and a weakly interacting particle.}
\label{fig:disappearingtrack}
\end{figure}

\subsubsection{Disappearing Tracks}
\label{sec:disappearing-track}

If a CLLP lives long enough to enter deep into the ID yet decays at some point within it, the track that it forms seems to disappear midway through the ID. The identification of such \emph{disappearing tracks} or \emph{tracklets} often involves a veto on ID hits at radii that are larger than the apparent point of disappearance. This signature is particularly important when any electrically charged products of the decay are too soft to be reconstructed, so that there is no displaced-vertex signature (see Sec.~\ref{sec:DVs}). 
Since disappearing tracks are necessarily short and have few hits, their momentum-measurement resolution is poorer than that of standard tracks. They are also more susceptible to combinatoric backgrounds, particularly at high luminosity hadron colliders where the track multiplicity is high. This necessitates dedicated optimization and trade-off between background levels and signal efficiencies, particularly at low lifetimes.

When the CLLP decay gives rise to one charged particle that is hard enough to be tracked, the combined signature is two connected tracks at an angle, known as a \emph{kinked track}.  A kinked-track search uses more information than a disappearing-track search, and is thus more effective in suppressing background. However, this comes at a cost of lower efficiency and sensitivity to a more restricted parameter space.

\subsubsection{Anomalous-Trajectory Tracks}
\label{sec:anomalous-tracking}

Depending on their properties, some track-forming LLPs may bend in the $z$-oriented magnetic field of an ID differently from a charged particle of the SM.

In particular, a magnetic monopole feels a force along the direction of the magnetic field, rather than perpendicularly to it. This leads to a track that bends parabolically in the $z$ direction for most trackers in solenoidal magnetic fields. Identifying such a track requires a dedicated tracking algorithm in three dimensions, with a resulting  signature that is strikingly different from that of any SM particle.
Alternatively, one can also track a magnetic monopole only in the $(x,y)$ plane perpendicular to the magnetic field. In this plane, the trajectory of a magnetic monopole that has no electric charge appears as a straight line, corresponding to infinite-momentum electrically charged particle. Utilizing less information than 3-dimensional tracking, this approach suffers from higher background, yet is simpler to execute.

Quirks represent another example of anomalous trajectories, as discussed in Sec.~\ref{sec:neutralnaturalness}. In addition to each electrically charged quirk feeling a standard Lorentz force through the magnetic field of the ID, an additional force arises from the dark-gluon flux tube between the pair of quirks. This spring-like coupling gives rise to very complex trajectories through the detector. This again requires dedicated tracking algorithms that have not yet been used for LLP searches.

\subsubsection{Displaced Tracks}
\label{sec:displaced-tracks}

\begin{figure}[tb]
\centering
\includegraphics[width=5in]{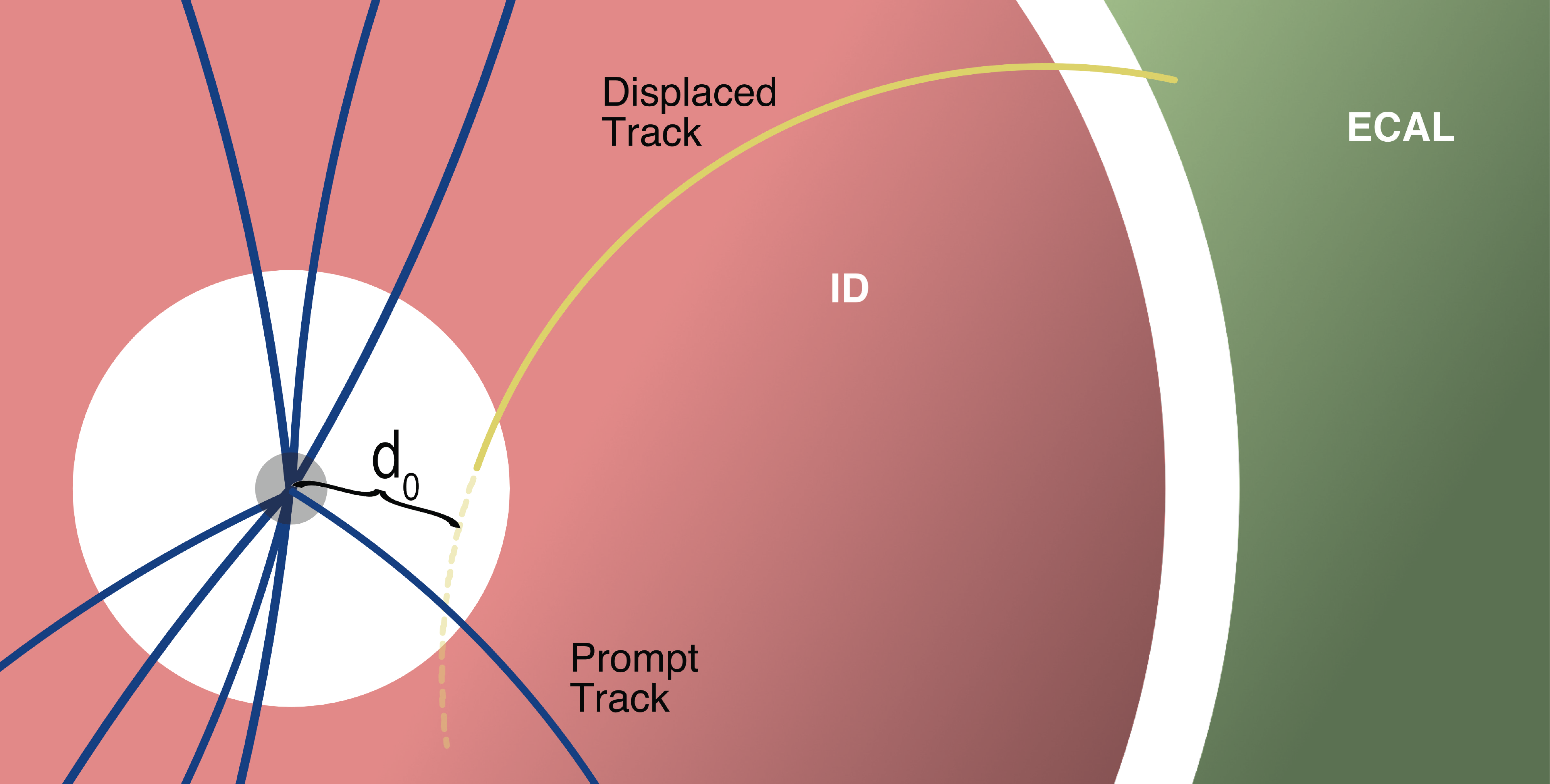}
\caption{In addition to a standard prompt track, a displaced track with a large transverse impact parameter $d_{0}$ is shown.
}
\label{fig:displacedtrack}
\end{figure}

Tracks of charged particles emitted in the decay of a LLP are often measurably inconsistent with originating from the \emph{beam spot}, the spatial region where beam-particle collisions take place. Such a track is illustrated in Fig.~\ref{fig:displacedtrack}. The degree of consistency is typically determined from the track's \emph{transverse impact parameter} $d_0$. This is the shortest distance, measured in the $(x,y)$ plane transverse to the beams, between the track and the hypothesized position of the collision. This position is taken to be either the \emph{interaction point} (IP) at the center of the beam spot or the \emph{primary vertex} (PV), which is the point from which reconstructed tracks originating from the collision appear to emanate in a particular event. For the sake of simplicity, our discussion does not make a distinction between these two methods of $d_0$ calculation.

At hadron colliders, a PV generally exists due to the composite nature of the colliding particles. In particular, LHC typically produces tens of PVs per proton-proton bunch crossing. In this case, the most energetic PV, measured using its tracks, is usually used for calculation of $d_0$. Whether or not a PV exists in a particular $e^+e^-$-collider analysis depends on the LLP production mode under study.

Accounting for detector resolution and beam-spot size, a large value of the ratio between $d_0$ and its uncertainty $\sigma_{d_0}$ defines a \emph{displaced track}, \emph{i.e.} one produced far from the beam spot. Use of $d_0/\sigma_{d_0}$ takes advantage of the small transverse beam spot size, which is of order a few microns to a fraction of a millimeter in recent and current colliders, as well as the tens-of-micron resolution of the PV position measurement at modern detectors.

The corresponding impact parameter in the $z$ direction (along the beams) measured with respect to the PV, denoted $z_0$, is also sometimes used for determining whether a track is displaced. However, at colliders with long interaction regions and multiple interactions possible per beam crossing, large $z_0$ may denote that a track originated from another beam interaction. Thus, it is often less useful than $d_0$ for identifying the decay of a LLP.

Current collider detectors and their reconstruction software were designed to study prompt objects. Because of the additional needed resources, tracks with large values of $d_0$ or $z_0$ are often not reconstructed by default, in order to limit data acquisition and computing resources. This leads to loss of efficiency in the reconstruction of LLP decays far from the PV, particularly when the decaying LLP is not highly boosted. Recent searches overcome this limitation by performing dedicated reconstruction of highly displaced tracks for a small part of the data set, selected based on signatures related to the search analysis.

\subsubsection{Displaced Vertices}
\label{sec:DVs}

\begin{figure}[tb]
\centering
\includegraphics[width=5in]{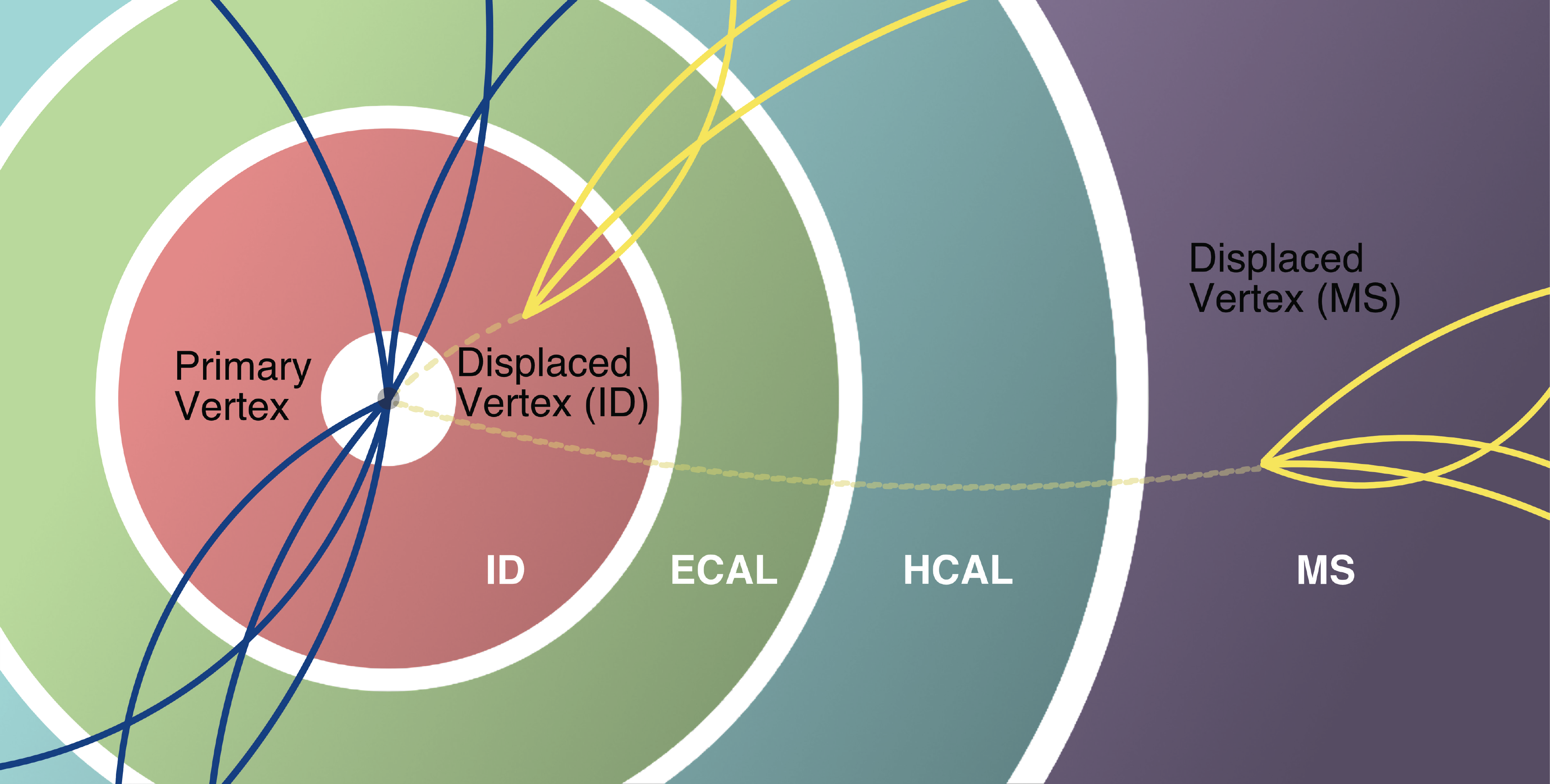}
\caption{Primary vertices formed with primary tracks are shown on a generic detector. Displaced vertices in both the ID and MS are shown.}
\label{fig:displacedvertex}
\end{figure}

When several LLP-daughter tracks are detected, their common point of origin constitutes a \emph{displaced vertex} (DV), with a position $\vec r_{\rm DV}$ and corresponding covariance matrix that can be determined by a vertex-fitting algorithm. Such a DV is illustrated in Fig.~\ref{fig:displacedvertex}. Since the vertex involves  several tracks, the distance $|\vec r_{\rm DV}|$ of the vertex from the IP or from the PV is determined more precisely than $d_0$ and directly represents the relevant decay length. The length of the transverse-plane projection of $\vec r_{\rm DV}$, denoted $\rho_{DV}$, is also used to separate signal from prompt background, as is the ratio between $\rho_{DV}$ and its uncertainty $\sigma_{\rho_{DV}}$. 

Another often-used variable is the collinearity angle $\alpha_{\rm col}\equiv \cos^{-1}(\hat r_{\rm DV} \cdot \hat p_{\rm DV})$ between $\vec r_{\rm DV}$ and the sum $\vec p_{\rm DV}$ of the momentum vectors of the tracks composing the DV. The transverse collinearity angle $\phi_{\rm col}\equiv \cos^{-1}(\hat \rho_{\rm DV} \cdot \hat p^T_{\rm DV})$ is defined analogously, with the transverse-plane projections of $\vec r_{\rm DV}$ and $\vec p_{\rm DV}$.

In addition, kinematic variables of the DV can be used to suppress background. Typical variables include the invariant mass $m_{\rm DV}$ of the tracks forming the DV (often assuming the tracks have the mass of a charged pion, by convention), track multiplicity, and their combined momentum $p_{\rm DV}$ or transverse momentum $p^{\rm DV}_{\rm T}$. 

Background from SM particles undergoing hadronic interactions with the detector material or support structures is often suppressed by rejecting DV candidates that are found in geometric volumes known to be dominated by dense material. Due to the effort required for accurate mapping of the detector material, the level of detail with which this is done varies, and generally leads to improved sensitivity as the analyses mature with time.

\begin{figure}[tb]
\centering
\includegraphics[width=5in]{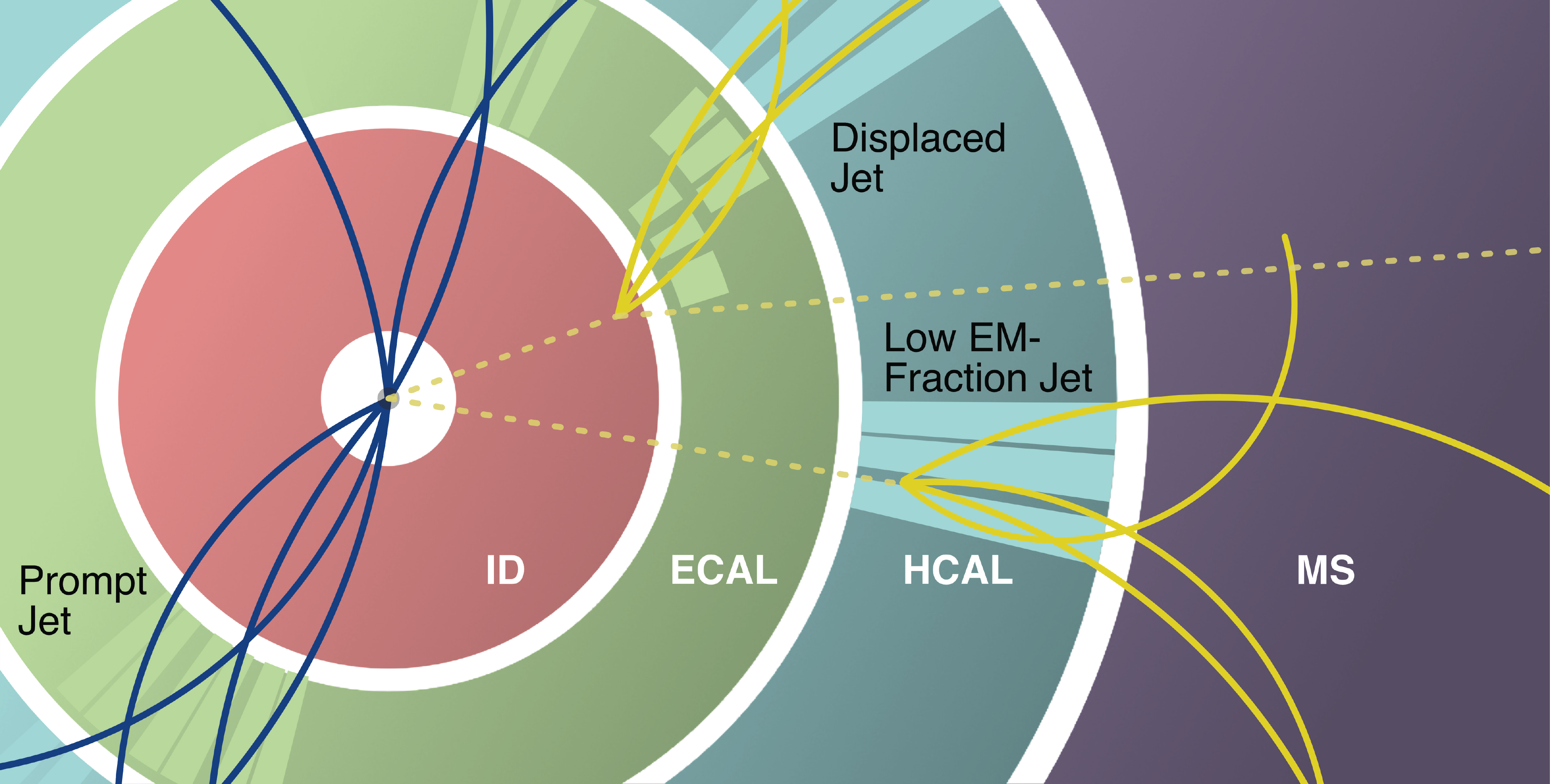}
\caption{Prompt jet activity is shown along with two types of displaced calorimeter deposits are shown in yellow. One depicts the use of calorimeter segmentation to determine the pointing direction of the incoming particles while the other shows the production of particles within the calorimeter system thus leaving relatively small amount of energy in the ECAL.}
\label{fig:displacedcalodeposits}
\end{figure}

\subsubsection{Displaced Calorimeter Deposits}
\label{sec:displaced-cal}

The precise spatial measurements obtained from the ID are usually limited to relatively small displacements of up to tens of
cm and can be performed only for charged-particle tracks. However,
detectors at high energy colliders have deep calorimeters, and these
have been used to search for LLP decays at larger distances. Calorimeters with 3-dimensional segmentation can measure the direction of an incoming particle, yielding its $d_0$ and $z_0$, which are used as handles for identifying LLPs.
These measurements are used to identify \emph{non-pointing photons} that do originate from the IP, and to find the vertex of a multi-photon decay.

If a LLP decay occurs within the calorimeter volume itself, the longitudinal shower shapes can provide a handle on SM backgrounds. A simple shower-shape observable is the \emph{EM ratio} between the energy deposited in the ECAL and that deposited in the HCAL in an angular region that defines a jet. For a LLP decay that takes place inside the calorimeter, this ratio is anomalously low relative to that of a SM jet produced near the IP. 

These signatures are illustrated in Fig.~\ref{fig:displacedcalodeposits}.

\subsubsection{Aggregate Signatures}
\label{sec:ddcllps}

While some LLPs give rise to the individual detector signatures described above, others produce multiple signatures that can be used simultaneously. In many such cases, using multiple detector subsystems allows for independent, uncorrelated handles on a single particle property, allowing for powerful rejection of SM backgrounds.

\begin{figure}[tb]
\centering
\includegraphics[width=5in]{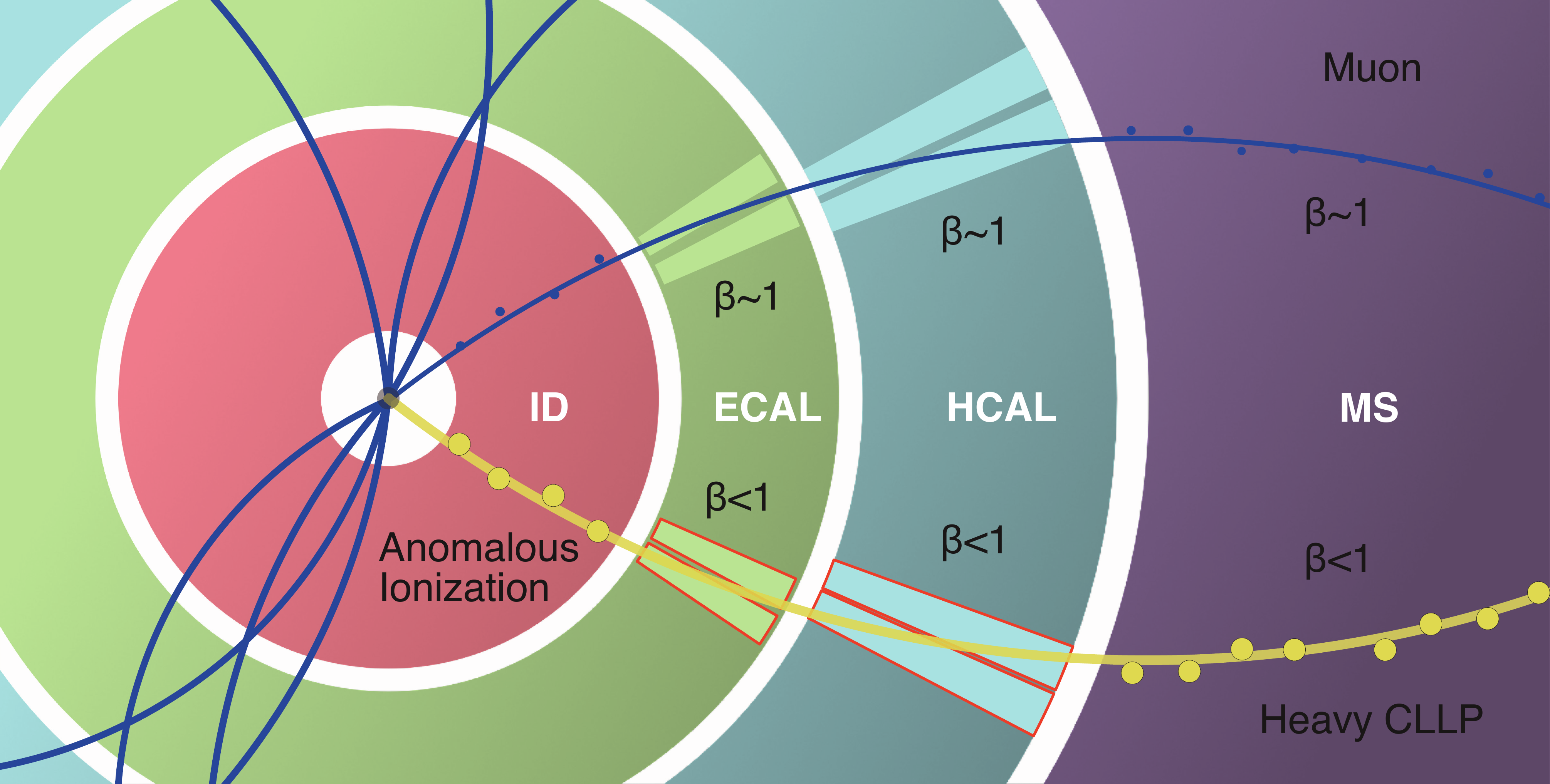}
\caption{A heavy, charged LLP is shown at the bottom-right of the figure traversing an example detector. Its signature would include anomalously high levels of ionization in the various detector subsystems. In addition, if the LLP is sufficiently slow, detectors with sufficient timing resolution can be used to measure its speed. By contrast, a muon with the same momentum, shown at the top-right of the diagram, is minimum-ionizing and highly relativistic.}
\label{fig:cllp}
\end{figure}

When a charged LLP (CLLP) passes through multiple detector subsystems, the signatures described above can be used in concert. One may simultaneously look for anomalous ionization in the ID, ECAL, HCAL, and the MS. If the CLLP is also slow-moving, it will give rise to delayed time of arrival at the ECAL, HCAL, and MS. These signatures are illustrated in Fig.~\ref{fig:cllp}.

Similarly, the signature of a magnetic monopole has multi-subsystem characteristics. As mentioned above, a magnetic monopole produces an anomalously shaped track and is usually also highly ionizing, properties that can be measured in the ID and the calorimeter. In addition, if heavy enough, it is slow-moving and leads to delayed signatures. 

A CLLP that decays after passing through (part of) the inner tracker gives rise to a track with anomalously high $dE/dx$ plus an additional signature that depends on the decay position. When the decay occurs inside the ID, the track can be a disappearing track, possibly with a displaced vertex at its endpoint. Decays inside the calorimeter system have a high-EM-fraction energy-deposition pattern, and decays outside the calorimeters produce a DV inside the MS. Depending on the speed of the CLLP, all these detected signals may in addition be delayed.

When searching for the decay products of the LLP, where assuming that the decay products are detectable allows a search to be relatively agnostic to the charge of the LLP itself, multiple signatures may be used as well. When a decay occurs inside a particular subsystem, use of signals from other subsystems can help provide further background rejection. For example, for a decay inside the ID, a DV signature may be augmented by simultaneously searching for displaced calorimeter deposits, delayed signals in both the calorimeter and the MS, and other such signatures.

As the lifetime approaches very small values, reconstruction efficiencies from standard prompt BSM decays searches increase. As a result, prompt searches can retain a tail of sensitivity for the small lifetime regime. However, the use of standard reconstruction on displaced objects can lead to additional systematic uncertainties as biases are introduced in the reconstruction, identification, and calibration techniques.

\subsection{Comparison of Detector Subsystem Acceptances}
\label{sec:acceptance-comparison}

It is useful to gain a basic understanding of the relative sensitivities of search analyses that rely on different detector subsystems for LLP detection with a volume-based acceptance study. For this purpose, we ignore the effects of background which is usually small for all displacements. Furthermore, we assume that the \emph{efficiency}, defined as the probability to trigger on the event and identify the LLP if indeed it decayed within the relevant detector subsystem, is 100\%. With these simplifications, the sensitivity for each detector subsystem can be estimated based on the subsystem \emph{acceptance}, which we define as the probability for the LLP decay to occur within that subsystem, given the LLP lifetime and boost distribution. 
This narrow definition of acceptance is useful for estimating the sensitivity in a range of search analysis methods, particularly those aimed at hadronic LLP decays. However, relating it to sensitivity fails for other analysis techniques, such as those aimed at reconstructing LLP decays into muons, which penetrate the calorimeter.

\begin{figure}[tb]
\centering
\includegraphics[width=0.48\textwidth]{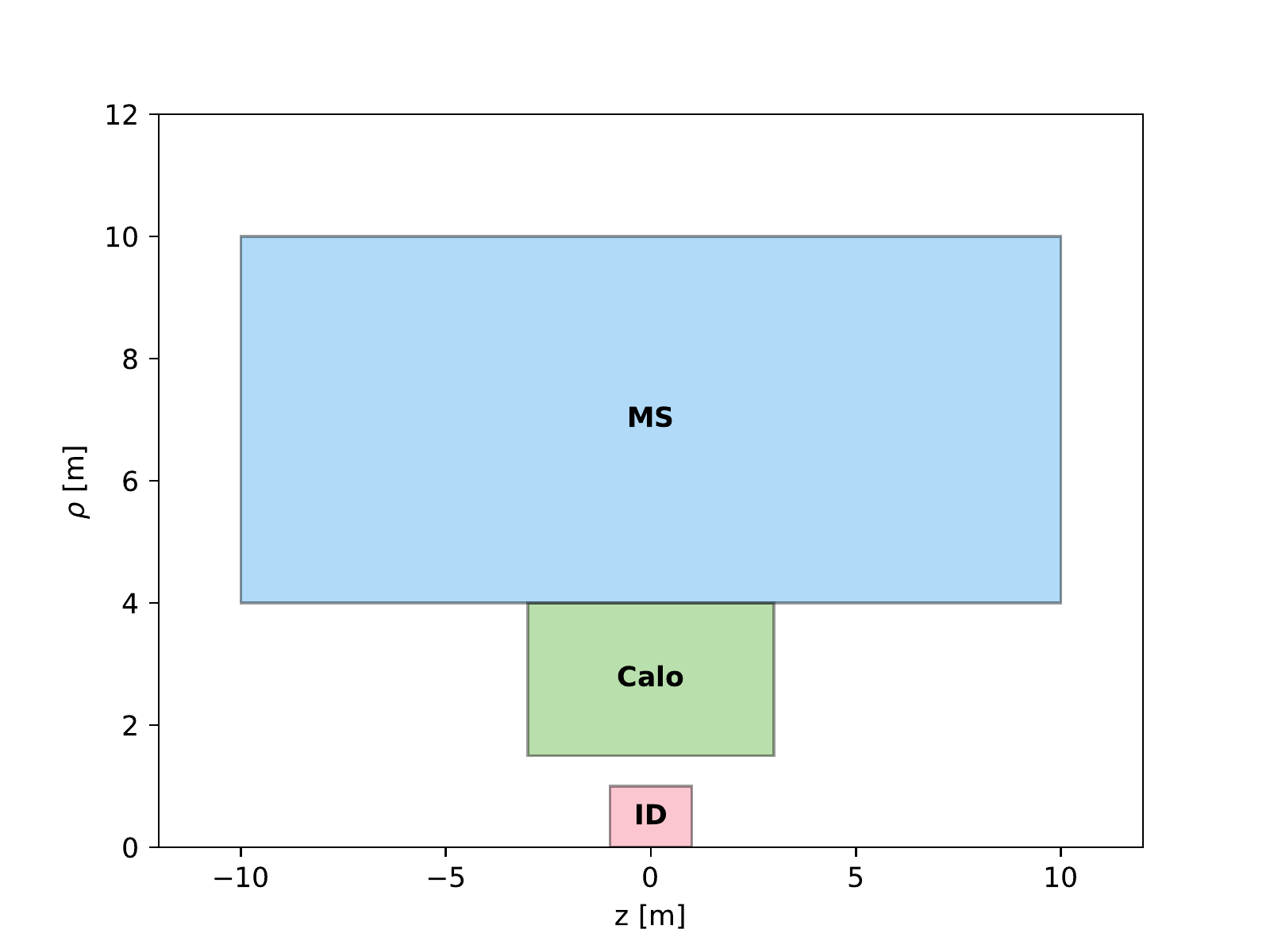}
\includegraphics[width=0.48\textwidth]{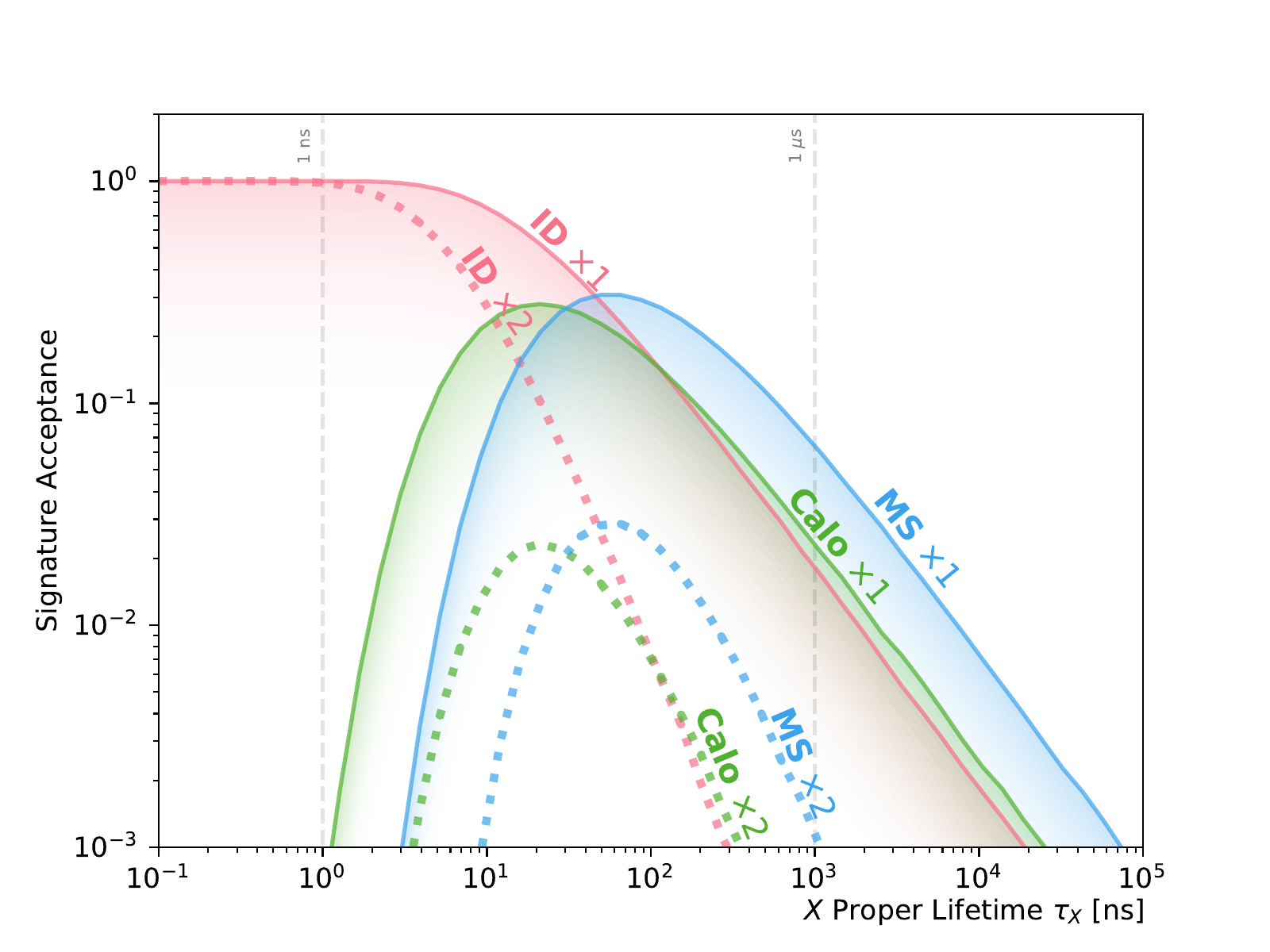}
\caption{An example detector is shown (left) containing an ID, a calorimeter system, and a MS represented in a $z$ vs. $\rho$ space. For pair-produced particles with kinematics described in the text, the volume acceptance for LLPs is shown as a function of lifetime (right). The fraction of events containing one LLP decay in each system is shown as solid lines. The fraction of events with \emph{both} LLP decays contained in a single system is shown in dashed lines.}
\label{fig:toystudy}
\end{figure}

With this caveat in mind, we proceed to calculate the acceptances of typical LHC detector subsystems. We define in Fig.~\ref{fig:toystudy} an example detector with three subsystems, defined by radial and longitudinal barrel-region extents: an ID ($0<\rho<1~\mbox{m}$, $|z|<1~\mbox{m}$), a calorimeter system ($1.5<\rho<4~\mbox{m}$, $|z|<4~\mbox{m}$), and a MS ($4<\rho<10~\mbox{m}$, $|z|<10~\mbox{m}$). These volumes are roughly representative of the detectors at the LHC.

For this detector, we determine the acceptance for pair-produced LLPs, with kinematics taken from a simulated sample of gluino pairs produced with \textsc{MadGraph5\_aMC@NLO}~\cite{madgraph} for $13$~TeV proton-proton collisions, with a gluino mass of $m_{\tilde{g}}=2$~TeV. For a given value of the gluino lifetime, the proper decay time for each gluino is sampled from an exponential distribution, and the decay position in the detector is calculated given the gluino velocity. Since our purpose is only to calculate the acceptance, it is assumed that the gluino does not undergo significant interaction with the detector material. The fractions of events that contain at least one LLP decay in each of the ID, calorimeter, or MS are shown as solid lines in Fig.~\ref{fig:toystudy}. Requiring two LLPs to decay in a particular detector subsystem results in reduced acceptance, as shown by the dashed curves in Fig.~\ref{fig:toystudy}. These curves do not represent the associated reduction in reconstruction efficiency. 

The actual search sensitivities depend on the trigger and reconstruction efficiencies, as well as on background levels, while the exercise presented here simply evaluates the spatial acceptance. However, since many searches in the ID have very low background even when requiring just a single LLP, Fig.~\ref{fig:toystudy} can be interpreted to demonstrate that the ID provides the best sensitivity for a wide range of lifetimes in these scenarios. In the case of a very large MS, such as that of the ATLAS detector, the MS acceptance  overtakes that of the ID in the very long lifetime ($\tau>100$~ns) regime only when requiring at least one LLP per event. Therefore, except when an ID analysis suffers from low reconstruction efficiency or particularly high background, it is likely most advantageous to search for a single decay in the ID, even in the case of pair-produced LLPs. 
We note that for analyses with high background levels, requiring two LLPs is usually a necessary measure for effective background reduction, despite the loss of acceptance.

\subsection{Considerations When Using Standard-Object Reconstruction}
\label{sec:recoconsiderations}

LLP searches often utilize standard reconstruction algorithms designed for prompt objects, such as jets, photons, and leptons. Depending on the LLP parameters, standard algorithms can be sensitive to the detector signatures of LLPs, often with significant efficiency. Nonetheless, considerable effort must be spent by analysts on understanding systematic effects that arise from the displacement and/or delay of the detector signals of a LLP. As an example, jets produced in a significantly displaced decay of a heavy LLP include hadrons that impinge on the calorimeter face at a significant grazing angle. This results in calorimeter energy deposits that are both delayed and have different cluster shapes from those of promptly produced jets. Accounting for these systematic effects on the jet energy scale and resolution requires dedicated studies. 

As displaced and delayed signatures are generally reconstructed with degraded efficiency compared to prompt standard objects, the interplay between LLP detection and the determination of MET can be non-trivial. For example, an electrically charged LLP could cross the calorimeter while depositing relatively little energy. As it enters the MS, it could be reconstructed as a muon with the correct momentum. However, if it is sufficiently slow, it might arrive in the MS too late for the MS signature to be associated with the correct bunch crossing. The muon signature would thus be missed, leading to a measurement of significant MET. 
Thus, on the one hand, LLP searches can sometimes utilize MET as a selection criterion. However, this requires careful study of the experimental effects that lead to the MET measurement. A further complication arises from the fact that MET calculation algorithms at the trigger level are often different from those used offline.

These aspects are important to consider when reinterpreting (or \emph{recasting}) results of a particular search for application to a theoretical model that was considered in the original experimental analysis. In particular, when considering how sensitive a search targeting prompt signals would be to for a model with displaced decays, these additional uncertainties should not be neglected.

\section{Review of Published Searches}
\label{sec:searches}

In this section, we  review experimental searches for the types of signatures described in Sec.~\ref{sec:signatures} at the LHC and, at times, other colliders. The program of direct-detection searches for LLPs is summarized first, followed by the program of indirect detection searches.

\subsection{Direct Detection: Signatures from LLPs Interacting with the Detector}
\label{sec:direct-detection}

A sufficiently long-lived particle created in a collider experiment could directly interact with the detector while traversing it. Depending on its lifetime, the LLP can deposit energy in only the innermost subdetectors, or even travel through all layers and leave signals throughout.

\subsubsection{Detector-Stable Charged LLPs}
\label{sec:detectorstablecllps}

A cornerstone of the LLP search programs at the current and most recent colliders is the suite of searches targeting heavy charged particles with decay lengths long enough that they interact directly with significant parts of the detector. Being heavy, these particles move at a speed significantly lower than that of a SM particle of the same momentum. This yields two unique experimental signatures: high ionization energy loss (Sec.~\ref{sec:dedx}) and delayed arrival in distant detector subsystems (Sec.~\ref{sec:ddcllps}). Since the particle's time of flight and specific ionization are measured independently, the tails of their distributions are uncorrelated for the background. Therefore, combining requirements on these two variables provides powerful background rejection.

Searches for charged long-lived particles (CLLPs) are often optimized for SUSY models, where the CLLPs are sleptons and $R$-hadrons, described in Sec.~\ref{sec:susytheory}. A heavy, detector-stable slepton would appear similar to a high-momentum muon, in that it would leave a stiff track in the ID, pass through the calorimeter, and continue as a track in the MS. However, unlike a high-$\pT$ muon, the CLPPs would be slow and highly ionizing. An $R$-hadron also interacts hadronically as it moves slowly through the detector. Its heavy parton is mostly a spectator in these processes, acting as a reservoir of kinetic energy for the bound light quarks and gluons which undergo low-energy scattering with nuclei in the detector material. Therefore, rather than a high-energy hadronic shower, the calorimeter signature is that of a penetrating particle with little energy loss compared to that of a SM hadron.\footnote{The case of very slow $R$-hadrons that stop in the detector is discussed in Sec.~\ref{sec:stoppedllps}.} 
The hadronic scattering processes can change the composition of the light-quark system, so that the electric charge of the $R$-hadron can vary during its flight. When electrically charged, an $R$-hadron also loses energy via ionization.

While SUSY serves as a strong motivation for CLLP searches, these searches are sensitive to other theoretical frameworks. Model-independent limits are often provided by the experiments, along with enough information about the acceptance and the efficiency of the analysis to enable reinterpretation in other models. 

\paragraph{Pre-LHC Searches.}

At LEP, the ALEPH, DELPHI, L3, and OPAL collaborations have all performed CLLP searches in $e^+e^-$ collisions in data sets with $\sqrt{s}$ ranging between 130 and 209~GeV. In the context of SUSY, the searches primarily targeted long-lived sleptons and charginos. GMSB scenarios were often used for interpreting the results. 

The ALEPH collaboration performed a CLLP search using $dE/dx$ measurements in their time-projection chamber (TPC) in data sets with $\sqrt{s}$ up to 172~GeV. With 0.3 events expected from background and none observed, exclusion limits were calculated. Long-lived staus and smuons below 67~GeV and charginos below 86~GeV were excluded. More model-independent cross-section upper limits in the range 0.2-0.4~pb were set for CLLPs with masses up to 86~GeV~\cite{Barate:1997dr}. ALEPH also included constraints from CLLP searches in their summary statement on GMSB in Ref.~\cite{Heister:2002vh}. The DELPHI detector was equipped with a ring imaging Cherenkov detector and a TPC, which together provided powerful particle identification capabilities and were exploited for searching for gluino-based $R$-hadrons. In the LEP1 data set with $\sqrt{s} = m_Z$, a search for long-lived gluinos, assumed to be pair-produced through final-state radiation of a gluon in $Z\rightarrow q\bar{q}$ events, excluded the mass range $2< m_{\tilde{g}} < 18$~GeV. Subsequently, using 609~pb$^{-1}$ of LEP2 data with $\sqrt{s}$ of 189-209~GeV, a search for $R$-hadrons produced in decays of pair-produced squarks was performed and excluded long-lived gluinos up to around 90~GeV~\cite{Abdallah:2002qi}. The OPAL experiment searched for pair-produced CLLPs using its jet chamber, which provided up to 159 $dE/dx$ measurements for each CLLP candidate track. Several data sets with $\sqrt{s} = $130-209~GeV were used, and the lack of observed excess over the expected background was interpreted as exclusions of smuons and staus with $m < 98$~GeV, as well as charginos with $m < 102$~GeV in constrained MSSM models~\cite{Abbiendi:2003yd}. OPAL expanded this search to include event topologies with large track multiplicities in order to improve sensitivity also to signal scenarios with production of color-charged particles~\cite{Abbiendi:2005gc}. This paper summarized constraints on GMSB and used a data set of 693.1~pb$^{-1}$ to set similar mass limits to the ones of Ref.~\cite{Abbiendi:2003yd} in the GMSB framework. The L3 collaboration produced exclusion limits for heavy charged leptons using $dE/dx$ measurements in their tracking chamber. Using their full LEP2 data set of 450~pb$^{-1}$ at $\sqrt{s} = $192-208~GeV, they excluded masses below 102.6~GeV~\cite{Achard:2001qw}. 

In connection with a measurement of anti-deuteron production, The H1 experiment~\cite{Abt:1996hi} at the HERA collider at the Deutsches Elektronen-Synchrotron laboratory in Hamburg, Germany performed a search for CLLPs~\cite{Aktas:2004pq}. The search used $dE/dx$ measurements in the jet chamber and track $\pT$ measurements in the tracker. An upper limit of 0.19~nb was set on the cross section for photoproduction of positively (negatively) charged LLPs with mass larger than that of the triton (anti-deuteron) in a given kinematic range in collisions of positrons and protons with energies 27.6~GeV and 820~GeV, respectively. 

In Run I of the Tevatron, several searches for CLLPs were performed by the CDF collaboration in $p\bar{p}$ collisions at $\sqrt{s} = 1.8$~TeV. In the first two iterations~\cite{Abe:1989es,Abe:1992vr}, time-of-flight measurements performed with the hadronic calorimeter were used. The third analysis, conducted with 90~pb$^{-1}$ of data, $dE/dx$ measurements in the central tracking chamber and silicon vertex detector were used as well~\cite{Acosta:2002ju}. With yields compatible with background expectations, cross-section upper limits around 1~pb were obtained for strong production of fourth-generation quarks with mass up to 270~GeV and Drell-Yan production of sleptons with 80~GeV$ < m < $120~GeV.
With the increased energy of $\sqrt{s} = 1.96$~TeV in Run 2, The D0 collaboration published two searches for CLLPs using 1.1~\cite{Abazov:2008qu} and 5.2~fb$^{-1}$~\cite{Abazov:2011pf}. 
A follow-up paper combined these results and provided additional interpretations for them~\cite{Abazov:2012ina}. 
Using time-of-flight measurements in the drift tubes of the muon detector and $dE/dx$ in the silicon microstrip tracker, D0 excluded long-lived gaugino-like and higgsino-like charginos below masses of 278~GeV and 244~GeV, respectively, and set cross-section upper limits for pair-production of staus with 100~GeV$ < m < $300~GeV in the $\mathcal{O}(10)$~fb range. Top squarks were excluded below a mass in the rage 285-305~GeV, depending on the assumptions of the interactions between the $R$-hadron and the detector material.
The final statement on CLLPs from the CDF collaboration used 1.0~fb$^{-1}$ of 1.96~TeV $p\bar{p}$ data and included timing measurements from a new dedicated TOF detector~\cite{Aaltonen:2009kea}. 
The result was interpreted as exclusions for pair-produced stop-based $R$-hadrons with $m < 249$~GeV, corresponding to upper cross-section limits in the vicinity of 50~fb.

\paragraph{Searches at LHC.}

The first CLLP searches at LHC were performed with the 7~TeV $pp$ data collected in 2010. The CMS collaboration used 3.1~pb$^{-1}$ to look for ID tracks with high $dE/dx$ measured in the silicon-strip tracking detector, with and without compatible tracks in the MS. Exclusion limits for $R$-hadrons were established for gluinos at $m < 398$~GeV (311~GeV for $R$-hadrons assumed to be neutral in the MS) and stops at $m < 202$~GeV~\cite{Khachatryan:2011ts}. Shortly afterwards, the ATLAS collaboration published a CLLP search using 34~pb$^{-1}$. They employed a combination of $dE/dx$ in the silicon pixel detector and time-of-flight measurements in the hadronic calorimeter, without requiring a track in the MS. This search raised the mass limits to up to 586~GeV for gluinos and 309~GeV for stops~\cite{Aad:2011yf}. 
A separate ATLAS search, performed with the same data set, used a muon-like signature based on time-of-flight in both the MS and the calorimeter. This search yielded limits of up to 544~GeV for gluino-based $R$-hadrons and up to 120~GeV for direct pair-production of long-lived sleptons~\cite{Aad:2011hz}.  The CMS measurement was repeated with 5.0~fb$^{-1}$ and additional use of time-of-flight information from the MS, extending the mass limits to up to 1098 (737)~GeV for gluinos (stops) and 223~GeV for staus~\cite{Chatrchyan:2012sp}. The following result from ATLAS used 4.7~fb$^{-1}$, and now combined all three discriminants from the ID, calorimeter, and MS. $R$-hadrons formed from gluinos, stops, and sbottoms were excluded up to a mass of 985, 683, and 612~GeV, respectively, and direct pair-production of long-lived sleptons with $m < 278$~GeV as well~\cite{Aad:2012pra}. 

Using nearly 20~fb$^{-1}$ of $pp$ data at $\sqrt{s} = 8$~TeV, CMS extended the limit on the mass of gluino (stop) $R$-hadrons to up to 1322 (935)~GeV, and excluded directly pair-produced staus with mass below 339~GeV~\cite{Chatrchyan:2013oca}. These results were later reinterpreted by the collaboration in the frameworks of the phenomenological MSSM and AMSB in Ref.~\cite{Khachatryan:2015lla}. With approximately the same amount of 8~TeV data, ATLAS repeated the CLLP search and reached gluino, stop and sbottom mass limits of up to 1270, 900 and 845 GeV. Direct tau pair-production was excluded up to 290~GeV and interpretations were included for GMSB and LeptoSUSY models~\cite{ATLAS:2014fka}. With this data set, ATLAS also performed a search sensitive to lifetimes as low as $\tau=0.6$~ns, by requiring only that the CLLP leave an  ID track with anomalous $dE/dx$~\cite{Aad:2015qfa}. For this signature, the lower-mass limits for gluino-based $R$-hadrons reached up to 750 (1250)~GeV for $\tau_{\tilde{g}} = 0.6~(10)$~ns, considering several different $R$-hadron decay possibilities and $\tilde{\chi}_1^0$ masses. Similarly, long-lived charginos nearly mass-degenerate with the $\tilde{\chi}_1^0$ LSP and with mass up to 239 (482)~GeV were excluded for $\tau_{\tilde{\chi}_1^\pm} = 1~(15)$~ns. The LHCb collaboration also performed a CLLP search with 3.0~fb$^{-1}$ of 7 and 8~TeV data, using their ring imaging Cherenkov detectors as a primary tool for identifying CLLPs. With zero events observed in the signal region, upper limits on the cross section were derived as a function of CLLP mass, assuming Drell-Yan-like pair-production kinematics in the pseudorapidity range $1.8 < |\eta| < 4.9$. 
The resulting limits range from 3.4~fb at $m = 124$~GeV to 5.7~fb at $m = 309$~GeV~\cite{Aaij:2015ica}. 

In Run-2 of the LHC (with $\sqrt{s} =13$~TeV), the full-detector CLLP search at ATLAS was performed with 3.2~fb$^{-1}$. With yields again matching those expected for the background-only hypothesis, the limits for gluino, stop, and sbottom $R$-hadrons were extended to 1580, 805 and 890 GeV, respectively~\cite{Aaboud:2016uth}.  
CMS published a corresponding result using 2.5~fb$^{-1}$, reaching mass exclusions of up to 1610, 1040, and 240~GeV for long-lived gluinos, stops and directly 
pair-produced sleptons~\cite{Khachatryan:2016sfv}. The ATLAS search using the ID-only signature was also performed with the 3.2~fb$^{-1}$ dataset~\cite{Aaboud:2016dgf}, and was recently repeated with 36~fb$^{-1}$ recorded in 2015-2016~\cite{Aaboud:2018hdl}. No significant excess was observed in the data, and gluino $R$-hadrons were excluded for a lifetime of 1 (10)~ns and mass of up to 1300 (2060)~GeV.

\subsubsection{Disappearing Tracks}

When a CLLP decays deep within the ID, and any charged particles produced in the decay are too soft to be reliably tracked, it can produce a disappearing-track signature. As discussed in Sec.~{\ref{sec:amsb}}, this signature is particularly motivated by AMSB SUSY models, where a heavy neutralino takes up most of the energy in the decay of the slightly heavier chargino. 

Searches for disappearing tracks have been performed in Run-1 and Run-2 of the LHC by the ATLAS \cite{ATLAS:DT7TeV1,ATLAS:DT7TeV2,ATLAS:DT8TeV,ATLAS:DT13TeV} and CMS \cite{CMS:DT8TeV,CMS:DT13TeV} experiments. The first such search was performed by ATLAS~\cite{ATLAS:DT7TeV1} using $1.01~\invfb$ of $\sqrt{s}=7~\tev$ data. This and other Run-1 searches from the ATLAS collaboration look for tracks that have limited hits in the outermost layers of the ID. This approach gives maximal sensitivity at a characteristic length scale of about $0.5$~m. The trigger strategy was to require at least one radiated jet that gives rise to moderate missing transverse energy.

In Run-2 of the LHC, exploiting a newly inserted tracking layer at short radius, ATLAS was able to also pursue shorter lifetimes, with a characteristic scale of $0.3$~m~\cite{ATLAS:DT13TeV}. Along with the increase in energy to $13~\tev$ and the larger luminosity of $36.1~\invfb$, this resulted in significant improvement in the signal sensitivity over a range of model parameters.

CMS also pursued this signature in both Run-1~\cite{CMS:DT8TeV} and Run-2~\cite{CMS:DT13TeV}, the latter analysis using $38.4~\invfb$. Reflecting the fact that CMS has an all-silicon tracker, the characteristic length scale in this analysis was about $0.8$~m. A veto on calorimeter energy was used to help reduce SM backgrounds.

For these searches, the SM backgrounds tend to be hadrons scattering in material, charged leptons undergoing a large momentum change due to bremsstrahlung, and spurious tracks formed from uncorrelated detector hits. Control regions were used to obtain template momentum distributions for the background. These distributions were then fit to data in signal regions to test for the presence of signal.
The expected background yield in the Run-2 analyses was typically in the tens of events, largely dominated by the hadron background. The analyses observed no significant deviation from the SM expectation, and limits were set on the chargino lifetime and mass. 

The LHC collaborations have not yet exploited a kinked-track signature. However, this has been done by the ALEPH experiment at LEP, ~\cite{Heister:2002vh}, and used for setting limits on the stau mass and lifetime.

\subsubsection{Particles with Anomalous Electric Charge}
\label{sec:anomalous-charge}

\paragraph{Fractional Electric Charge.}

Since all free charges in the SM are integer multiples of the electron charge $Q_e$, identification of a LLP with a non-integer charge would be a clear sign of BSM. Various searches have been performed to search for such particles~\cite{Perl:2009zz}.
At LEP, limits were set on CLLPs with charges   $\frac{2}{3}Q_e$, $\frac{4}{3}Q_e$, and $\frac{5}{3}Q_e$ at a range of collision energies from $91$~GeV to $209$~GeV~\cite{Abbiendi:2003yd,Abreu:1996py,Akers:1995az,Buskulic:1992mr}. At the Tevatron, the CDF experiment set limits on CLLPs with charges $\frac{2}{3}Q_e$ and $\frac{1}{3}Q_e$, excluding masses up to $250~\gev$~\cite{Acosta:2002ju}. 

The CMS experiment performed a search for fractionally charged particles using $5.0~\invfb$ of $\sqrt{s}=7$~TeV data~\cite{CMS:fraccharge}. The search used tracks in the ID that appeared to have significantly lower ionization than minimum-ionizing particles. A background ionization template was obtained using a sample of $Z\rightarrow\mu^+\mu^-$ events to model the low-side tail of the SM ionization distribution. Since selected events were triggered by the presence of a muon, cosmic-ray muons were also considered as a potential source of background. Given the analysis criteria, a total SM background of $0.012\pm0.007$ events was expected. No events were observed, and limits were set on the mass of fractional-charge CLLPs. For charges $\frac{1}{3}Q_e$ and $\frac{2}{3}Q_e$, masses below $140$~GeV and $310$~GeV were excluded, respectively. 

This search was augmented with $18.8~\invfb$ of $\sqrt{s}=8~\tev$ data, extending the mass bounds on $\frac{1}{3}Q_e$ and $\frac{2}{3}Q_e$ CLLPs to about $200$~GeV and $480$~GeV, respectively~\cite{Chatrchyan:2013oca}.

\paragraph{Multiple Electric Charge.}

In contrast to fractionally charged particles having unusually low ionization signatures, particles with large electric charge would leave strikingly large ionization signatures in particle detectors.

The ATLAS collaboration has performed multiple searches for such particles in Run-1 of the LHC with sensitivity to charges ranging from $2Q_e$ to $17Q_e$~\cite{atlas:multicharge1,atlas:multicharge2,atlas:multicharge3}. In Ref.~\cite{atlas:multicharge1}, the variables  used to identify signal were ionization in the transition radiation tracking detector (TRT) and the electromagnetic shower shape in the ECAL. Multiply charged particles would be expected to leave an unusually large number of high-threshold hits in the TRT and very narrow showers in the ECAL. These properties were exploited in this search to construct a signal region with an expected background yield of $0.019\pm0.005$ events in $3.1~\invpb$ of $\sqrt{s}=7$~TeV collision data. No events were observed, and cross section limits were set between $1$ and $12$~pb for various charges and masses.

ATLAS also published a set of searches for multicharged CLLPs using ionization information from the MS~\cite{atlas:multicharge2,atlas:multicharge3}. In these searches, independent measurements of a reconstructed muon candidate were performed in the ID and the MS. Two different signal region selections are used, one for doubly charged particles and another for larger charges. These regions were expected to contain $0.013\pm0.002\textrm{(stat.)}\pm0.003\textrm{(stat.)}$ and $0.026\pm0.003\textrm{(stat.)}\pm0.007\textrm{(stat.)}$ background events, respectively, in $20.3~\invfb$ of $\sqrt{s}=8$~TeV data. No events were observed, and limits were placed on the allowed mass of such multicharged states. Lower mass limits range from $660$~GeV for a charge of $2Q_e$ to $760$~GeV for a charge of $6Q_e$. 

A Run-1 search from ATLAS that targeted magnetic monopoles~\cite{Aad:2015kta} also excluded particles with with charges in the range 10--60$Q_e$ and is described in Sec.~\ref{sec:monopole-searches}. The Run-1 CLLP search from CMS described in Sec.~\ref{sec:detectorstablecllps} also set limits on multiply charged particles, extending the mass bounds on $2Q_e$ and $6Q_e$ charges to about $690$~GeV and $780$~GeV, respectively~\cite{Chatrchyan:2013oca}.

\subsubsection{Magnetic Monopoles}
\label{sec:monopole-searches}

Magnetic monopoles have inspired searches at many colliders in the past decades, as well as in non-collider experiments~\cite{Tanabashi:2018oca,Tanabashi:2018oca,Eberhard:1971re,Ross:1973it,GRAF1991463}. In this section we review the energy-frontier searches performed at the Tevatron and at the LHC.

A general comment is in order about simulations used for interpreting the results of magnetic monopole searches. Pair production of monopoles is typically simulated as an electromagnetic process, most commonly Drell-Yan or photon fusion. This enables generation of events that can be used to study the detector response and determine the reconstruction and selection efficiencies. However, due to the large charge of magnetic monopoles, calculations and simulations that rely on perturbation theory are unreliable. Therefore, interpretation of searches in terms of model parameters is not trivial. 

\paragraph{Searches with General-Purpose Detectors.}

The CDF experiment has searched for magnetic monopoles created in $p\bar p$ collisions at the Tevatron and traversing the ID tracker~\cite{Abulencia:2005hb}. Using the plastic scintillator time-of-flight counters surrounding the central outer tracker of the detector, a dedicated trigger was implemented based on the extreme ionization signature expected for a magnetic monopole. The events recorded by this trigger were scrutinized for tracks that do not bend in the azimuthal direction. No signal was found, and an upper limit on the production cross section for monopoles with mass in the range $200 < m < 800$~GeV was set at 0.2~pb.

At the LHC, the ATLAS experiment has published searches for magnetic monopoles in $pp$ collisions at $\sqrt{s} = 7$~TeV~\cite{Aad:2012qi} and 8~TeV~\cite{Aad:2015kta}. Similar to the searches for highly charged particles reported in Sec.~\ref{sec:anomalous-charge}, the monopole searches used the fraction of high-threshold hits in the transition radiation tracker and the electromagnetic shower shape in the ECAL to identify magnetic monopoles using their property of high ionization. The observed signal yields were compatible with the expectations for the background-only hypothesis. Two types of upper limits on the production cross section were extracted. In the first type, monopole pair-production with kinematics generated from a Drell-Yan process was assumed. The second type was model-independent limits for a single monopole produced in fiducial regions defined by transverse kinetic energy and pseudorapidity, in which the selection efficiency was uniform and high. For the fiducial volume analysis, the most stringent upper cross section limit was set at 0.5~fb for the production of a monopole with a magnetic charge between $0.5Q_{D}$ and $2Q_{D}$ and mass in the range 200~GeV~$< m<$~2500~GeV~\cite{Aad:2015kta}.
For the case of Drell-Yan kinematics, cross section limits were quoted for both spin-0 and a spin-1/2 monopole hypotheses, excluding masses below 430~GeV and 700~GeV for a Dirac monopole, respectively.

\paragraph{Searches with Dedicated Detectors.}

Another magnetic monopole-search method exploited at colliders involves removing material exposed to collision particles and examining it offline. Such searches have used two main techniques. The first is to examine the material for the presence of magnetic monopoles that have stopped within it, using a superconducting quantum interference device (SQuID), in which a captured monopole would induce a permanent current. The second involves searching for the characteristic tracks caused by the passage of highly ionizing particles through the material.

Parts of the CDF (lead from the Forward EM calorimeter and an aluminum cylinder) and D0 (beryllium beam pipe and aluminum cylinders) detectors were examined with the dedicated E-882 SQuID experiment~\cite{Kalbfleisch:2003yt}. The analyzed parts had been exposed to approximately 175~pb$^{-1}$ of $p\bar{p}$ collisions at $\sqrt{s} = 1.8$~TeV. No signal was seen. Assuming monopole kinematics of Drell-Yan-like pair-production, upper limits on the cross-section in the range $0.07$ to $0.2$~pb were determined for magnetic charges of 1, 2, 3, and 6 times $Q_{D}$.

The track-based technique was used in a search performed at the E0 interaction region of the Tevatron~\cite{0295-5075-12-7-007}. This search used stacks of plastic sheets, in which a highly ionizing particle would damage the chemical bonds around its trajectory, leaving a permanent track. After exposure to particles created in $p\bar{p}$ collisions at $\sqrt{s} = 1.8$~TeV, the plastic sheets were etched with NaOH, creating a visible hole where a highly ionizing particle had passed through. No coinciding holes between layers were observed, and an upper cross section limit of $0.2$~nb was set for monopoles with magnetic charge of $0.5Q_D$ or higher and mass up to 850~GeV. 

The \textit{Monopole and Exotics Detector At the LHC} (MoEDAL) is a current experiment specifically designed to search for monopoles and other highly ionizing particles produced at the LHC and entering material surrounding the IP~\cite{Acharya:2014nyr}. Exploiting the fact that the LHCb detector is a single-arm spectrometer, MoEDAL is situated at the other side of the IP. The experiment consists primarily of two passive detector subsystems. 
The first subsystem is the Magnetic Monopole Trapper (MMT), consisting of blocks made of aluminum, chosen for its anomalously large nuclear magnetic moment and hence expected ability to capture a magnetic monopole. MMT elements are examined by a SQuID system with a sensitivity of $0.1Q_D$. Subsequently, they are to be stored in a deep underground detector to search for late annihilations or decays of any heavy trapped particles. 
The second subsystem, the Nuclear Track Detector (NTD), consists of stacks of aluminum-housed plastic mounted around the LHCb Vertex Locator (VELO) detector. The NTDs can be removed and inspected by scanning microscopes to search for tracks created by highly ionizing particles. 

MoEDAL has published three sets of results with increasing exposure to LHC collisions~\cite{MoEDAL:2016jlb, Acharya:2016ukt, Acharya:2017cio}. The results were obtained only from SQuID analysis of the MMT. As in the case of the ATLAS analysis described above, the MoEDAL results were interpreted under the assumption of a Drell-Yan-like production process, and also presented as model-independent upper limits on production cross sections in fiducial volumes. 
Their last paper~\cite{Acharya:2017cio}, based on exposure of 222~kg to $2.11~\invfb$ of data collected at $\sqrt{s}=13~\tev$, provides the most stringent results. No monopole candidates were found, and monopole-pair production cross-section upper limits between 40 and 105~fb were set for magnetic charges up to $5Q_D$ and masses up to 6~TeV. Monopole mass limits between 490 and 1790~GeV were obtained.

\subsection{Indirect Detection: Reconstruction of LLP Decays}
This section reviews existing indirect detection searches, categorized by the detector system in which the LLP decay takes place: the tracking system (Sec.~\ref{sec:ID-based}), the calorimeters (\ref{sec:calo-based}), and the muon system (\ref{sec:ms-based}). We refer the reader to Sec.~\ref{sec:detectors} for details on the use of these subsystems for LLP searches.

\subsubsection{Searches Based on Inner Detector Signatures}
\label{sec:ID-based}

The ID provides precise tracking of charged particles, allowing one to measure their displacement relative to the IP and to identify displaced vertices (DVs). As a result, the largest number of existing indirect searches rely on ID signatures.

\paragraph{Tevatron and LEP Searches.}

We begin with searches conducted before the turn-on of the LHC. These include analyses performed by the CDF and D0 collaborations at the Tevatron $p\bar p$ collider at a center-of-mass energy of $\sqrt{s}=1.96~\tev$, as well as by the ALEPH, DELPHI, L3, and OPAL collaborations at the LEP $e^+e^-$ collider.

CDF has searched for displaced $Z$ bosons in the $Z\to e^+e^-$ decay signature~\cite{Abe:1998ee}. The invariant mass of the $e^+e^-$ pair was required to be consistent with that of the $Z$. The signal yield was extracted by examining the distribution of the radial position $\rho_{\rm DV}$ of the displaced vertices. For $\rho_{\rm DV}>0.1~\cm$, the background yield expected from the $\rho_{\rm DV}$ uncertainty distribution was 1 event, and 4 events were observed. Upper limits on the cross section for production of a $Z$ boson were calculated as a function of $\lambda_{xy}\equiv \gamma\beta_T c\tau$, where $\gamma\beta_T$ is the transverse Lorentz boost of the LLP parent of the $Z$ and $\tau$ is its lifetime. 

The D0 collaboration has searched for a LLP that decays into a $\mu^+\mu^-$ pair and potentially an additional neutrino~\cite{Abazov:2006as}. 
The analysis used a data sample of $0.38~\invfb$ collected at $\sqrt{s}=1.96~\tev$.  
Muons were required to have an impact parameter of at least 0.1~mm. The di-muon DV was required to be at a radius of $5<\rho_{\rm DV}<20~\cm$. The expected background was determined to be $0.75\pm 1.1$ events by linear extrapolation of the numbers of events in sideband regions where signal contamination is small relative to the signal region. No events passing the final criteria were observed, and limits on the production cross section as a function of LLP lifetime were set in the context of RPV neutralino decays. 

Using a data sample of $3.6~\invfb$ at $\sqrt{s}=1.96~\tev$, D0 has searched for pair production of LLPs, with each decaying into a $b\bar b$ quark pair~\cite{Abazov:2009ik}. The use of $b$ quarks was motivated by the presence of muons from bottom hadron  decays, which were used for triggering. Events were required to contain two DVs, each with at least four tracks and a radius $\rho_{\rm DV}>1.6~\cm$. The signal yield was determined from the distributions of the minimal invariant mass of the tracks originating from each of the DVs, and the minimal collinearity. The expected background yield was about 5 events, consistent with the observed yield. Limits on the cross section for production of a standard-model Higgs boson that decays into two long-lived hidden-valley scalars, each of which decays into a $b\bar b$ pair.

The DELPHI collaboration has searched for a LLP in a sample containing $3.3\times 10^6$ $e^+ e^- \to Z$ events, with the $Z$ decaying hadronically~\cite{Abreu:1996pa}. The  LLP signature was a DV formed from at least two tracks at a radius of $\rho_{\rm DV}>12~\cm$. The DV was required to be isolated, pass a loose collinearity cut, and have a momentum of at least 3~GeV. No events passed the full set of criteria, and limits on the branching fraction for $Z\to \nu N$, where $N$ is a heavy neutrino-like LLP, where obtained.

The experiments at LEP2 employed a series of techniques targeting long-lived sleptons as motivated in GMSB models~\cite{lep2:gmsbsleptons}. An array of prompt and long-lived techniques, including displaced lepton tracks, kinked tracks, and CLLP signatures, was used to exclude long-lived sleptons in a statistical combination of the four main LEP2 experiments. Over a wide range of lifetimes $\mathcal{O}(10^{-3}-10^{3})$~ns, selectrons below about $66$~GeV were excluded, with limits strengthening up to about $90$~GeV at longer lifetimes. Smuons below about $96$~GeV were excluded across this same lifetime range. Long-lived staus were also excluded below about $87$~GeV for lifetimes around $\mathcal{O}(10^{-3})$~ns with strengthened limits to approximately $97$~GeV at larger lifetimes.  The small lifetime range in particular remains relatively unexplored at the LHC. 

\paragraph{Searches at LHC.}

The CMS collaboration at LHC has searched for resonances that decay into two long-lived particles, each decaying into a pair of leptons~\cite{Chatrchyan:2012jna}. The analysis was performed with $5.1~\invfb$ of $\sqrt{s}=7~\tev$ data. An event was required to contain two DVs, each reconstructed from two opposite-charge leptons. Each DV was required to satisfy a collinearity-angle requirement and to have invariant mass greater than $15~\gev$. The DV radial position had to satisfy $\rho_{\rm DV}/\sigma_{\rho_{\rm DV}}>8$ for electrons and $\rho_{\rm DV}/\sigma_{\rho_{\rm DV}}>5$ for muons. A smooth-function fit to the $\rho_{\rm DV}/\sigma_{\rho_{\rm DV}}$ distribution of simulated background events was used to estimate the background yield. The background estimates were $0.02^{+0.09}_{-0.02}$ in the muon channel and $1.4^{+1.8}_{-1.2}$ in the electron channel, consistent with the observed yields in data of 0 and 4 events, respectively. Limits were extracted on the production cross of a heavy scalar times the branching fractions for its decay into two LLPs, each decaying into a lepton pair. The limits were calculated for several benchmark values of the scalar and LLP masses.

CMS refined this technique in a $\sqrt{s}=8~\tev$, $20.5~\invfb$ search for events that may contain only one LLP, which decays into a pair of electrons or muons~\cite{CMS:2014hka}. With higher background expected in the case of a single-DV search, the leptons were required to satisfy a tighter displacement requirement, $|d_0|/\sigma_{d_0}>12$. The sample of events with negative values of the transverse collinearity angle $\phi_{\rm col}$ was used as a signal-free control sample with which the $|d_0|/\sigma_{d_0}$ distribution of background events was modeled. Zero background event were expected, and no events were observed. The method was validated in simulated events and with data events having $|d_0|/\sigma_{d_0}<4.5$. Limits on the branching fractions of scalars to two LLPs were set.

In a separate study~\cite{Khachatryan:2014mea}, CMS used $19.7~\invfb$ of $\sqrt{s}=8~\tev$ data to search for DVs composed of an electron and a muon. Several lepton-$d_0$ requirements, ranging from $0.02~\cm$ to $0.1~\cm$, were used to define different signal regions. A data-driven method was used to estimate the background yield from heavy-flavor decays, while other background sources were determined from simulation. Background predictions ranged from 18 events to a fraction of an event, depending on the signal region, and were consistent with the observed numbers of events. Limits were extracted in the context of a supersymmetric model, with long-lived stop squarks with proper decay distances in the range $0.02<c\tau_{\tilde t} <100~\cm$.

Using $20.3~\invfb$ of $\sqrt{s}=8~\tev$ data, ATLAS searched for events with a DV composed of two leptons~\cite{Aad:2015rba}. Muon and electron candidates were required to satisfy $|d_0|>2~\mm$ and $|d_0|>2.5~\mm$, respectively. The DV position was required to be in the radial range $4<\rho_{\rm DV}<300~\mm$, and the DV invariant mass had to satisfy $m_{\rm DV}>10~\gev$. The dominant background was determined to arise from accidental crossing of prompt leptons, and was evaluated by vertex-fitting leptons from different events. The expected background yield was of order $10^{-3}$ events. No events were observed, and limits were calculated in the context of supersymmetry with gauge mediation or R-parity violation. 

ATLAS and CMS have conducted a number of searches involving LLPs that decay hadronically or to a combination of hadrons and leptons, which we describe below.

Ref.~\cite{Aad:2015rba} reports an ATLAS search for events with a DV formed from at least 5 tracks, which may be hadrons or leptons. The analysis, performed with $20.3~\invfb$ of $\sqrt{s}=8~\tev$ data, is a refinement of earlier searches performed with smaller data samples~\cite{Aad:2011zb, Aad:2012zx}. 
In particular, starting with Ref.~\cite{Aad:2012zx} ATLAS began reconstructing tracks with impact parameter as large as $300~\mm$ for its ID-based LLP searches~\cite{Lutz:2018gir,ATL-PHYS-PUB-2017-014}. This greatly increased the efficiency for highly displaced tracks relative to the that of the standard track reconstruction, which required $|d_0|<10~\mm$. 
Tracks were required to satisfy $|d_0|>2~\mm$, and the vertices had to satisfy the requirements $4<\rho_{\rm DV}<300~\mm$ and $m_{\rm DV}>10~\gev$ were applied to DV candidates. Events were triggered by requiring a high-$p_T$ muon, electron, jets, or large MET, resulting in four different signatures. The dominant background was determined to arise when a high-$p_T$ track that accidentally passes close to the position of a low-mass, low-multiplicity DV, typically originating from material interactions. The background level was obtained by combining DVs with tracks from other events. Background from accidental combination of nearby, low-mass DVs was determined to be subdominant. The total background estimate was between about $10^{-3}$ events for the muon-trigger signature and 0.4 events for the jet-trigger signature. No events were seen, and limits were placed on scenarios in the context of Split-SUSY, GMSB, and R-parity violation. 
An improved version of the MET analysis was carried out with $32.8~\invfb$ of $\sqrt{s}=13~\tev$ data~\cite{Aaboud:2017iio}. The expected background yield was determined to be of order $10^{-2}$ events. No events were observed, and limits were calculated for a Split-SUSY scenario.

CMS has searched for a signature of two jets that originate from a DV in $18.5~\invfb$ of $\sqrt{s}=8~\tev$ data~\cite{CMS:2014wda}. A DV was formed from at least 4 tracks, with at least one track from each of the candidate jets. The DV radial distance $\rho_{\rm DV}$ was required to be at least 8 times greater than its uncertainty $\sigma_{\rho_{\rm DV}}$. The invariant mass of the DV tracks and their combined transverse momentum were required to satisfy $m_{\rm DV}>4~\gev$ and $p^T_{\rm DV}>8~\gev$. The final selection of events in two signal regions was based on the number of prompt tracks (defined as those with $|d_0|<500~\mum$) in the jets, the fractions of jet energies carried by these tracks, the number of displaced tracks in the jets, the number of tracks consistent with originating from a point along the direction defined by the dijet momentum vector, and signed impact parameters of these tracks relative to this vector. Inverting some of the criteria results in 8 categories of events. Ratios between the numbers of events in the different categories are used for final background prediction and validation. The observed data yield was 2 events in one of the signal regions and 1 event in the other, consistent with the background expectation. Limits were calculated for a long-lived scalar model and for supersymmetry with R-parity violation. 
This study was refined with a $2.6~\invfb$, $\sqrt{s}=13~\tev$ sample~\cite{Sirunyan:2017jdo}. Displaced jet candidates were selected based on the $|d_0|/\sigma_{d_0}$ of the tracks, the angle between each jet track's transverse-momentum vector and the transverse-plane line between the PV and the position of the track's lowest-radius detector hit, and the relative energy of jet tracks originating from the PV. Events are required to have two jets that are identified as displaced. The background was estimated from events with one displaced jets, by parameterizing the probability for misidentifying a jet as displaced as a function of the jet track multiplicity. The procedure was validated with simulated multi-jet events. The expected background yield was 1 event, and 1 event was observed in the data sample. Limits were calculated for a model in which a pair of long-lived scalars are produced from by a new vector boson, or R-parity-violating decays of a long-lived stop squark into a $b$ quark and a lepton. 

CMS has searched for 
long-lived particles giving rise to a DV signature in  events containing at least four hadronic jets~\cite{Sirunyan:2018pwn}.
The analysis, which was an improvement over the search reported in Ref.~\cite{Khachatryan:2016unx}, used $38.5~\invfb$ of $\sqrt{s}=13~\tev$ data. 
Events were required to have at least 2 DVs, each containing at least 5 tracks satisfying impact-parameter requirement $|d_0|/\sigma_{d_0}>4$. The DV radial position was required to be in the range $0.1<r_{\rm DV}<20~\mm$. This small range resulted in limited sensitivity to long lifetimes relative to the sensitivity of other searches. However, it allowed the analysis to do away with the need to model the detector material. The background and potential yields were determined from a fit to the distribution of the distance $d_{\rm 2DV}$ between the two DVs that had the largest numbers of tracks or largest masses. In the fit, the $d_{\rm 2DV}$ distribution of the background was modeled from data using DVs taken from different single-DV events. One event was observed in the data, with a $d_{\rm 2DV}$ value consistent with background. Upper limits were extracted on the cross section for production of pairs of neutralinos, gluinos, or stop squarks that decay into multijet final states.

ATLAS has searched for pair-produced LLPs that decay in the ID or the muon spectrometer~\cite{Aad:2015uaa}. We describe the analysis in Sec.~\ref{sec:ms-based}.

\paragraph{High-energy Searches at LHCb.}

The first LHCb LLP search~\cite{Aaij:2014nma} involved a displaced dijet
signature, and used $0.62~\invfb$ at $\sqrt{s}=7~\tev$. We report on
the updated analysis~\cite{Aaij:2017mic}, performed with $2.0~\invfb$
at $\sqrt{s}=7$ and $8~\tev$. The search was sensitive to DVs with
$\rho_{\rm DV}<30~\mm$ and $z_{\rm DV}<200~\mm$. Events were required
to have two jets containing tracks associated with a DV and having
total momentum consistent with the direction of the DV relative to the
PV. The DV had to satisfy $\rho_{\rm DV}>0.4~\mm$, as well as
$\rho_{\rm DV}$-dependent requirements on the number of tracks and
invariant mass.
The final analysis step was a fit to the invariant-mass
spectrum. The spectrum of the dominant background, which arose from
heavy-flavor decays or material interactions, was modeled with an
analytic function. The spectrum of SM dijet background was modeled from
events with large angular separation between the jets. No significant signal contribution was found in the fit, and limits were calculated for SM
Higgs decays into two long-lived scalars, each decaying into a $q\bar
q$ pair.

An LHCb search based on two DVs in the same event and without an
associated jet requirement is reported in Ref.~\cite{Aaij:2016isa}. The analysis used $0.62~\invfb$ of data collected at $\sqrt{s}=7~\tev$. Each DV
was required to be composed of at least tracks, have a mass $m_{\rm
  DV}>6~\gev$, and have a radial displacements of $\rho_{\rm
  DV}>0.4~\mm$. A fit to the combined invariant mass of the two DVs was
used to test for the presence of signal. The background distribution was
obtained from control-region data events satisfying loose cuts.  The
method was validated using simulated events and data validation
regions. No significant signal was detected, and limits were calculated
for benchmark models of a scalar that decays into two long-lived
fermions.

In Ref.~\cite{Aaij:2016xmb}, LHCb reports a search for a LLP that
decays to a muon and hadrons, using $3~\invfb$ of $\sqrt{s}=7$ and
$8~\tev$ data. The DV was required to have at least 4 tracks,
including the muon, and satisfy $\rho_{\rm DV}>0.55~\mm$ and $m_{\rm
  DV}>4.5~\gev$. A multivariate discriminator, calculated from the
muon $p_T$, the number of tracks in the DV, $\rho_{\rm DV}$, and the
uncertainties on the DV position, was used to further suppress
background, which was dominated by $b\bar b$ production.
A fit to the $m_{\rm DV}$ distribution was used to search for signal.  The $m_{\rm DV}$ distribution of background events was obtained from a
sample of events in which only loose isolation criteria were applied
to the muon, and fitting it simultaneously with the distribution of the signal-region
sample.  The resulting signal yield was consistent with zero, and
limits on single- and pair-production of neutralino in RPV scenarios
were calculated.

\paragraph{GeV-scale Searches at LHCb and $\boldsymbol{e^+e^-}$ $\boldsymbol{B}$ factories.}

The final searches described in this section are aimed at the case of
a LLP with mass of up to 10~\gev, searched for by LHCb or in $e^+e^-$ colliders running at the $\Upsilon$ energy range.

LHCb has searched for a long-lived dark photon that decays via $A'\to \mu^+\mu^-$ in $1.6~\invfb$ of $\sqrt{s}=13~\tev$ data~\cite{Aaij:2017rft}. Each muon was required to be inconsistent with originating from the PV. Consistency with the PV was required, however, for the $A'$ candidate trajectory, obtained from the dimuon DV and momentum vector. Background from photon conversions in material was reduced to a negligible level by excluding material regions, which were mapped out with hadronic interactions. The dimuon invariant-mass spectrum was fit with a smooth background model plus a signal peak function, which was moved throughout the fit range to scan for signal. No significant signal was observed, and small regions of parameter space were excluded. The excluded regions covered values of the mixing parameter $\epsilon^2$ in the range $4\times 10^{-10}$ to $2\times 10^{-9}$ for several dark-photon mass values between about 220 and 320~GeV. The low mass values reflect the high boost needed for observation of a long-lived $A'$ with $\epsilon^2$ large enough for significant production cross section.

A scalar LLP that decays to a $\mu^+ \mu^-$ pair was searched for by
LHCb in the penguin decays $B^+\to K^+ \mu^+
\mu^-$~\cite{Aaij:2016qsm} and $B^0\to K^{*0} \mu^+
\mu^-$~\cite{Aaij:2015tna}, with $K^{*0}\to K^+\pi^-$. The analyses
used $3~\invfb$ of $\sqrt{s}=7$ and $8~\tev$ data. $B$-meson
candidates were identified based on their invariant mass and a
multivariate discriminator designed to suppress non-$B$
background. Specific peaking backgrounds, such as those involving
$B\to K^{(*)}V$, where $V=\omega,\, \phi$ and $\psi$ vector-mesons,
were removed with cuts on the $\mu^+ \mu^-$ invariant-mass
$m_{\mu^+\mu^-}$. The lifetime of the dimuon DV was required to be 3
times larger than its estimated uncertainty. A fit to the
$m_{\mu^+\mu^-}$ distribution was used to test for the presence of signal as
a function of the LLP mass. In the fit, the background was modeled
as an exponential, and a signal-peak component was moved throughout the
mass range in small steps to scan for a signal peak. Limits on the
branching fractions as a function of the scalar LLP mass were
extracted.

LHCb has also searched for a LLP as part of a search for the
lepton-number-violating decay $B^- \to \pi+ \mu^-
\mu^-$~\cite{Aaij:2014aba}. The search was performed with $3~\invfb$ of data collected at $\sqrt{s}=7$ and $8~\tev$.  A DV was reconstructed from the pion and one of
the muons, forming a heavy, neutrino-like LLP candidate. The distance from the
DV to the PV was required to be 10 times larger than its uncertainty.
The level of background from specific $B$ decays was obtained by fully
reconstructing these decays, and combinatorial background from random
track combinations was estimated by fitting the $B$-candidate
invariant-mass distribution outside the invariant-mass signal region.
The event yield was consistent with the expected background, which was at the level of a few tens of events. Limits on the branching fraction of the process were extracted as a
function of the LLP mass.

A neutrino-like LLP has also been searched for by the Belle
experiment, using $711~\invfb$ of $\sqrt{s}=10.59~\gev$ $e^+e^-$
collisions produced by the KEKB collider at KEK~\cite{Liventsev:2013zz}. The LLP was assumed
to be produced in semileptonic $B$-meson decays along with a lepton
(an electron or muon), as well as a hadronic state that was not
reconstructed, and to decay to a pion and a lepton.  Thus, the
observed final state was two leptons, which were allowed to be of the
same charge, and a pion. A DV, formed from the pion and one of the
leptons, was required to satisfy different displacement and
collinearity requirements depending on its location and detector hits
associated with the daughter tracks. The observed yield was consistent
with the background expectation of a few events, obtained from simulation. Limits were extracted
on the electron and muon couplings of the neutrino-like LLP as a
function of its mass.

The BABAR experiment at the PEP-II $e^+e^-$ collider at SLAC searched
for a LLP that decays into any of the combinations $e^+e^-$,
$\mu^+\mu^-$, $e^\pm \mu^\mp$, $\pi^+\pi^-$, $K^+K^-$ or $K^\pm
\pi^\mp$~\cite{Lees:2015rxq}. The analysis used $448~\invfb$ of data
collected at and just below the $\Upsilon(4S)$ resonance, and
$42~\invfb$ collected at the $\Upsilon(2S)$ and $\Upsilon(3S)$
resonances. The two samples were analyzed separately, as they may
involve different LLP production mechanisms. However, no assumption
was made regarding the production mechanism. The DV was required to
satisfy a collinearity requirement and to be positioned within $1 <
\rho_{\rm DV} < 50~\cm$. The $m_{\rm DV}$ distribution was fit to a
spline representing the background component, and a signal mass-peak was
used to scan the $m_{\rm DV}$ range for a signal contribution. No
significant signal was seen, and limits were extracted in a
production-mechanism-independent way as well as for a scalar LLP produced
in penguin $B$ decays.

\subsubsection{Searches Based on Calorimeter Signatures}
\label{sec:calo-based}

The D0 collaboration has searched for a LLP that decays into two
photons or electrons observed in the electromagnetic
calorimeter~\cite{Abazov:2008zm}.  The analysis used $1.1\invfb$ of
$p\bar p$ collision data. The photon candidates were required to have
transverse energies of at least 20~\gev. The segmentation of the
calorimeter in the radial direction provided 5 measurements along the
electromagnetic shower, from which the direction of the photon
momentum was obtained. The directions of the two photons were used to
extract a common vertex for their origin. For signal events, the
vertex position relative to the PV was expected to be consistent with
the momentum of the diphoton candidate. Events for which the vertex
position was in the opposite direction were used to model the
background distribution. The background expectation was a few tens of events, consistent with the observed yield. Limits were extracted on the cross section
for production of a LLP times the branching fraction of its decay into
two electrons as a function of its lifetime, as well as on the mass
vs. lifetime of a long-lived fourth generation quark.

ATLAS has searched for LLP decays into photons that originate away
from the PV~\cite{Aad:2014gfa}. The analysis, which used $20.3~\invfb$
of $\sqrt{s}=7~\tev$ data, was an improvement of an earlier,
$4.8~\invfb$ search at $\sqrt{s}=7~\tev$~\cite{Aad:2013oua}.  Selected
events were required to have two photons with transverse energies of
25 and $35~\gev$, as well as total missing transverse energy of at
least 75~\gev. The longitudinal displacement $z_0$ of one of the
photons relative to the PV was measured by exploiting the segmentation of the calorimeter in the
radial and pseudorapidity directions in order to measure the direction of the
photon momentum. The arrival time $t$ of this
photon in the calorimeter, relative to that expected for a photon
originating from the PV, was also used to detect whether it originated
from the decay of a slow-moving LLP. The $t$ distributions of events
in several $z_0$ bins were simultaneously fit to obtain the background and possible signal yields in each bin. This approach exploited the relative
independence of the $t$ distribution on the background composition.
Background was studied from $Z\to e^+e^-$ events and events with low
missing transverse energy. The signal region contained 386 events. No signal was observed over the background expectation, and limits were calculated for a supersymmetry model with gauge mediation.

ATLAS has also searched for LLPs that decay within the calorimeter
system~\cite{Aad:2015asa}. The ``CalRatio'' technique used in this
search exploited the broad segmentation of the calorimeter into an
inner electromagnetic calorimeter and an outer hadronic calorimeter.
Events were required to have two energetic jets. The jets were
identified with the calorimeters and were required to be isolated from
charged tracks. Each jet was also required to satisfy $\log_{10}
(E_{\rm H}/E_{\rm EM})<1.2$, where $E_{\rm H}$ and $E_{\rm EM}$ are
the energies deposited in the hadronic calorimeter and electromagnetic
calorimeter, respectively. A data sample containing two back-to-back
jets was used to estimate the multijet background.  Only one jet was
required to pass the $\log_{10} (E_{\rm H}/E_{\rm EM})$ cut, and was used
to evaluate the probabilities for passing the trigger and jet-energy
cuts. Fits to these probabilities as functions of jet energy were used
to estimate the multijet background.  A smaller level of background
from cosmic rays was estimated with events triggered outside of
beam-crossing times. Background from beam-halo muons that underwent
hard bremsstrahlung in the calorimeter was suppressed with 
timing cuts. Using events triggered when only one beam passed through
the detector, the level of this background was determined to be
negligible. The observed yield of 24 events was consistent with the expected background. Limits were extracted for a scalar boson that decays into two LLPs.

ATLAS has searched for ``lepton jets'' from a LLP decay in the calorimeter or
the MS~\cite{Aad:2012kw,Aad:2014yea}. We report on these searches in
Sec.~\ref{sec:ms-based}.

\subsubsection{Searches Based on Muon System Signatures}
\label{sec:ms-based}

DELPHI has searched for a LLP creating a narrow cluster of hits in the MS or the HCAL, using a sample of $3.3\times 10^6$ hadronic $Z$ decays~\cite{Abreu:1996pa}. The HCAL energy deposition was required to be consistent with that of a hadronic shower rather than a muon, while MS hits were required to point back to the IP to within 40~cm. Events were required to have no more than 3 tracks, all starting at a radius of at least 12~cm from the IP. 
Background from diphoton and dilepton background was rejected by exploiting the back-to-back topology of such events. No events survived the final selection, and limits were extracted on the branching fraction of the decay $Z\to \nu N$, where $N$ is a neutrino-like LLP.

Using $1.9~\invfb$ of $\sqrt{s}=7~\tev$ data, ATLAS has searched for
``lepton jets'' from the decay of a long-lived hidden-sector photon,
identified in the MS or the
calorimeter~\cite{Aad:2012kw,Aad:2014yea}. We describe only
Ref.~\cite{Aad:2014yea}, which uses more final states.
A lepton jet was defined as two or more muons, two calorimeter
clusters consistent with electrons or hadronic jets, or two muons and
a calorimeter cluster, with the objects fitting within a narrow
cone. Lepton-jet candidates were required to be isolated from ID tracks. An event
was required to contain two lepton jets with an azimuthal separation
of $\Delta\phi>1$. Background from cosmic rays was estimated from `empty
bunch crossings, when no $pp$ collisions occur. Background from multijet events was estimated from
sidebands of the $\Delta\phi$ and $\sum p_T$ signal region. The
observed yield was consistent with the background expectation of several tens of events. Limits
were computed for a model in which the Higgs boson decays into two or
four hidden-sector photons and two stable hidden-sector particles.

ATLAS has searched for pair-produced LLPs, each giving rise to a DV in
either the ID or the MS~\cite{Aad:2015uaa}. The
analysis used $20.3~\invfb$ of $\sqrt{s}=8~\tev$ data. Events were
required to contain two DVs. DVs in the ID were formed from at
least 5 or 7 tracks (depending on the trigger), each satisfying
$10<|d_0|<500~\mm$. These DVs were further required to have a nearby
jet, potentially originating from the LLP decay products. DVs in the
MS were intended to find LLP decays that occur outside the
calorimeter, where background arises due to jets with some particles
that punch through the calorimeter. Therefore, such DVs were required
to have MS tracks with a large number of hits, and to be isolated from
tracks in the ID and from jets in the calorimeter.
The dominant background was determined to originate from multijet
events. Its level was estimated from the probability of a jet to form
a DV, obtained from data events selected with trigger criteria that
were different from those of the signal region. Two events were found
in the signal region, consistent with the background
expectation. Limits were extracted for LLPs produced in scenarios of
stealth supersymmetry, hidden-valley, and decays of the Higgs boson or
other scalars.

ATLAS has searched for displaced $\mu^+\mu^-$ pairs identified only in
the MS~\cite{Aaboud:2018jbr}, thus providing sensitivity
to LLP decays occurring as far as the outer edge of the
calorimeter. The analysis used $32.9~\invfb$ of $\sqrt{s}=13~\tev$
data. Muon candidates were required to be isolated from jets and from
tracks, and to not have a corresponding track in the ID. Each
muon was linearly extrapolated backwards, and the midpoint along the
shortest line between the two extrapolations was taken as the dimuon
DV. Angular cuts were used to remove cosmic-ray and beam-halo
background. The level of background was estimated by studying events
in which one or both muons did have a corresponding track, events with
muons that failed the isolation cuts, and events in which the two
muons had the same electric charge. The observed yield was consistent
with the background expectation of 14 events in one signal region and about 1 event in the other. Limits on GMSB and Higgs decays to two long-lived dark photons were
extracted.

\subsubsection{Out-of-Time Decays of Particles Stopped in Detectors}
\label{sec:stoppedllps}
In the case that a LLP has a long lifetime ($\tau\gtrsim10$~ns) and interacts with SM particles via strong or electromagnetic interactions, such a particle can lose momentum to interactions with dense detector material. If it loses enough momentum via nuclear or electromagnetic interactions, it can come to rest within the detector volume. If the particle is not stable, it may decay well outside of the detector trigger and readout timing windows for the collision in which it was produced. Such a decay can give rise to significant detector activity, especially in the calorimeter system, in a pair of RF buckets that are not filled in the collider. Several searches for stopped particles have been performed by the D0 \cite{PhysRevLett.99.131801}, CMS \cite{Chatrchyan:1458954,Khachatryan:2015jha,Sirunyan:2017sbs}, and ATLAS \cite{Aad:2013gva,Aad:2012zn} collaborations, looking for calorimeter and MS activity in empty bunch crossings, where no collision backgrounds are expected. 

Such searches require a careful understanding of the bunch train structure of the collider (in these cases, the Tevatron and the LHC) as well as non-collision backgrounds such as beam halo, cosmic rays, and detector noise. These searches are able to set limits on LLPs with lifetimes in the range of $100$~ns to beyond years due to the unique nature of the search. In most cases, these searches for out-of-time decays of stopped particles have significant overlap in sensitivity with direct detection searches as the particle is expected the have passed through some portion of the detector before coming to a stop. 

For Run-1 of the LHC, ATLAS searched for out-of-time calorimeter activity with dedicated triggers \cite{Aad:2013gva}. For gluino and squark $R$-hadron models, between about $5\%$ and $12\%$ of $R$-hadrons will come to rest within the ATLAS detector if sufficiently long-lived. This range represents the spread across $R$-hadron species and interaction models. This particular search uses two signal regions with two different requirements on the momentum scale of jets reconstructed in the calorimeters, at $100$~GeV and $300$~GeV. The different regions are sensitive to different portions of signal parameter space where the latter is more sensitive provided the decay products have large enough momentum. With these requirements, as well as additional vetoes on cosmic muons, detector noise, and beam background, no significant deviation is observed from the expected signal region yields. The $100$~GeV and $300$~GeV signal regions expect $6.4\pm2.9$ and $2.9\pm2.4$ events and observe $5$ and $0$ events, respectively. In Run-1, CMS performed similar searches~\cite{Khachatryan:2015jha,Chatrchyan:1458954} utilizing a dedicated background sample from data-taking runs performed before the first LHC collisions. Both experiments have also analyzed post-beam-dump data for additional sensitivity for decays that may happen minutes after there are no more beam backgrounds. 

Building upon their own Run-1 LHC results, CMS also performed a similar search using Run-2 data recorded in 2015 and 2016, looking for out-of-time decays with the calorimeter and MS \cite{Sirunyan:2017sbs}. For the calorimeter-based search, signal regions are optimized for different lifetime ranges with background levels from about $0$ to $11$ events with no significant deviation from this observed in data. For the search channel using the MS, the same procedure is performed with background levels of $0$ to about $0.5$ events expected with no events observed in any region, consistent with SM expectations.

\section{Summary of Model Constraints}
\label{sec:constraints}

The result space spanned by the detector signatures reported in Sec.~\ref{sec:searches} has a non-trivial mapping onto the models described in Sec.~\ref{sec:theory}. Even a particular BSM final state may give rise to a mixture of detector signatures that depend on the LLP lifetime and boost. Conversely, a given experimental study often implies limits on the parameters of a variety of models. 

In this section and the figures contained therein, we summarize the current limits for a selection of LLP scenarios as a function of lifetime. When possible, we show not only the observed limits, which can be subject to statistical fluctuations, but also the limits expected for an average measurement given the sensitivity of the analysis. 
Given the multi-parameter nature of the models, these limits include assumptions made by the analysts regarding the values of parameters that are not shown in the figures. %
We also note that this summary reflects a current snapshot. In particular,  when comparing the sensitivities of different search methods, one should account for the different integrated luminosity and center-of-mass energy of the data used to obtain each result.

\subsection{Long-Lived \texorpdfstring{$\tilde{g}$}{g}}

\begin{figure}[bht]
\centering
\includegraphics[width=6in]{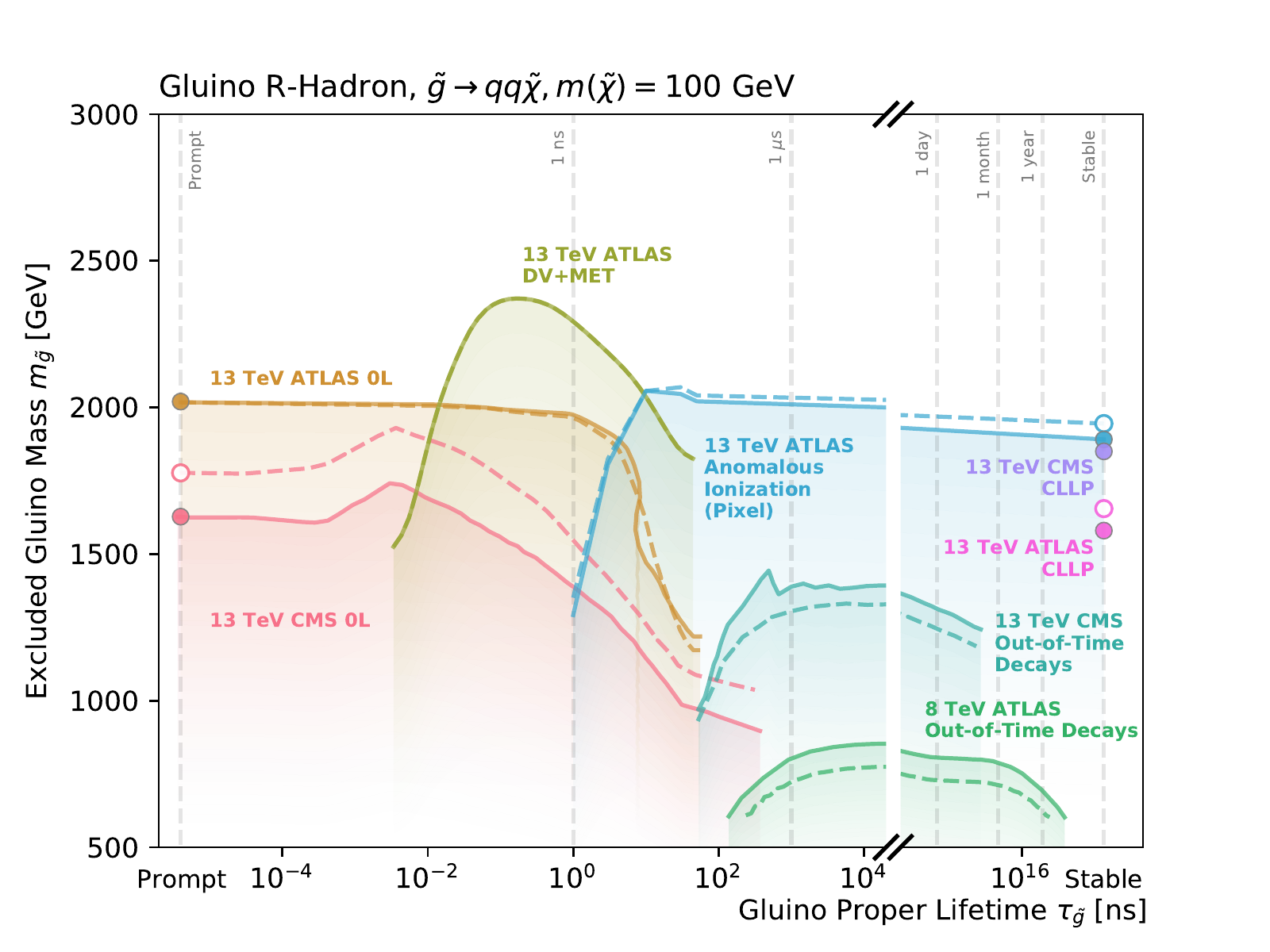}
\caption{A broad range of limits on the mass vs. lifetime of the gluino is obtained from a number of searches~\cite{ATLAS-CONF-2018-003,Sirunyan:2018vjp,Aaboud:2017iio,Aaboud:2018hdl,Sirunyan:2017sbs,Aad:2013gva,Aaboud:2016uth,Khachatryan:2016sfv}. When available, dashed lines and open circles denote the expected limits given the experimental sensitivity, while solid lines and filled circles represent the limits that were actually observed in the experiment. Circles at lifetime values labeled as ``prompt'' denote a search based on a prompt signature, rather than a long-lived one.
}
\label{fig:gluinosummary}
\end{figure}

In Split-SUSY (see Sec.~\ref{sec:split-susy}), long-lived gluinos hadronize to form color-singlet $R$-hadrons which, if metastable, can decay to hadronic jets and the lightest neutralino via a virtual intermediate squark. Various detector signatures are sensitive to this signal for different gluino lifetimes, as shown in Fig.~\ref{fig:gluinosummary}. 

In the small lifetime region, searches for prompt decays of gluinos set the tightest limits. Their sensitivity decreases at moderate lifetimes, as hadronic jet reconstruction breaks down due to jet-quality requirements that are optimized for prompt jets. Both the ATLAS and CMS collaborations have produced results to this effect~\cite{ATLAS-CONF-2018-003,Sirunyan:2018vjp}. If the decays predominantly occur within the ID, a striking DV signature together with significant MET allows for a very sensitive search, excluding gluino masses up to $2.4$~TeV for lifetimes around $100$~ps~\cite{Aaboud:2017iio}. At longer lifetimes, sensitivity is provided by searches for anomalous-ionization, stopped particles decaying out of time, and slow-moving CLLPs~\cite{Aaboud:2018hdl,Sirunyan:2017sbs,Aad:2013gva,Aaboud:2016uth,Khachatryan:2016sfv}.

\subsection{Long-Lived \texorpdfstring{$\tilde{t}$}{t}}

\begin{figure}[tb]
\centering
\includegraphics[width=6in]{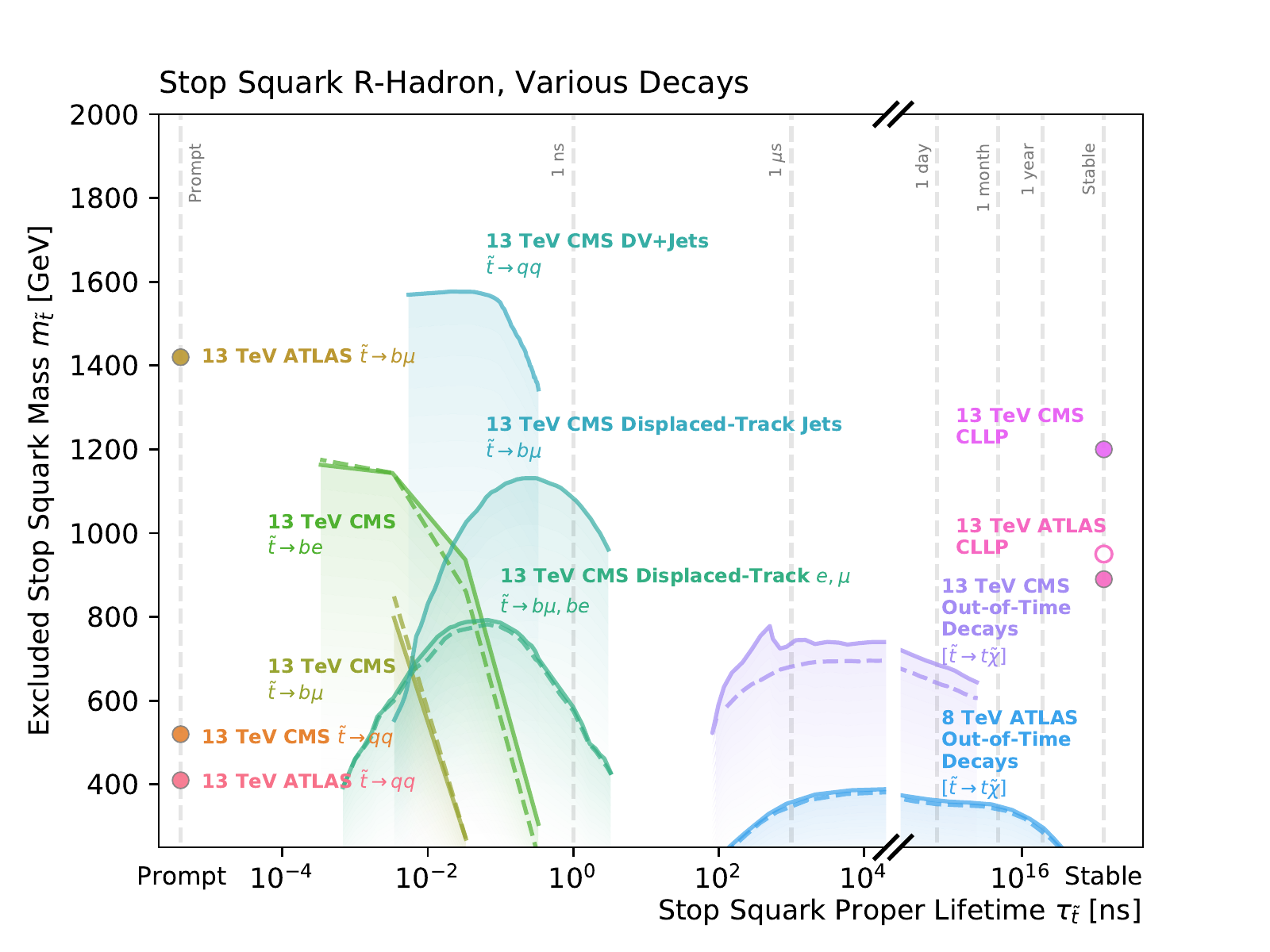}
\caption{Limits on the mass vs. lifetime of a long-lived stop squark decaying via RPV couplings, obtained from LLP searches at LHC~\cite{Aaboud:2017opj,Sirunyan:2018rlj,Aaboud:2017nmi,CMS-PAS-EXO-17-009,Sirunyan:2018ryt,Sirunyan:2018pwn,Sirunyan:2017jdo,Khachatryan:2014mea,Sirunyan:2017sbs,Khachatryan:2015jha,Khachatryan:2016sfv,Aaboud:2016uth}. When available, dashed lines and open circles denote the expected limits while solid lines and closed circles represent the observed limits. If no LLP signature is labeled, the contours show the sensitivity from a search for prompt decays.}
\label{fig:rpvstopsummary}
\end{figure}

Various models allow for a stop squark LSP that may decay via RPV couplings, and many searches have been performed by ATLAS and CMS for different RPV couplings. A summary of relevant limits is shown in Fig.~\ref{fig:rpvstopsummary}. 

Prompt searches and short-lifetime reinterpretations of prompt searches have coverage up to lifetimes of roughly $100$~ps, especially for leptonic decays of the stop~\cite{Aaboud:2017opj,Sirunyan:2018rlj,Aaboud:2017nmi,CMS-PAS-EXO-17-009,Sirunyan:2018ryt}. Dedicated LLP searches provide significantly stronger limits for a range of lifetimes from about $10$~ps to $1$~ns~\cite{Sirunyan:2018pwn,Sirunyan:2017jdo,Khachatryan:2014mea,Khachatryan:2016sfv,Aaboud:2016uth}.
Searches for out-of-time decays of stopped particles are sensitive to long-lived stop squarks provided the decay products deposit sufficient energy in the calorimeter. Existing stopped particle searches have set limits in a relatively unmotivated model of long-lived stop squarks decaying via a gauge coupling to $t\tilde{\chi}$. These limits should, however, apply to other decay signatures given enough calorimeter energy deposition. 
In the limit that the stop is detector-stable, CLLP searches~\cite{Sirunyan:2017sbs,Khachatryan:2015jha} have significant sensitivity excluding stop masses below about $1200$~GeV.

\subsection{AMSB SUSY}

\begin{figure}[tb]
\centering
\includegraphics[width=6in]{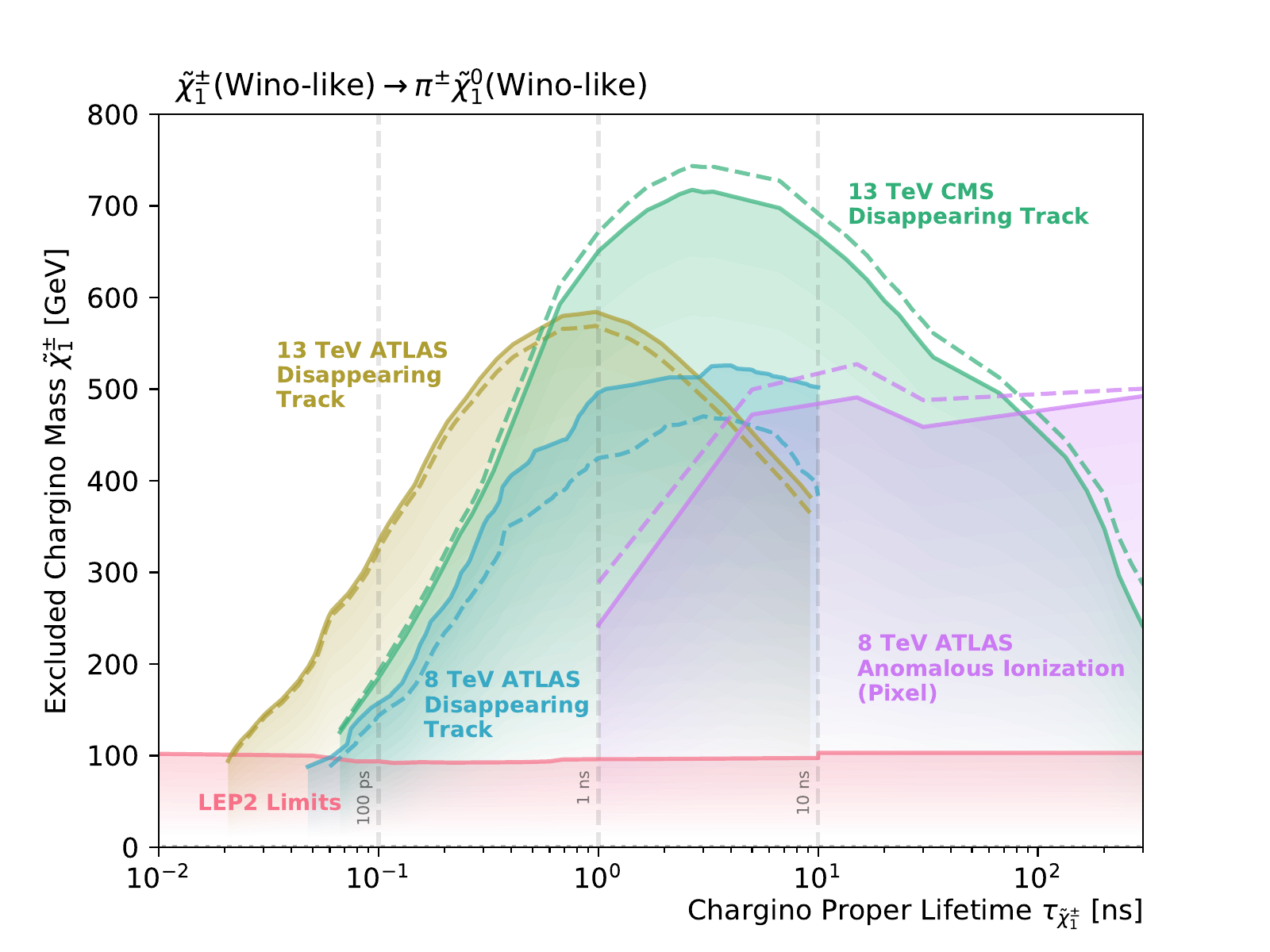}
\caption{Limits on the chargino mass as a function of its lifetime in AMSB SUSY scenarios, obtained from LHC and LEP2 searches~\cite{ATLAS:DT8TeV,ATLAS:DT13TeV,CMS:DT13TeV,Aaboud:2018hdl,lep2:charginolowdM}. 
The chargino is assumed to be largely wino-like. When available, dashed lines denote the expected limits while solid lines represent the observed limits. The limits from LEP2 use a combination of prompt analyses, CLLP searches, and radiation-based searches~\cite{lep2:charginolowdM}.}
\label{fig:amsb}
\end{figure}

As described in Sec.~\ref{sec:amsb}, AMSB SUSY can give rise to a small mass splitting between the lightest chargino and lightest neutralino. A summary of relevant searches is shown in Fig.~\ref{fig:amsb}. 

The experiments at LEP2 set combined limits using multiple techniques, and exclude chargino masses up to around $100$~GeV across lifetime space~\cite{lep2:charginolowdM}.
Relying on the fact that the charged chargino daughter is too soft to be tracked, dedicated disappearing-track searches from the LHC set tighter limits, up to around $700$~GeV, for lifetimes between about $20$~ps and several hundred ns~\cite{ATLAS:DT7TeV1,ATLAS:DT7TeV2,ATLAS:DT8TeV,ATLAS:DT13TeV,CMS:DT8TeV,CMS:DT13TeV}. Searches for anomalous ionization are sensitive to longer tracks, corresponding to longer lifetimes. The Run-1 iteration of this search from ATLAS sets limits on long-lived charginos with lifetimes from $1$~ns up to the stable case. Charginos are excluded up to about $480$~GeV for the entirety of the lifetime range of $0.2$~ns to stable.

While not motivated by AMSB, models with a pure-Higgsino LSP also obtain small mass splittings between the lightest chargino and the neutralino LSP. Such models predict lifetimes of order $10$~ps. This low-lifetime region is particularly challenging to search in, as evidenced by the weak exclusion in the left side of Fig.~\ref{fig:amsb}. Charginos with lifetimes below $20$~ps with masses above about $100$~GeV remain unexcluded.

\subsection{Scalar Portal LLP Production}

\begin{figure}[tb]
\centering
\includegraphics[width=6in]{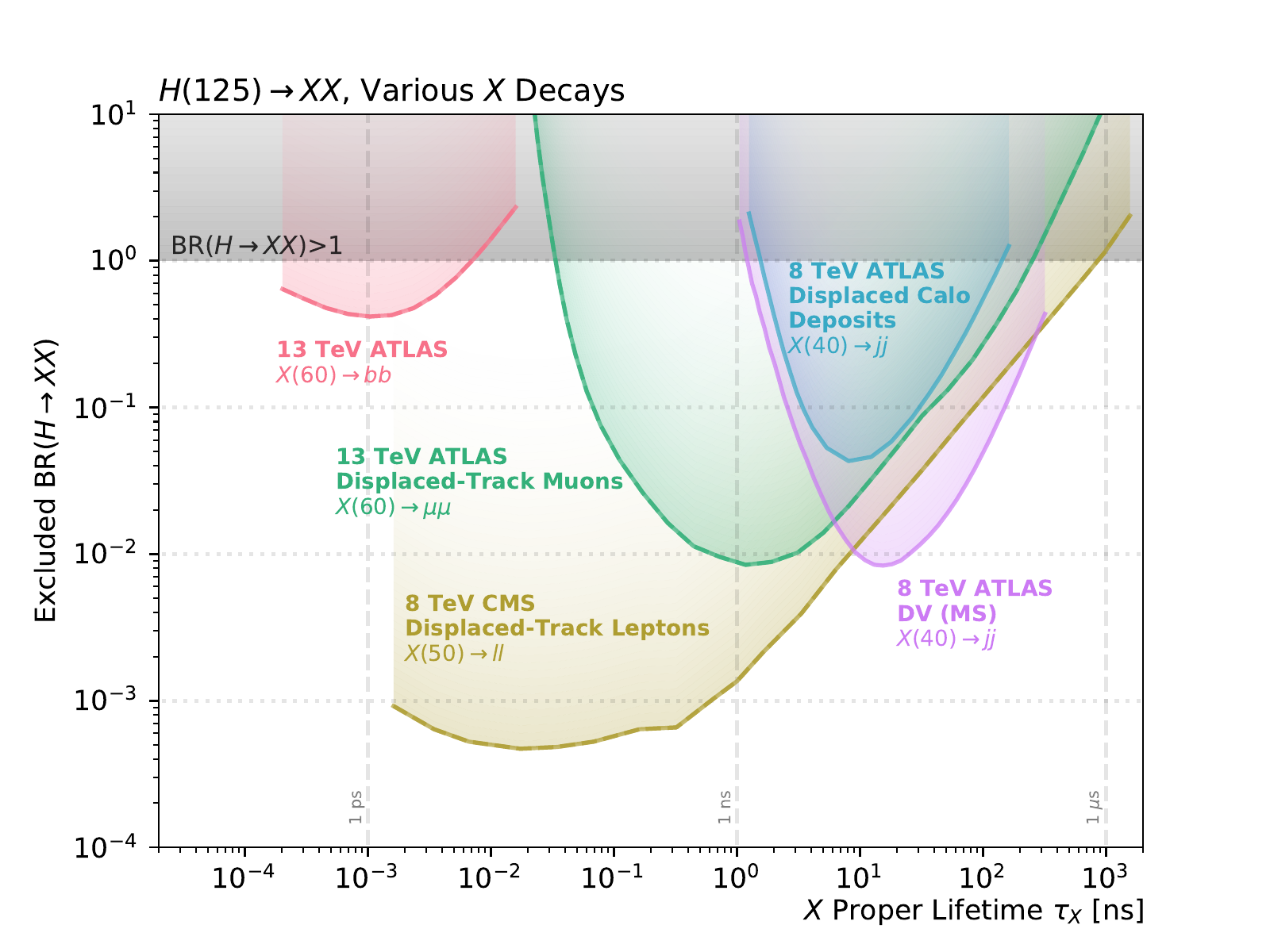}
\caption{LLP-lifetime-dependent limits on the branching fraction for the decay $H\to XX$ of the Higgs boson into two LLPs. The LLP mass and probed decay mode, assumed to have a branching fraction of 100\%,
are indicated by $X(m_X/\mbox{GeV})\to YY$. All limits are obtained from LHC searches~\cite{Aaboud:2018iil,CMS:2014hka,Aaboud:2018jbr,Aad:2015asa,Aad:2015uaa}. When available, dashed lines denote the expected limits, while solid lines represent the observed limits. The region where the $H\rightarrow XX$ branching ratio is larger than $1$ is also shown. 
The contours labeled ``$X(60)\rightarrow bb$'' show the sensitivity from a search for prompt decays.
}
\label{fig:higgsportalsummary}

\end{figure}

Production of LLPs via a portal mechanism has been
studied in several sensitive searches. These are summarized in Fig.~\ref{fig:higgsportalsummary}, which shows the limits on the branching ratio for di-LLP production in $125$~GeV Higgs decays as a function of LLP lifetime for multiple LLP masses and decay modes.

Different decay modes of the LLP lead to significantly different signatures, with limits having been extracted for LLP decays to light-flavor jets, $b$-quark jets, and light leptons. In the roughly $1$~ps regime, a reinterpretation of a prompt search for Higgs decays to four $b$-quarks excludes exclude branching ratios of order $10\%$~\cite{Aaboud:2018iil}. For leptonic decays of the LLP, displaced track techniques have been used to set limits on branching ratios below $0.1\%$ for a lifetimes between about $1$~ps and $1$~ns~\cite{CMS:2014hka,Aaboud:2018jbr}. Larger lifetimes have been probed by dedicated searches in the ATLAS calorimeters and MS, with unique sensitivity to hadronic branching fractions at the $1\%$ level~\cite{Aad:2015asa,Aad:2015uaa}. For LLP lifetimes below order $1$~ns with hadronic decays, branching ratios below $30\%$ remain unprobed.

\subsection{Magnetic Monopoles}

\begin{figure}[tb]
\centering
\includegraphics[width=6in]{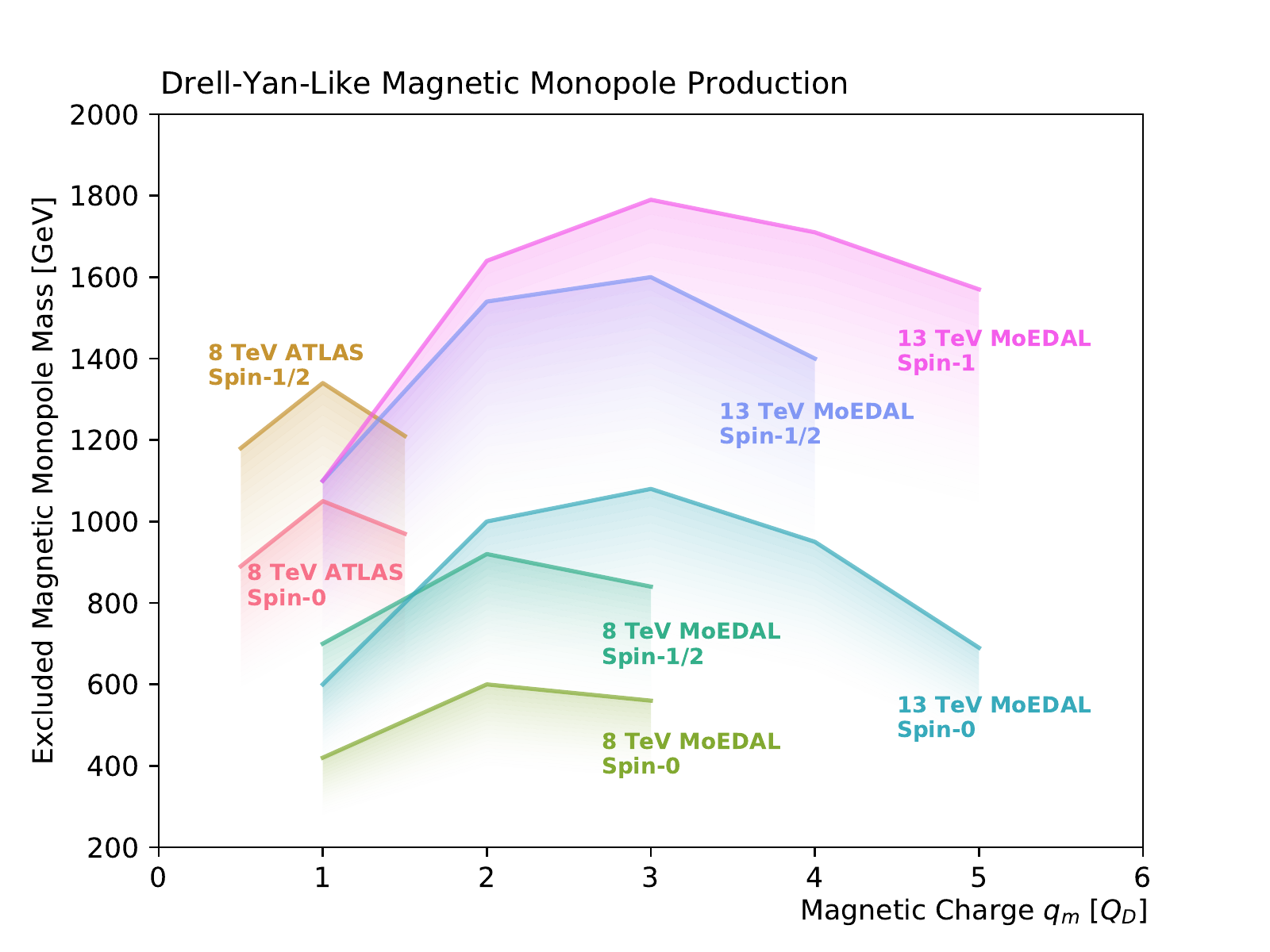}
\caption{Magnetic monopole mass limits from ATLAS and MOeDAL searches~\cite{Aad:2015kta,Acharya:2017cio} are shown as a function of magnetic charge for various spins, under the assumption of a Drell-Yan-like pair-production mechanism. These interpretations are primarily useful for comparing experimental results, but are otherwise unreliable, as the large coupling makes perturbative calculations diverge.
}
\label{fig:monopolesummary}
\end{figure}

A summary of searches for magnetic monopoles can be found in Fig.~\ref{fig:monopolesummary}  for various spin and magnetic charge assumptions. Despite the non-perturbative nature of monopole production, limits are obtained assuming Drell-Yan-like production. Results shown were obtained by searches at ATLAS~\cite{Aad:2015kta} and MoEDAL~\cite{Acharya:2017cio} with mass limits as large as $1790$~GeV for a magnetic charge of $3Q_D$ and spin $1$.

The two experiments used very different data set sizes and techniques. Nonetheless, with ATLAS limits dominating at small values of $q_m$ and the higher-charge regime being covered by MoEDAL, this current snapshot shows complementarity between the searches performed at a general-purpose detector and those at a dedicated experiment.

\section{Future Searches and Experiments}
\label{sec:future}

As seen from the review of results in Sec.~\ref{sec:searches}, interest in LLPs as a means for probing and discovering BSM physics has been growing rapidly. While LLP searches are being carried out, methods for exploiting new experimental signatures are constantly being developed, and the data samples available for analysis keep growing. Therefore, we conclude that the coming years will see significant expansion in LLP physics. In this section we go beyond the current searches and discuss the outlook for LLP searches at future facilities.

\subsection{Future Searches at Collider Experiments}

Approved and proposed collider experiments generally feature a large increase in the integrated luminosity relative to past or currently operating facilities of a similar nature. The large samples collected will lead to significant improvements in the sensitivity of BSM searches. In searches that are dominated by SM background, the signal and background yields grow together, so that the resulting sensitivity to the production rates of BSM particles is proportional to the square root of the integrated luminosity. As seen in Sec.~\ref{sec:searches}, many LLP searches have very low background levels, since they are often able to exploit experimental signatures (reviewed in Sec.~\ref{sec:split-susy}) that are not available to prompt BSM searches. In zero-background cases, the sensitivity grows roughly linearly with the integrated luminosity. Since LLP searches frequently target spectacular signatures with tiny irreducible backgrounds, they are therefore particularly interesting to pursue at future collider facilities. Details of these facilities and their LLP capabilities are discussed in this section.

\subsubsection{ATLAS and CMS}

As mentioned in Sec.~\ref{sec:intro}, only a small part of the roughly $150~\invfb$ data samples so far collected by each of ATLAS and CMS has been used for LLP searches. Full exploitation of these samples is expected by the start of Run-3 in 2021 which will be at $\sqrt{s} = 14$~TeV. After this running period, each experiment will have collected about $300~\invfb$ by 2024, approximately evenly split between 13 and 14~TeV. In addition to the slight energy increase and doubling of integrated luminosity, analysis techniques can improve and mature significantly during this period, leading to both increased sensitivity and significant expansion of the model-parameter space explored by LLP searches.
The subsequent High-Luminosity phase of the LHC (HL-LHC) is approved and funded. ATLAS and CMS are each to collect an integrated luminosity of about $3000~\invfb$ over a dozen years of operation starting with Run 4 in 2026~\cite{hl-lhc}. 

The increased luminosity of the HL-LHC comes at the price of dramatically increased pileup, and the existing LHC experiments will undergo comprehensive upgrades for Run 4, both to withstand the high particle rates and to improve the sensitivity of precision measurements and searches for new physics in this challenging experimental environment. Some of the upgrades will provide enhanced capabilities for detecting LLPs, and we briefly comment here on some important aspects of these upgrades.

Improved ID tracking detectors will be installed, with higher granularity and improved resolutions to enable effective measurements of charged-particle tracks at luminosities nearly an order of magnitude larger than that of the present-day LHC. The layouts of these detectors can allow for improved tracking efficiencies at large displacements and less strict disappearing-track conditions, likely resulting in significant gains in LLP sensitivity. For example, these advances may make it possible for ID DV searches to loosen vertex mass and track multiplicity requirements, thus expanding the model space covered by such searches.

Despite these improvements, some LLP signatures will be negatively affected by the ID upgrades. For example, both ATLAS and CMS are building their trackers with only a few bits of digital information allocated to the measured charge deposition. This will degrade the resolution of ionization measurements in these new detectors relative to their current incarnations~\cite{atlasitkpixels,atlasitkstrips,cmsphaseiitracker}.

The barrel calorimeter systems will be upgraded with new readout electronics aiming to provide improved timing resolution down to 30~ps for photons with $\pT > 25$~GeV in the CMS ECAL barrel at the start of the LHC~\cite{cmsphaseiibarrelcalo}. Anomalous calorimeter deposit signatures can be greatly improved in the endcap of CMS with the High-Granularity Calorimeter proposed for the HL-LHC upgrade program~\cite{Collaboration:2293646}. With calorimeter cell areas of order 1~\si{\centi\metre\square} and an expected per-cell timing resolution on the order of tens of ps, measurements could identify showers that do not point back to the PV or are delayed. In ATLAS, the readout electronics for the electromagnetic and hadronic calorimeters will also be updated to improve performance at the HL-LHC~\cite{atlasphaseiilar,atlasphaseiitile}.

In addition, both ATLAS and CMS have recently added very precise MIP timing detectors to their Phase-II upgrade plans, and these can have a significant effect on LLP searches. They are primarily motivated by increased pileup and the need to exploit the time dimension of the beam spot to discern individual vertices and enable accurate track-to-vertex association despite the high vertex density. Aiming to provide timing measurements with 30~ps accuracy for all MIPs in their acceptance, they offer timing measurements of passing LLPs or their decay products with a resolution that is orders of magnitude better than what is currently achievable. As this track-to-vertex association is most challenging for tracks at shallow angle with the beam line, the High-Granularity Timing Detector will cover $2.4 < |\eta| < 4.0$ in ATLAS~\cite{atlashgtd}. Instead prioritizing coverage in the barrel region, the CMS MIP Timing Detector will cover $|\eta| < 3.0$~\cite{Collaboration:2296612}. These detectors will provide exciting new capabilities for LLP searches. The improvements may allow for reconstruction of the mass of a heavy LLP decaying with a displaced-vertices signature by using measurements of both momenta and times for charged tracks and photons, for example~\cite{Liu:2018wte}. For these applications, CMS should have an advantage given that LLPs are primarily produced centrally in many models.

A challenge faced by modern LLP searches is that the trigger systems of the detectors are largely designed for prompt particle production. As a result, current and past experiments may be blind to particular regions of model space. Some upgrades for the HL-LHC can directly address this challenge.
ATLAS and CMS plan on major redesigns of the trigger and data acquisition systems for the HL-LHC, introducing tracking abilities to their hardware-level trigger systems. The upgraded CMS tracker design includes a series of double-layers that will provide very fast reconstruction of short track stubs~\cite{cmsphaseiitrigger}. This will enable global-event track-stub tracking down to $\pt>2$~GeV at the HL-LHC collision rate of $40$~MHz, providing this information for the first time as input to the first-level trigger decision. The ATLAS upgrade program includes hardware tracking systems that will be usable in the trigger. In regions of interest, tracking may be performed with a high efficiency for charged particles with $\pt>4$~GeV at an input rate of about $1$~MHz~\cite{atlasphaseiitdaq}. The ATLAS hardware-based tracking system will do fast track-finding through pattern matching, and though the memory banks will primarily be populated with patterns that optimize the performance for prompt high-$\pT$ tracks, a portion could be dedicated to tracks with large impact parameters, \emph{i.e.} specifically targeting tracks from displaced decays. This could allow triggering directly on the decays products from a displaced vertex and could bring valuable sensitivity gains for event topologies that trigger limitations so far prevented searches from examining for LLPs. The CMS hardware tracking can similarly relax the requirements placed on the track-stubs to allow for addition displaced sensitivity at the trigger level.

\subsubsection{LHCb}

The LHCb detector will undergo extensive upgrades to enable it to collect $50~\invfb$ by 2028~\cite{Bediaga:2012uyd, Piucci:2017kih}. Of particular relevance to LLP searches, the current silicon microstrip sensors of the vertex locator (VELO) will be replaced with pixel sensors, to withstand higher track multiplicity, simplify reconstruction, and improve resolution. The distance of the VELO from the IP, which is currently $8.4~\mm$, will be reduced to $5.1~\mm$. The amount of material traversed by a particle before the first VELO hit will be reduced from 4.6\% to 1.7\% of a radiation length. These measures will lead to a 40\% improvement in the track impact-parameter  resolution, leading to better prompt-background rejection and hence increased sensitivity for LLPs with small lifetimes. A new silicon-microstrip upstream tracker (UT) will improve reconstruction of LLP decays that occur after the VELO. Lastly, the hardware trigger system will be removed, and events will be selected by a software-only trigger system at the LHC collision rate. 

The LHCb collaboration is interested in conducting further upgrades to allow the experiment to collect a data sample of $300~\invfb$ by the currently foreseen end of the LHC program in the 2030s. Maintaining or improving the performance of the detector given the high instantaneous luminosity will require tracking detectors with precise timing information. Similarly, increased use of FPGA and GPU technology is being explored for meeting the challenging trigger performance requirements.

\subsubsection{Belle~II}
The Belle~II experiment~\cite{Abe:2010gxa} will begin taking physics data with the full detector in 2019, and is planned to collect about $50,000~\invfb$ by the year 2025, with potential subsequent upgrades. In addition to the large luminosity increase over previous $B$~factories, Belle~II features a factor-of-2 improvement in the spatial resolution of the vertex detector. Coupled with a very small beamspot, this will increase the sensitivity of LLP searches at small distances. 

Due to the relatively low particle multiplicity at a $B$~factory, the trigger requirements are much looser than at the LHC. In particular, Belle~II will even accept events with a single photon and no tracks. This increases the likelihood of retaining sensitivity to searches that have not yet been conceived. In addition, reconstruction of tracks with large $d_0$ values is much less difficult than at a high-multiplicity hadron collider. Similarly to LHCb, Belle~II is designed to cleanly identify charged particle types based on  $dE/dx$ and Cherenkov-radiation measurements, a capability that can be used for CLLP detection. 

Predicting the sensitivity of Belle~II to a range of LLP-predicting models is hampered by the fact that very few LLP searches have been conducted at a $B$~factory. Nonetheless, one can expect that Belle~II will have an advantage when it comes to light particles that would be difficult to trigger on at ATLAS and CMS, as well as final states involving hadrons plus missing particles or photons, which would be difficult to identify at LHCb. Studies involving $\tau$ leptons are also of interest, exploiting the clean environment and the roughly 1~nb cross section for $e^+e^-\to\tau^+\tau^-$ at $B$-factory energies. 

\subsubsection{Proposed Colliders}

Beyond the timescale of the approved colliders and their experiments, proposals have been made for future colliders at the energy frontier.

In the LHC tunnel, the proposed \emph{High-Energy LHC} (HE-LHC) would operate with dipole magnetic fields of $20$~T~\cite{Assmann:1284326}. With $pp$ collisions taking place at a center-of-mass energy of $\sqrt{s}=33$~TeV, production cross sections would be much larger than at the LHC, particularly for multi-TeV BSM particles. 
Further increase in heavy BSM sensitivity could come from the proposed \emph{Future Circular Collider} (FCC) at CERN~\cite{fcc-hh} or the \emph{Super proton-proton Collider} (SppC) at the Institute of High-Energy Physics (IHEP) in China~\cite{CEPC-SPPCStudyGroup:2015csa}. With a circumference of 80-100~km, these $pp$ colliders would reach energies of $\sqrt{s}=100~\tev$. 
LLP searches can be expected to be an important part of the physics that would be performed at these facilities~\cite{Arkani-Hamed:2015vfh}.

New electron-positron colliders have been proposed as well. 
The \emph{International Linear Collider} (ILC) is a linear $e^+e^-$ collider currently proposed for construction in Japan. Collisions would be detected in two experiments at center-of-mass energies of $\sqrt{s}=500$~GeV, with opportunities for upgrades up to $1$~TeV. A cost-saving $\sqrt{s}=250$~GeV~\cite{ilc} Higgs-factory configuration would enable precision studies of the Higgs boson, produced via $e^+e^-\to ZH$. While the cross section for this process is much smaller than that of Higgs production at LHC, the advantage of such a Higgs factory lies in the low-background environment and well understood cross sections of $e^+e^-$ collisions. However, since LLP searches typically have low backgrounds even at LHC, the case for LLP studies at a Higgs factory is largely limited to LLP decays that are difficult to reconstruct in hadron collider environments, such as low multiplicity decays or decays to weakly interacting particles. 

Operating at the same energy scale as ILC but with more than an order of magnitude higher luminosity,  FCC-ee is a circular $e^+e^-$ collider proposed for the 100~km FCC tunnel~\cite{Gomez-Ceballos:2013zzn}. With center-of-mass energies between $90$ and $400$~GeV, this machine could be used for high-precision $Z$, Higgs, and top-quark physics. Unlike the ILC, the circular configuration would not enable increasing the $e^+e^-$ collision energies beyond about $400~\gev$. A similar initiative, known as the Circular Electron-Positron Collider (CEPC) is also being proposed by IHEP~\cite{CEPCStudyGroup:2018rmc}.

On a longer timescale, an even higher-energy $e^+e^-$ collider has been proposed for potential installation at CERN. The \emph{Compact Linear Collider} (CLIC) would use very high-field radio-frequency technology to reach collision energies from $380$~GeV to $3$~TeV~\cite{clic}. 

\subsection{Proposed Dedicated LLP Experiments at the LHC}

Several experiments dedicated to the search for LLPs have been proposed for the LHC. Targeting longer lifetimes than those that can be accessed by the main detectors, these dedicated experiments tend to be located at a significant distance from the IP, cleverly taking advantage of existing open space. These proposals span a wide range of maturities, and some have already collected data with test stands to provide proof of concept and obtain background estimations as input to detector design. None of these projects are fully funded.

The MATHUSLA experiment~\cite{Chou:2016lxi,Evans:2017lvd,Curtin:2018mvb,Alpigiani:2631491} would be an enormous tracking detector, roughly $200\times200\times20~m^3$ in size, that would sit at the surface roughly $100$~m above either the CMS or ATLAS caverns. A modular array of trackers would fill this large volume, shielded by close to $100$~m of earth from almost all backgrounds produced in the HL-LHC collisions. 
Neutral LLPs with very large lifetimes produced in the collisions may decay within the volume of MATHUSLA, where displaced vertices could be reconstructed. The sensitive volume extends downward into the earth for decays into penetrating, energetic muons. Timing and pointing resolutions would allow for vetoes of cosmic backgrounds, as well as identification of promptly produced energetic muons, which could penetrate the earth shield and would be used for calibration and alignment. Initial estimates indicate that with $3000~\invfb$ of data, MATHUSLA would be sensitive to LLPs with lifetimes up to the $\tau\lesssim10~\mu$s limits obtained from Big-Bang Nucleosynthesis for some models.

The milliQan experiment~\cite{Ball:2016zrp} proposed for the HL-LHC would live in an unused underground tunnel near the CMS cavern with about $15$~m of rock shielding between the IP and the detector. The experiment would be sensitive to fractionally charged LLPs with electrical charge as low as $\mathcal{O}(10^{-3}-10^{-2})$. milliQan is estimated to be sensitive to LLP masses of up to $\mathcal{O}(1-10)$~GeV with $300~\invfb$ of collision data, significantly improving upon the reach of previous experiments.

The FASER experiment~\cite{Feng:2017uoz} would be situated hundreds of meters downstream from the IP of the ATLAS or CMS experiment, beyond the point at which the beams curve away.
Placed at 0~degrees (infinite pseudorapidity) relative to the collision axis, FASER would search for neutral LLPs with particular sensitivity to production of sub-GeV dark photons.

The CODEX-b~\cite{Gligorov:2017nwh} tracking detector would look for displaced vertices in a $10~\mbox{m}^3$ volume behind $3$~m of shielding about $25$~m from the LHCb IP. Placed at a large angle with respect to the beam line, CODEX-b would cover the uninstrumented low-pseudorapidity region of the LHCb IP. If the decommissioned DELPHI detector parked in that location could be removed, CODEX-b could potentially double in volume.

Another proposal, AL3X~\cite{Gligorov:2018vkc}, involves repurposing parts of the ALICE detector for Run-5 of the LHC. The ALICE magnet system, itself inherited from the L3 experiment, and the ALICE TPC would be used to measure the decays of LLPs produced in collisions that would take place at an IP shifted by about 11~m relative to the current ALICE IP. Moving the IP would allow insertion of shielding between the IP and the detector to remove SM backgrounds. The high-quality tracking provided by the ALICE TPC would provide active background rejection in addition to that provided by the passive shielding and roughly $100$~m of earth protecting the detector from most cosmic backgrounds. For detection of LLP signals from exotic Higgs decays, dark photons, and exotic $B$ decays, the estimated sensitivity of AL3X with only $100~\invfb$ is competitive with those of searches at ATLAS/CMS, CODEX-b, and MATHUSLA performed with $3000~\invfb$. This advantage of AL3X arises from its closer distance to the IP, the large solid-angle coverage provided by a large detector, and the low background provided by shielding and precise tracking. If the physics priorities of ALICE allow for this repurposing on the Run-5 timescale, the IP can be moved, and the LHC is able to deliver $100~\invfb$ to what is  currently a low-luminosity IP, the AL3X proposal is an attractive potential path forward in the search for LLPs.

We note another possibility for placing a shielded, large-scale tracker at a distance of order 20~m from a high-luminosity IP at the LHC. The ATLAS detector has a roughly 6-m-long gap between the farthest endcap muon trigger chambers and the last precision-tracking station of the MS, which is placed next to the cavern wall~\cite{1748-0221-3-08-S08003, ATLAS:1999uwa}. Used mainly for measuring the curvature of hard muons in the toroidal magnetic field, the precision-tracking station is equipped with azimuthally oriented drift tubes. Augmenting these with layers of radially oriented tubes of the same technology would make for a full tracker with precise three-dimensional vertexing capability. Protecting this tracker from hadronic background produced in the IP by partly filling the gap with shielding would make for a LLP detector at about 20~m from the ATLAS IP, in the approximate pseudorapidity range $1.3<\eta<2.7$ and with full azimuthal coverage. Searching for a DV signature, this setup would be particularly sensitive to LLPs that decay into final states containing muons inside the large volume of the shielding, but also to decays into hadrons in the air gap between the shielding and the detector. The full ATLAS endcap systems would provide a powerful veto against hard muons produced at the IP. Unless the tracker is equipped with a magnetic field, it could not directly measure the mass of the LLP. However, a lower limit on the mass could be obtained from the distance that the daughter tracks travel inside the shielding from their production point at the DV. Since LHC is already designed to provide high luminosity at this IP, no change to the collider would be required, and the luminosity integrated by the new detector would be equal to that collected by ATLAS. 
Negative impact on the prompt-physics program of ATLAS, if any, would be small. A similar configuration could also be constructed at CMS with its modular design. Some of the movable slices of detector can be moved further away from its IP, and shielding can be inserted similarly. The feasibility of these options in terms of cost, mechanical engineering, etc., has yet to be evaluated.

\section{Conclusion}
\label{sec:conclusions}

The particle physics community is rapidly defining the lifetime frontier as an important part of its BSM search program. The lack of discovery of BSM physics at the LHC thus far has motivated physicists to explore theoretically motivated and sometimes overlooked lifetime ranges. 
The number of searches targeting LLP signatures has greatly increased in recent years, as has that of LLP interpretations of standard analyses. These searches often employ new experimental techniques or address previously unexplored theoretical scenarios. In addition, new, dedicated LLP experiments have been proposed for both colliders and non-collider facilities. 
Reflecting the growing interest in LLPs, this paper reviews their theoretical motivations, detectable signatures, and the common analysis techniques used in searching for them at modern colliders. 
A comprehensive summary of experimental LLP results, particularly those from the LHC, is given.
Finally, promising new avenues and considerations regarding sensitivity to BSM via LLPs at future facilities are discussed.

In closing, we wish to emphasize the importance of this lifetime frontier in the development of the physics programs, accelerators, and detectors of the future. As LLP searches often have very low background, they are particularly promising as sensitive probes of BSM physics at future, high-luminosity colliders. For most proposed future facilities, the detector designs are in their infancy. With decades before the realization of these experiments, it is imperative that their detectors be built considering the needs of unconventional search signatures from the very beginning, ideally with enough flexibility to allow pursuing signatures that have not yet been possible to explore. Recent history has shown that multiple areas of interest and ideas regarding BSM physics can arise during the multi-decade lifetimes of modern collider facilities. When it comes to the design of individual sensors and their readout systems, the overall detector layout, trigger architectures, and reconstruction algorithms, future collaborations would do well to keep LLP searches in mind as a major driver of the detector design.

\section{Glossary}
\label{sec:glossary}
\begin{description}
\item[ALEPH:] 
an experiment at LEP.
\item[ALP:] axion-like particle.
\item[AMSB:] anomaly-mediated supersymmetry breaking.
\item[ATLAS:] an experiment at LHC.
\item[BaBAR:] an experiment at the SLAC laboratory, USA, 1999-2008.
\item[BELLE:] an experiment at the KEK laboratory, Japan, 1999-2010.
\item[BELLE~II:] an experiment at KEK, Japan, 2018-2025. 
\item[BSM:] beyond the Standard Model.
\item[CDF:] an experiment at the Tevatron.
\item[CERN:] 
a particle-physics laboratory, Switzerland/France. 
\item[CEPC:] 
a proposed $\sqrt{s}\sim 250$~GeV circular $e^+e^-$ collider in China.
\item[CLLP:] charged, long-lived particle.
\item[CMS:] an experiment at LHC.
\item[DELPHI:] an experiment at LEP.
\item[DM:] dark matter.
\item[DV:] displaced vertex.
\item[D0:] an experiment at the Tevatron.
\item[ECAL:] electromagnetic calorimeter
\item[FCC-ee] a proposed $\sqrt{s}=90-360$~GeV $e^+e^-$ collider at CERN
\item[FCC-hh] a proposed $\sqrt{s}=100$~TeV hadron collider at CERN.
\item[FPGA:] field programmable gate arrays.
\item[GMSB:] gauge-mediated supersymmetry breaking.
\item[GPU:] graphics processing unit.
\item[GUT:] grand unified theory.
\item[HCAL:] hadronic calorimeter.
\item[HERA:] a hadron-electron collider at the DESY laboratory, Germany.
\item[HL-LHC:] the high-luminosity phase of LHC, after 2026.
\item[HIP:] highly ionizing particle.
\item[ID:] inner detector (tracker).
\item[IP:] the average interaction point of colliding beams.
\item[L3:] an experiment at LEP.
\item[LEP:] Large Electron-Positron collider, a $\sqrt{s}=90-209$~GeV $e^+e^-$ collider at CERN, 1989-2000.
\item[LHC:] Large Hadron Collider, 
a $\sqrt{7-14}$~TeV $pp$ collider at CERN.
\item[LHCb:] an experiment at LHC.
\item[LLP:] long-lived particle.
\item[LSP:] lightest supersymmetric particle.
\item[MIP:] minimally ionizing particle.
\item[MoEDAL:] a magnetic-monopole and HIP search experiment at LHC.
\item[MS:] muon system.
\item[MSSM:] minimal supersymmetric standard model.
\item[NLSP:] next-to-lightest supersymmetric particle.
\item[OPAL:] an experiment at LEP.
\item[pNGB:] pseudo-Nambu-Goldstone Boson.
\item[PV:] primary vertex, point of beam-particle collision in a particular event.
\item[QCD:] quantum chromodynamics.
\item[RPV:] $R$-parity violation.
\item[SM:] Standard Model of particle physics.
\item[SppC] a proposed $\sqrt{s}=100$~TeV $pp$ collider in China.
\item[SQuID:] superconducting quantum interference device.
\item[SUSY:] supersymmetry.
\item[Tevatron:] a $1.8-1.96$~TeV $p\bar p$ collider at Fermilab, USA, 1983-2011.
\item[vev:] vacuum expectation value.
\item[WIMP:] weakly interacting massive particle.

\end{description}

\section*{Acknowledgments}

We thank Albert De Roeck, Zhen Liu, David Milstead, Hideyuki Oide, and Brian Shuve for useful comments on early drafts. We thank Zach Marshall for useful discussions during the preparation of this review, and Timothy Gershon for information on LHCb results.
AS is supported by the Israel Science Foundation, the United States-Israel Binational Science Foundation, and the German-Israeli Foundation for Scientific Research and Development. CO is supported by the Swedish Research Council. LL is supported by the US Department of Energy.

\bibliography{LLPReview.bib}

\end{document}